%% file: main.tex
\documentclass[conference]{IEEEtran}
\IEEEoverridecommandlockouts
\usepackage{amsmath,amssymb}   % Math + boxempty symbol
\usepackage[caption=false,font=footnotesize]{subfig}
\usepackage{float}   % Force [H] placement
\usepackage{cite}              % IEEE citation style
\usepackage{amsmath,amssymb}   % Math
\usepackage{newtxmath}         % Better math font (italic, elegant)
\usepackage{newtxtext}         % Matching text font
\usepackage{graphicx}          % Figures
\usepackage{booktabs}          % Professional tables
\usepackage[table]{xcolor}     % Colors + table row coloring
\usepackage{colortbl}          % Cell coloring
\usepackage{pifont}            % Circle symbols (\LEFTcircle etc.)
\usepackage{hyperref}          % URLs
\usepackage{listings}          % Code listings
\usepackage{subfig}            % Subfigures
\usepackage{tikz}              % Diagrams
\usetikzlibrary{shapes.geometric, shapes.symbols, arrows.meta, positioning, fit, backgrounds, decorations.pathreplacing}
\usepackage{float}             % Figure placement
\usepackage{array}             % Table columns
\usepackage{multirow}          % Multi-row cells
\usepackage{tabularx}          % Flexible tables
\usepackage{caption}           % Captions
\usepackage{enumitem}          % Custom lists

\definecolor{QCblue}{RGB}{30, 60, 95}        % Primary: deep navy
\definecolor{QCteal}{RGB}{0, 128, 128}       % Act 2 / Neutral atoms
\definecolor{QCviolet}{RGB}{123, 104, 238}   % Act 3 / Explore
\definecolor{QCrose}{RGB}{180, 50, 100}      % Act 4 / Compare
\definecolor{QCamber}{RGB}{212, 160, 23}     % Act 1 / Nobel
\definecolor{QCgreen}{RGB}{34, 139, 34}      % Success / Neutral atoms
\definecolor{QCorange}{RGB}{230, 126, 34}    % CDN / Warning
\definecolor{QCpurple}{RGB}{142, 68, 173}    % World Labs
\definecolor{QCgray}{RGB}{245, 245, 245}     % Background
\definecolor{QCdark}{RGB}{45, 45, 45}        % Text

\definecolor{T1}{RGB}{30, 60, 95}     % Tier 1: primary navy
\definecolor{T2}{RGB}{0, 128, 128}   % Tier 2: teal
\definecolor{T3}{RGB}{123, 104, 238} % Tier 3: violet
\definecolor{T4}{RGB}{180, 50, 100}  % Tier 4: rose
\definecolor{T5}{RGB}{212, 160, 23}  % Tier 5: amber
\colorlet{BestVal}{green!15}          % Best value cell shading
\colorlet{AltRow}{gray!8}            % Alternating row background

\hypersetup{
    colorlinks=true,
    linkcolor=QCblue,
    citecolor=QCblue,
    urlcolor=QCteal,
    pdftitle={Quantum Cinema: An Interactive Cinematic Exploration of Quantum Computing Hardware via Generative World Models},
    pdfauthor={QuantBlockchain Research},
    pdfsubject={Quantum Computing Visualization, Generative AI, Scientific Communication},
    pdfkeywords={quantum computing, generative world models, scientific visualization, human-computer interaction, quantum education}
}

\newcommand{\fullcirc}{\ding{108}}    % Filled circle (large, visible)
\newcommand{\halfcirc}{\ding{119}}    % Half-filled circle (LEFTcircle from pifont)
\newcommand{\emptycirc}{\ding{109}}   % Empty circle (large, visible)
\title{Quantum Cinema: An Interactive Cinematic Exploration of Quantum Computing Hardware via Generative World Models.}

\author{
  \IEEEauthorblockN{Aoyu Zhang\textsuperscript{\dag}}
  \IEEEauthorblockA{\textit{Amazon Web Services}\\
    Beijing, China\\}
  \and
  \IEEEauthorblockN{Dongping Liu\textsuperscript{\dag}\textsuperscript{\ddag}}
  \IEEEauthorblockA{\textit{Amazon Web Services}\\
    Hong Kong, China\\}
  \and
  \IEEEauthorblockN{Luyao Zhang\textsuperscript{\dag}\textsuperscript{*}}
  \IEEEauthorblockA{\textit{Duke Kunshan University}\\
    Suzhou, China\\}
  \thanks{\textsuperscript{\dag} Authors are listed in alphabetical order by first name. \textsuperscript{*}Corresponding author: Luyao Zhang (lz183@duke.edu), Digital Innovation Research Center and Social Science Division, Duke Kunshan University. Address: Duke Avenue No.8, Kunshan, Suzhou, Jiangsu, China, 215316.  \textsuperscript{\ddag} Work done while at Amazon Web Services. Dongping Liu is currently with Tenorshare, Hong Kong, China.}
}

\begin{document}

\maketitle
\input{figs/fig_teaser}  % <-- ADD THIS LINE
% ============================================================
%  ABSTRACT
% ============================================================
\begin{abstract}
Quantum computing promises transformative advances across science and industry, yet the physical hardware that enables these computations remains invisible to the public: quantum processors operate within inaccessible laboratory infrastructure---ranging from dilution refrigerators to ultra-high-vacuum ion traps and optically controlled neutral-atom chambers---making direct observation difficult.. This ``imagination gap'' between quantum computing's growing societal impact and the public's ability to visualize it represents a significant barrier to quantum literacy and workforce development. We present \textit{Quantum Cinema}, an open-source, browser-based interactive application that closes this gap by transforming invisible quantum hardware into explorable, cinematic experiences using generative world models. Quantum Cinema guides users through a four-act narrative---from the 2025 Nobel Prize in Physics as a contemporary
quantum-hardware anchor, with the 2022 Nobel Prize-winning science of entanglement situated within the broader historical timeline, through curated video introductions to three major quantum computing architectures (trapped-ion, neutral-atom, and superconducting systems), into immersive three-dimensional generative worlds that make invisible quantum phenomena observable, and finally to interactive radar-chart comparisons grounded in real quantum device specifications. All three-dimensional environments are generated using World Labs' generative world model platform and are scientifically grounded in curated metrics from Amazon Web Services (AWS) Braket quantum hardware. Quantum Cinema requires no installation, no specialized hardware, and no quantum computing background. It is designed to serve two distinct communities: scholars and developers seeking to replicate or extend the platform, and educators, researchers, and science communicators seeking an intuitive tool for explaining quantum hardware to diverse audiences. This paper describes the system architecture, the generative world model pipeline, use cases for both communities, and directions for future work.
\end{abstract}

\begin{IEEEkeywords}
quantum computing, generative world models, scientific visualization, human-computer interaction, quantum education, AI for science, immersive visualization
\end{IEEEkeywords}

% ============================================================
%  SECTIONS
% ============================================================
\input{secs/sec1_introduction}

\input{secs/sec2_related_work}
\input{secs/sec3_system_design}
\input{secs/sec4_world_model_pipeline}
\input{secs/sec5_use_cases}
\input{secs/sec6_conclusion}

% ============================================================
%  ACKNOWLEDGMENT
% ============================================================
\section*{Acknowledgment}

The authors thank the participants and organizers of the tutorials at The Web
Conference 2026 \cite{10.1145/3774905.3793916} and IEEE ICBC 2026
\cite{guo2026blockchaininfrastructureintelligentcyberphysicalsocial}, where
\textit{Quantum Cinema} was presented as an interactive demonstration and
subsequently refined based on participant feedback. The authors also acknowledge
the open-source communities supporting Next.js, React, and Amazon Web Services,
and World Labs for the generative world-model platform used in this work.
% ============================================================
%  DATA AND CODE AVAILABILITY
% ================================================

% ============================================================
%  BIBLIOGRAPHY
% ============================================================
\bibliographystyle{IEEEtran}
\bibliography{refs/references}

% ============================================================
%  APPENDIX
% ============================================================

\appendix
\input{secs/sec8_appendix}
\input{secs/sec9_appendix_acts}  % <-- ADD THIS LINE
\input{secs/sec10_appendix_worldmodels}  % <-- ADD THIS
\end{document}

%% file: figs/fig_teaser.tex
% Teaser Figure: Three Quantum Architectures as Generative World Models
% Double-column figure with 15 screenshots, compact layout
% Color palette: teal=Ion Trap, orange=Neutral Atoms, violet=Superconducting
\begin{figure*}[t]
\centering
\footnotesize
\setlength{\tabcolsep}{1.5pt}
\setlength{\fboxsep}{0.5pt}
\setlength{\fboxrule}{0.6pt}

% --- Row 1: Ion Trap (teal) ---
\begin{tabular}{@{}>{\centering\arraybackslash}m{1.1cm}@{}>{\centering\arraybackslash}p{2.9cm}@{}>{\centering\arraybackslash}p{2.9cm}@{}>{\centering\arraybackslash}p{2.9cm}@{}>{\centering\arraybackslash}p{2.9cm}@{}>{\centering\arraybackslash}p{2.9cm}@{}}
\cellcolor{QCteal!20}{\shortstack{\textcolor{QCteal}{$\blacklozenge$}\\[-1pt]\tiny\textbf{Ion}\\[-2pt]\tiny\textbf{Trap}}}
& \fcolorbox{QCteal}{white}{\includegraphics[width=2.8cm, height=1.4cm, keepaspectratio]{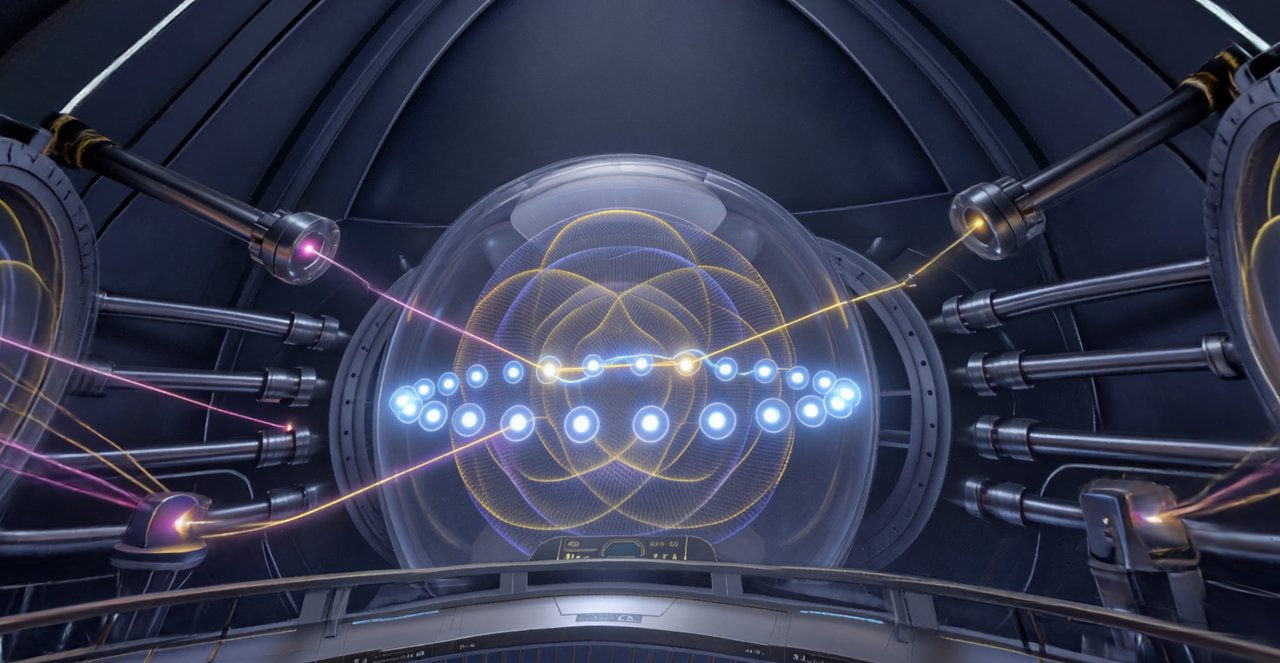}}
& \fcolorbox{QCteal}{white}{\includegraphics[width=2.8cm, height=1.4cm, keepaspectratio]{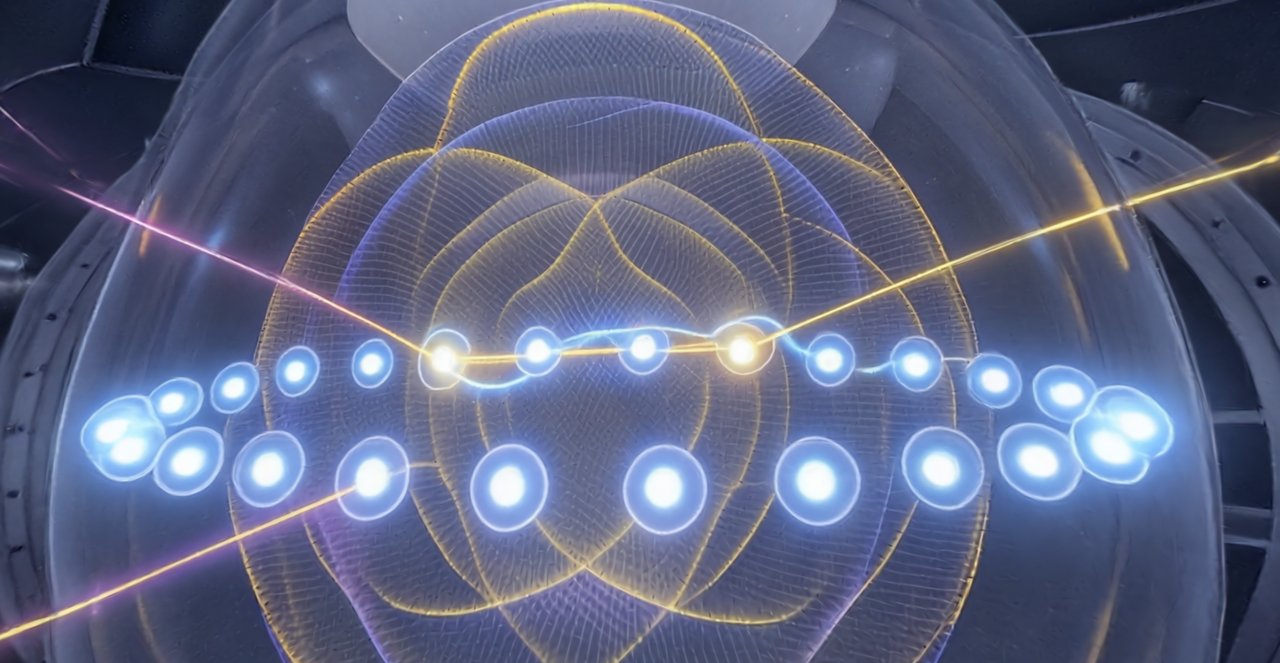}}
& \fcolorbox{QCteal}{white}{\includegraphics[width=2.8cm, height=1.4cm, keepaspectratio]{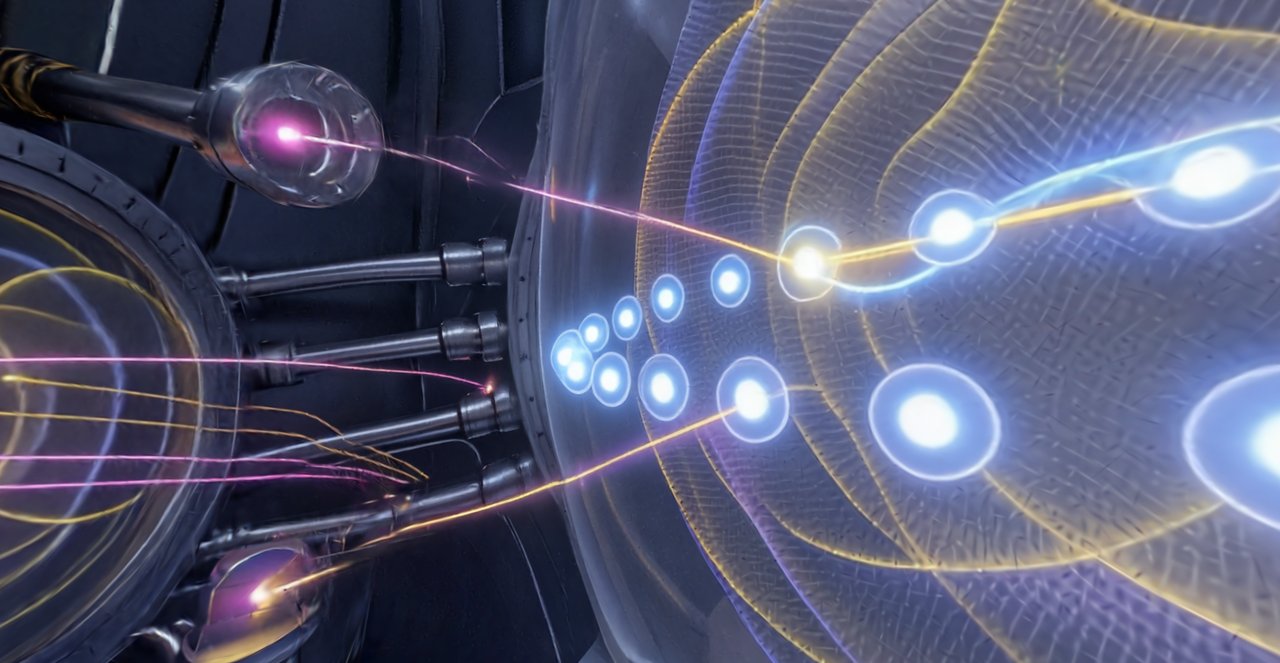}}
& \fcolorbox{QCteal}{white}{\includegraphics[width=2.8cm, height=1.4cm, keepaspectratio]{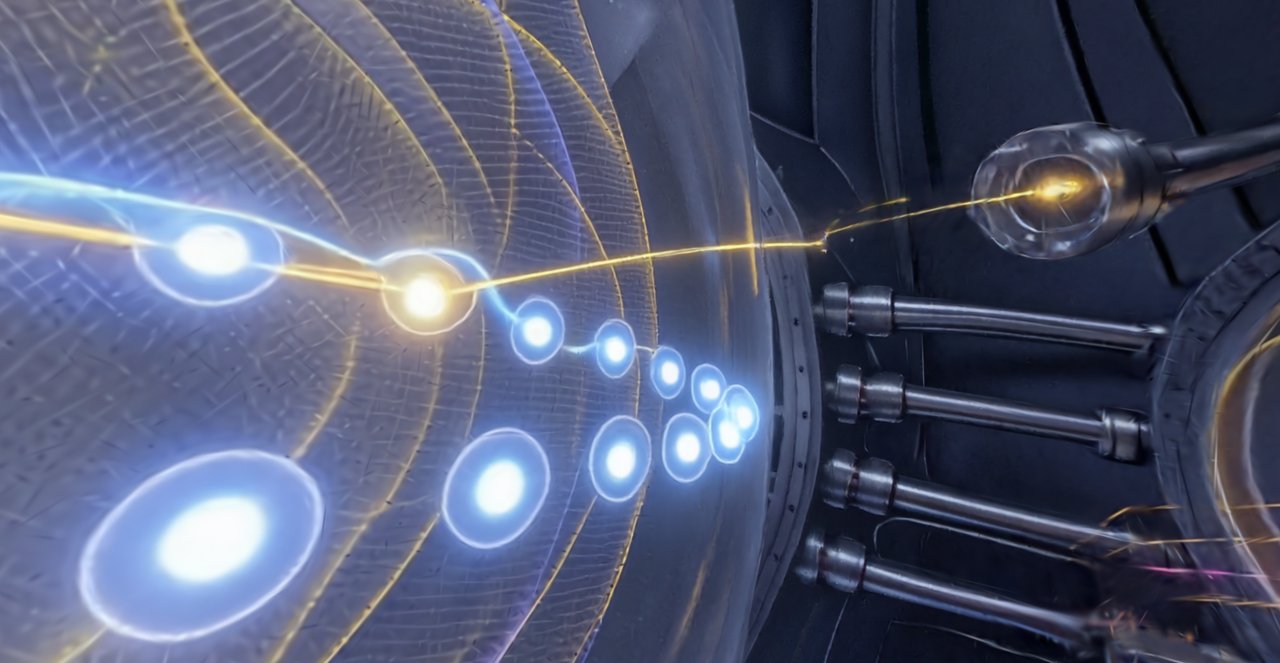}}
& \fcolorbox{QCteal}{white}{\includegraphics[width=2.8cm, height=1.4cm, keepaspectratio]{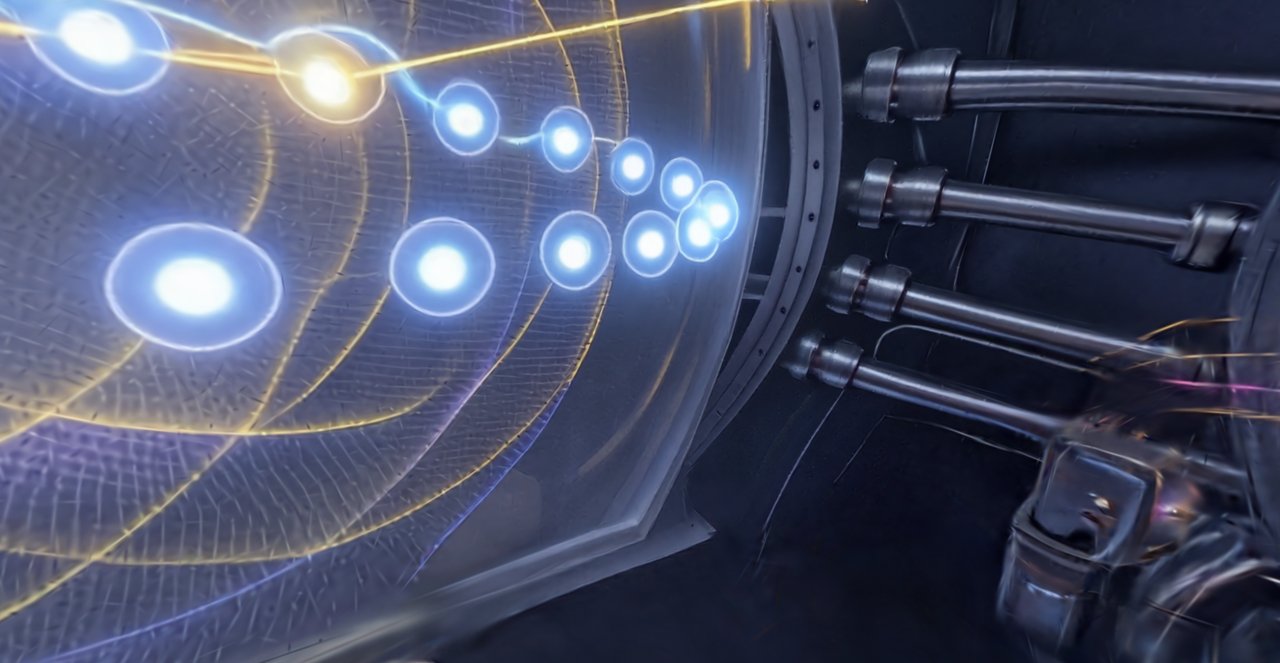}} \\[-2pt]
& \textcolor{QCteal}{\tiny$\triangleright$1} & \textcolor{QCteal}{\tiny$\triangleright$2} & \textcolor{QCteal}{\tiny$\triangleright$3} & \textcolor{QCteal}{\tiny$\triangleright$4} & \textcolor{QCteal}{\tiny$\triangleright$5}
\end{tabular}

\vspace{1pt}
{\color{QCteal}\rule{\textwidth}{0.3pt}}
\vspace{1pt}

% --- Row 2: Neutral Atoms (orange) ---
\begin{tabular}{@{}>{\centering\arraybackslash}m{1.1cm}@{}>{\centering\arraybackslash}p{2.9cm}@{}>{\centering\arraybackslash}p{2.9cm}@{}>{\centering\arraybackslash}p{2.9cm}@{}>{\centering\arraybackslash}p{2.9cm}@{}>{\centering\arraybackslash}p{2.9cm}@{}}
\cellcolor{QCorange!20}{\shortstack{\textcolor{QCorange}{$\circ$}\\[-1pt]\tiny\textbf{Neutral}\\[-2pt]\tiny\textbf{Atom}}}
& \fcolorbox{QCorange}{white}{\includegraphics[width=2.8cm, height=1.4cm, keepaspectratio]{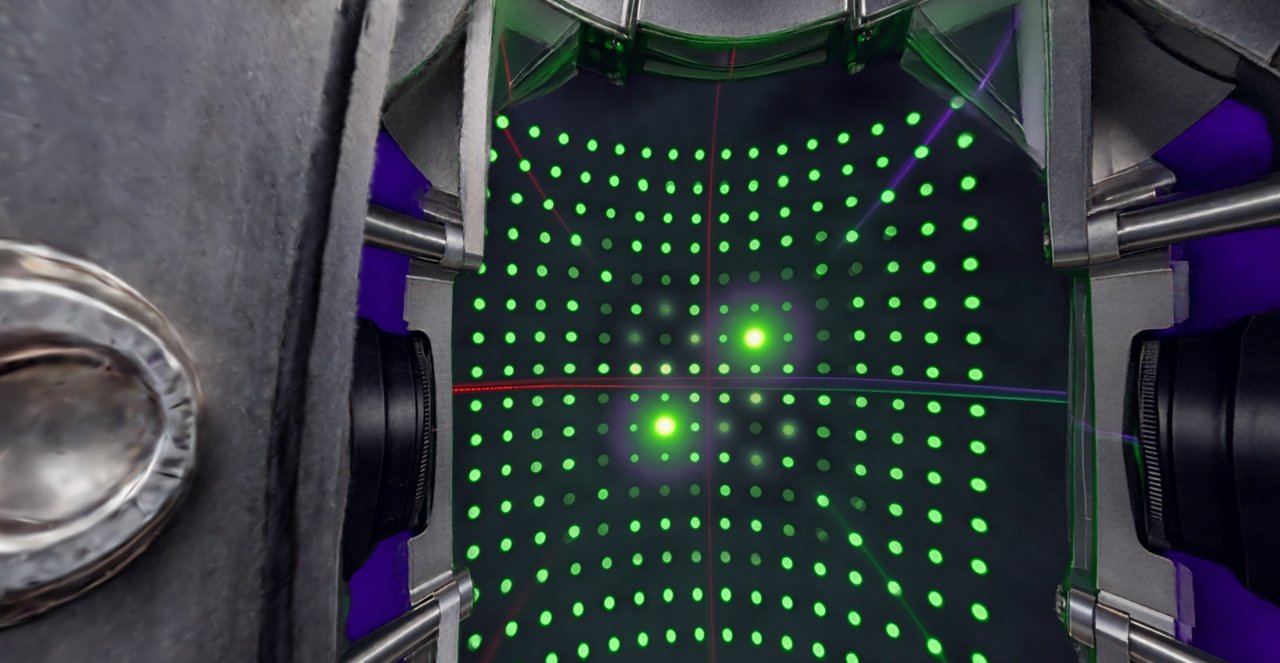}}
& \fcolorbox{QCorange}{white}{\includegraphics[width=2.8cm, height=1.4cm, keepaspectratio]{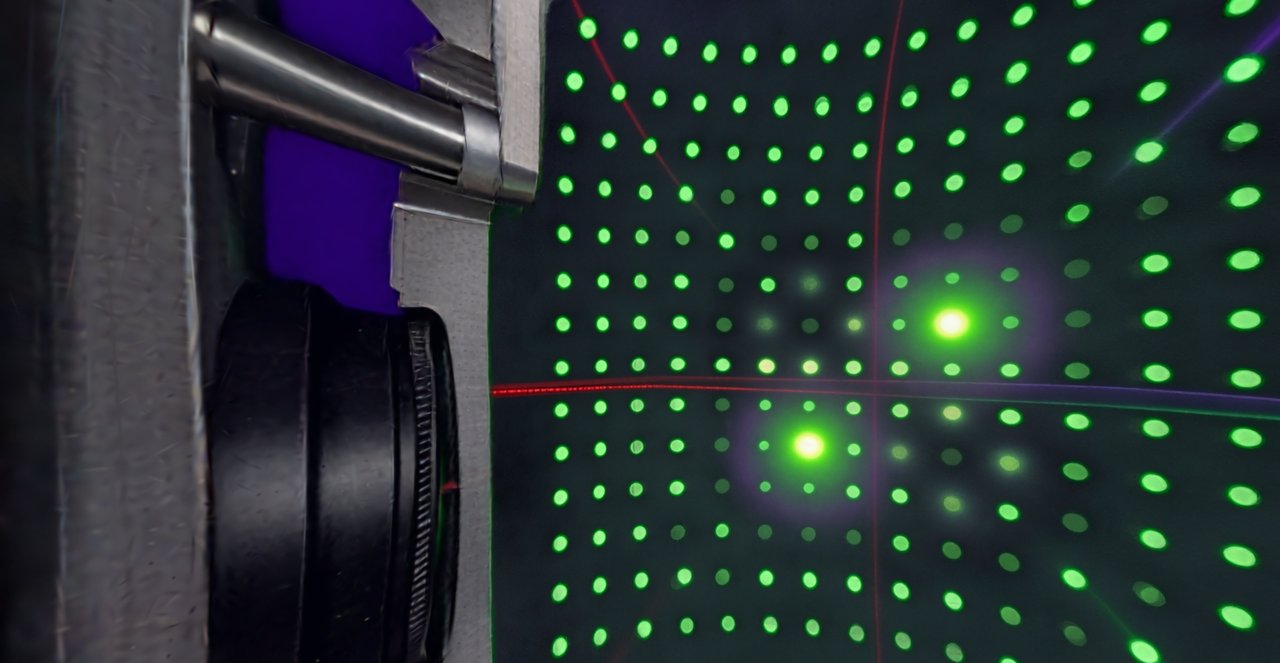}}
& \fcolorbox{QCorange}{white}{\includegraphics[width=2.8cm, height=1.4cm, keepaspectratio]{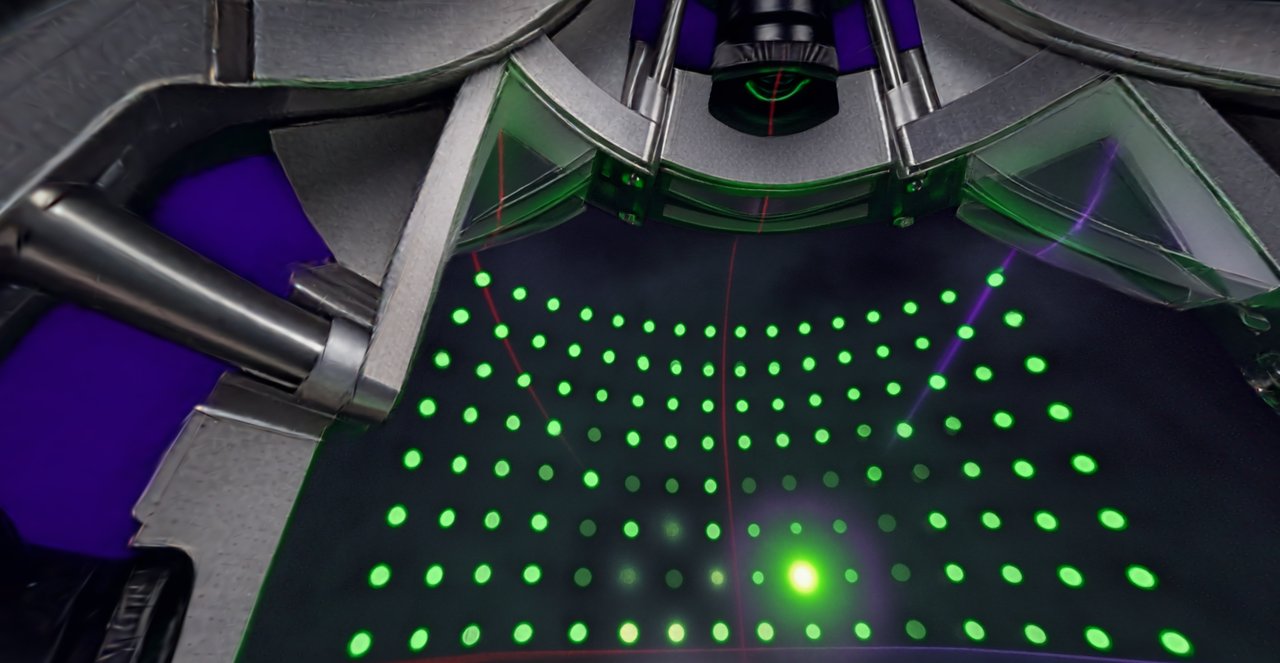}}
& \fcolorbox{QCorange}{white}{\includegraphics[width=2.8cm, height=1.4cm, keepaspectratio]{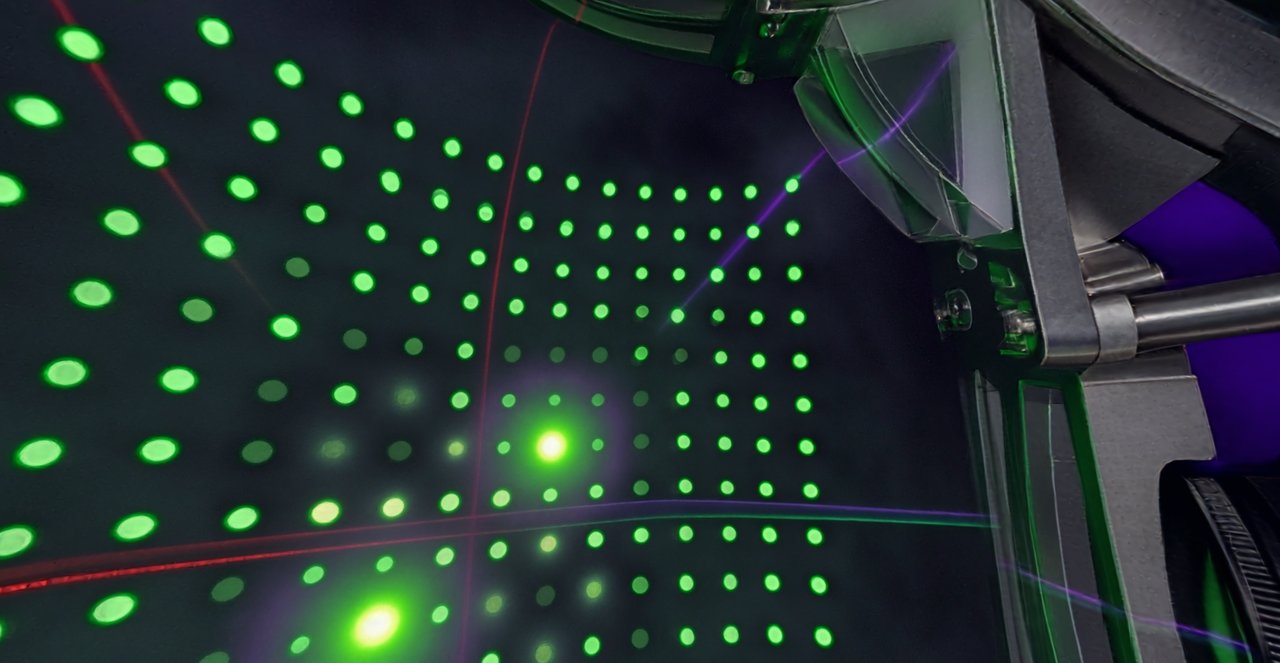}}
& \fcolorbox{QCorange}{white}{\includegraphics[width=2.8cm, height=1.4cm, keepaspectratio]{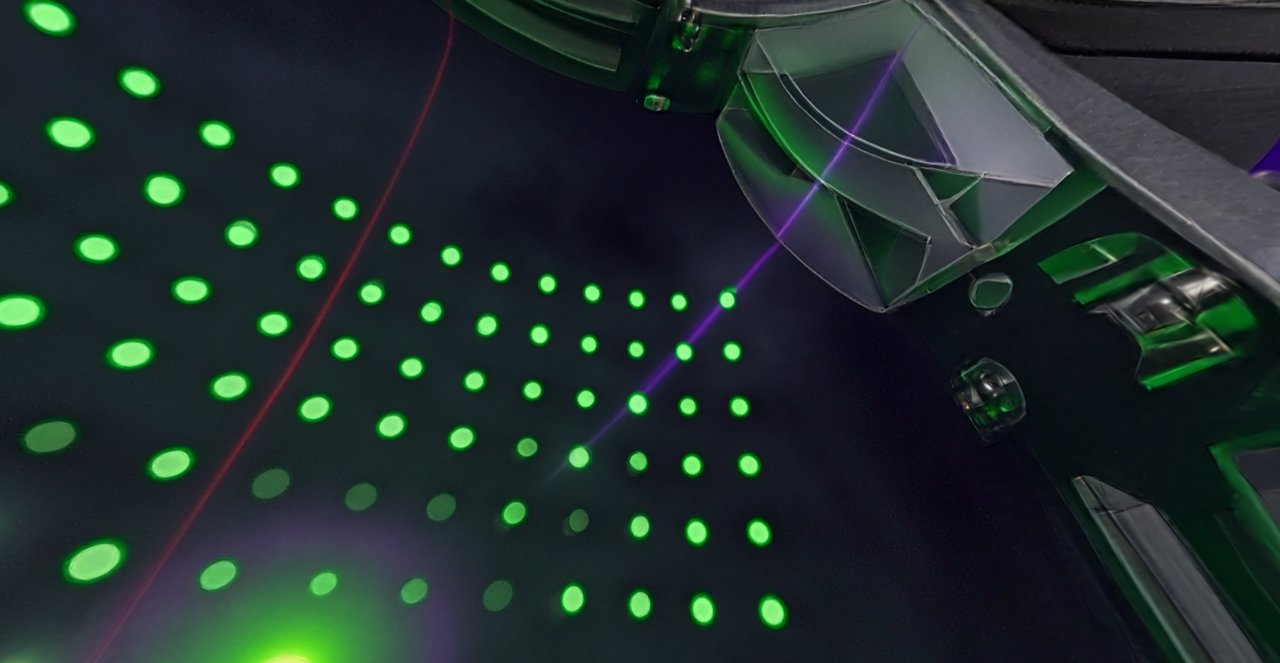}} \\[-2pt]
& \textcolor{QCorange}{\tiny$\triangleright$1} & \textcolor{QCorange}{\tiny$\triangleright$2} & \textcolor{QCorange}{\tiny$\triangleright$3} & \textcolor{QCorange}{\tiny$\triangleright$4} & \textcolor{QCorange}{\tiny$\triangleright$5}
\end{tabular}

\vspace{1pt}
{\color{QCorange}\rule{\textwidth}{0.3pt}}
\vspace{1pt}

% --- Row 3: Superconducting (violet) ---
\begin{tabular}{@{}>{\centering\arraybackslash}m{1.1cm}@{}>{\centering\arraybackslash}p{2.9cm}@{}>{\centering\arraybackslash}p{2.9cm}@{}>{\centering\arraybackslash}p{2.9cm}@{}>{\centering\arraybackslash}p{2.9cm}@{}>{\centering\arraybackslash}p{2.9cm}@{}}
\cellcolor{QCviolet!20}{\shortstack{\textcolor{QCviolet}{$\blacksquare$}\\[-1pt]\tiny\textbf{JJ}\\[-2pt]\tiny\textbf{Chip}}}
& \fcolorbox{QCviolet}{white}{\includegraphics[width=2.8cm, height=1.4cm, keepaspectratio]{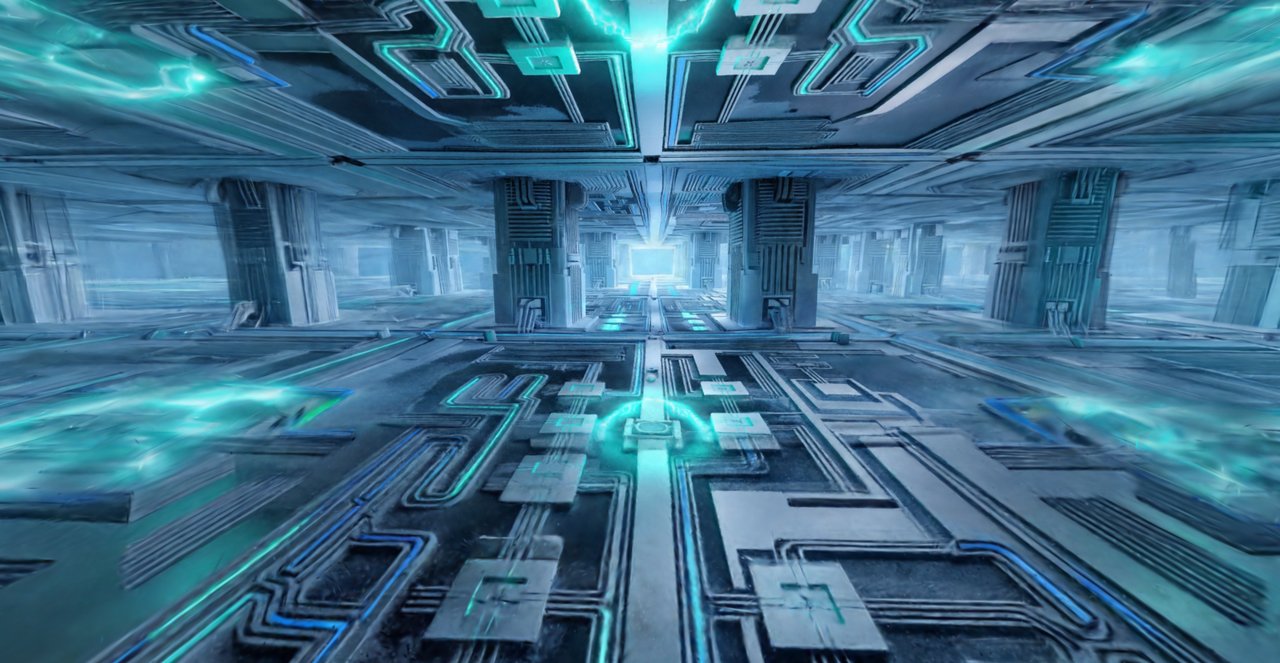}}
& \fcolorbox{QCviolet}{white}{\includegraphics[width=2.8cm, height=1.4cm, keepaspectratio]{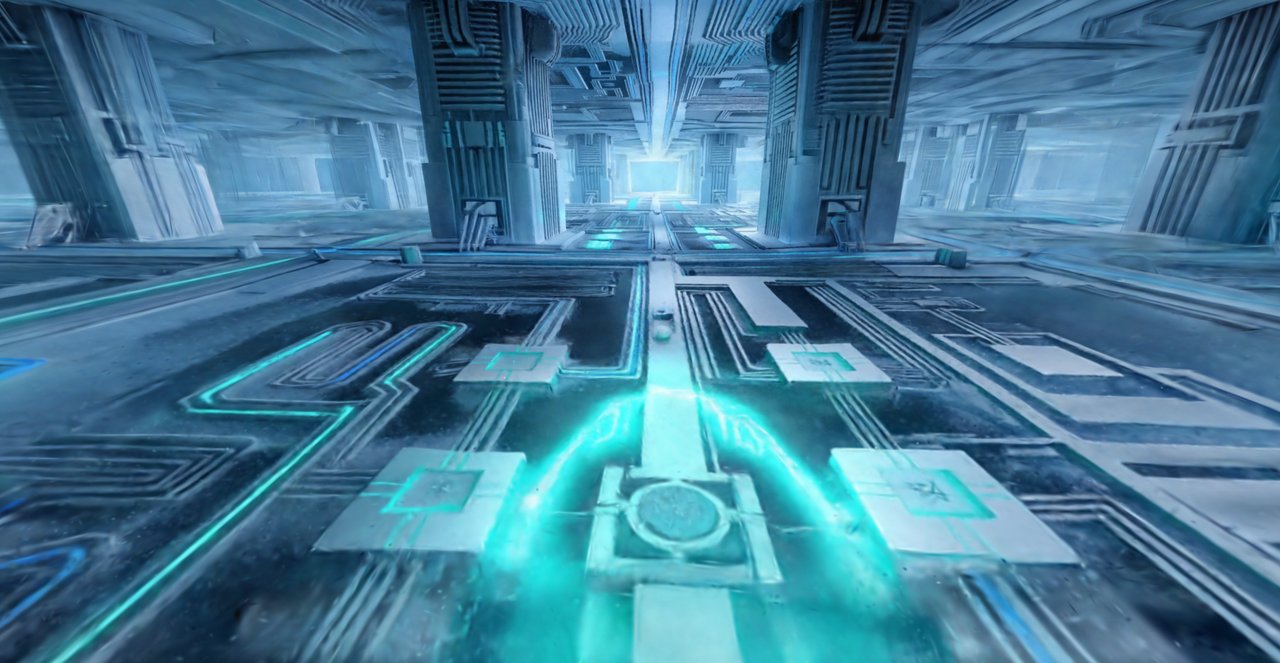}}
& \fcolorbox{QCviolet}{white}{\includegraphics[width=2.8cm, height=1.4cm, keepaspectratio]{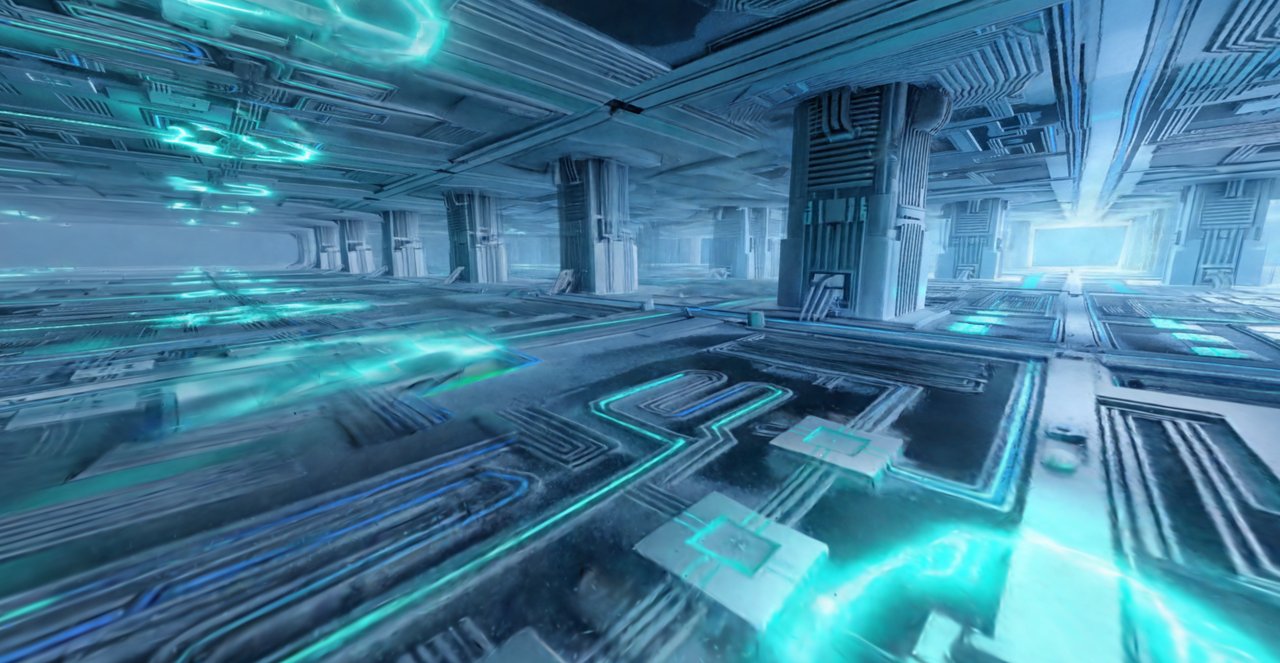}}
& \fcolorbox{QCviolet}{white}{\includegraphics[width=2.8cm, height=1.4cm, keepaspectratio]{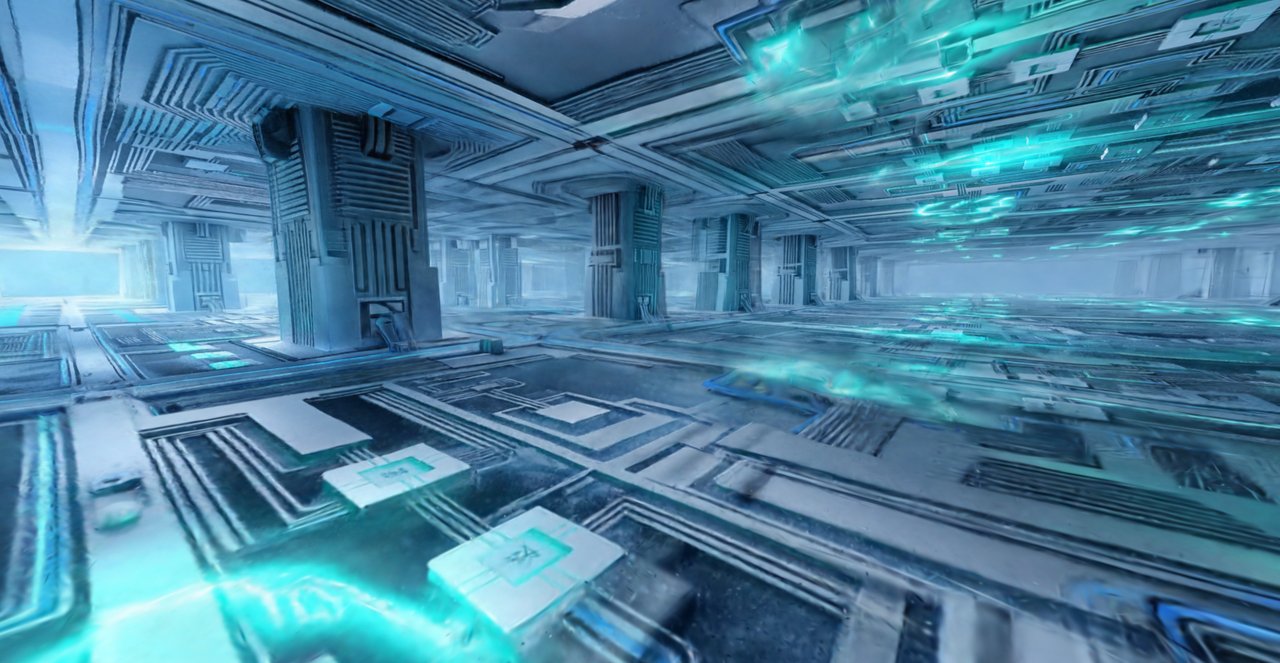}}
& \fcolorbox{QCviolet}{white}{\includegraphics[width=2.8cm, height=1.4cm, keepaspectratio]{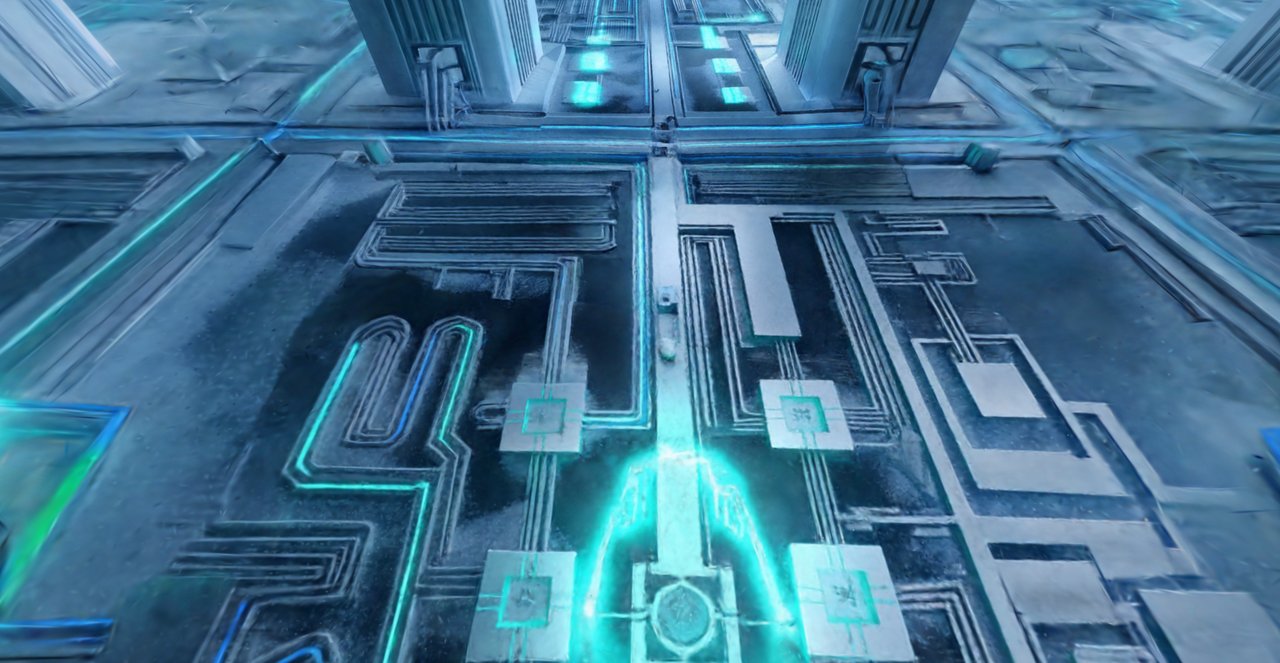}} \\[-2pt]
& \textcolor{QCviolet}{\tiny$\triangleright$1} & \textcolor{QCviolet}{\tiny$\triangleright$2} & \textcolor{QCviolet}{\tiny$\triangleright$3} & \textcolor{QCviolet}{\tiny$\triangleright$4} & \textcolor{QCviolet}{\tiny$\triangleright$5}
\end{tabular}

\caption{The three generative world models of \textit{Quantum Cinema}, each showing five navigable views. \textbf{Top:} trapped-ion (teal)---ytterbium ions in a Paul trap. \textbf{Middle:} neutral-atom (orange)---rubidium array via optical tweezers. \textbf{Bottom:} superconducting (violet)---Josephson-junction chip in a dilution refrigerator.}
\label{fig:teaser}
\end{figure*}

%% file: secs/sec1_introduction.tex
\section{Introduction}
\label{sec:introduction}

Quantum computing stands poised to transform science, industry, and society. From drug discovery and materials science to cryptography and financial modeling, the potential applications of quantum computational advantage span nearly every sector of the global economy~\cite{biamonte2017quantum}. Yet there exists a profound \textit{imagination gap}: while the 
\textit{software} layer of quantum computing---quantum circuits, algorithms, and gates---has become increasingly accessible through educational tools and cloud platforms, the \textit{hardware} itself remains fundamentally invisible to the vast majority of researchers, students, and the public. quantum processors operate within inaccessible laboratory infrastructure---ranging from dilution refrigerators to ultra-high-vacuum ion traps and optically controlled neutral-atom chambers---making direct observation difficult. The physical reality of quantum computing hardware---the golden coaxial cables, the superconducting \textit{quantum bits} (qubits) etched onto silicon chips, the layered cryogenic stages descending toward absolute zero---has remained locked behind laboratory walls and abstracted away into circuit diagrams and mathematical notation.

The scientific significance of quantum technologies has
received the highest levels of international recognition.
The 2025 Nobel Prize in Physics, awarded to John Clarke,
Michel H. Devoret, and John M. Martinis for macroscopic
quantum mechanical tunnelling and energy quantisation in
an electric circuit, provides the contemporary hardware
anchor for Quantum Cinema
\cite{nobel2025physics}. The experience situates this
milestone within the longer history of quantum
information science, including the 2022 Nobel Prize in
Physics awarded to Alain Aspect, John Clauser, and Anton
Zeilinger for foundational experiments on quantum
entanglement and Bell inequalities
\cite{nobel2022physics}. Together, these awards connect
the physical foundations of quantum hardware with the
entanglement principles that motivate quantum
computation and communication.

The intersection of artificial intelligence (AI) and quantum science represents one of the most promising frontiers in modern research. The 2024 Nobel Prize in Physics, awarded to John Hopfield and Geoffrey Hinton for foundational discoveries in neural networks and machine learning \cite{nobel2024physics}, underscored the transformative role of AI in scientific discovery. Parallel advances in quantum machine learning (QML) --- the use of quantum computers to enhance machine learning algorithms and vice versa --- have demonstrated potential advantages in areas ranging from molecular simulation to optimization \cite{biamonte2017quantum}. However, the inaccessibility of quantum hardware remains a bottleneck: even as AI techniques increasingly support the quantum-computing stack and the representation and characterization of quantum systems \cite{alexeev2025artificial,du2026artificial}, the physical reality of quantum processors remains hidden from the researchers, educators, and students who most need to understand them. This paper bridges that gap by applying generative world models --- a technology at the forefront of AI research \cite{worldlabs2024marble} --- to the specific scientific challenge of quantum hardware visualization.

For researchers outside quantum physics---including the artificial intelligence (AI) and computer science communities that this venue serves---this invisibility creates a significant barrier to engagement. Terms such as \textit{superconducting qubit}, \textit{Josephson junction}, \textit{cryogenic stage}, and \textit{quantum control electronics} remain opaque without a tangible mental model. The physical architecture of a quantum computer, from room-temperature control electronics to the mixing chamber plate at the base of the dilution refrigerator, follows a spatial logic that is difficult to convey through text or two-dimensional diagrams alone. Current visualization approaches fall into two categories, each with significant limitations. Circuit-level simulators such as Quirk~\cite{gidney2016quirk} and the IBM Quantum Experience~\cite{ibmquantum2024} provide excellent interactive environments for learning quantum logic gates and circuit construction, but they operate entirely at the abstract level of quantum information---the physical hardware that executes these circuits remains unseen. On the other end of the spectrum, immersive
virtual-reality (VR) approaches---such as the
Bloch-sphere environment studied by Zable et
al.~\cite{zable2020vr}---have demonstrated the
pedagogical potential of spatial immersion, but require
specialized headsets and additional setup. By contrast,
QuantumEyes is a two-dimensional interactive
visualization that maps multi-qubit states to a radial
``dandelion'' chart~\cite{ruan2023quantumeyes}, while the
Quantum Flytrap Virtual Lab is a browser-based simulator
for optical quantum circuits~\cite{migdal2022virtual}.
These systems improve access to quantum-state and circuit
abstractions, but do not render physical
quantum-computing hardware as an explorable spatial
environment. Broader reviews likewise identify both the
educational potential of interactive quantum tools and
the accessibility and scalability constraints associated
with headset-based immersive systems
\cite{seskir2022quantum,song2024immersive}.

We present \textit{Quantum Cinema}, the first interactive cinematic platform that leverages \textit{generative world models} to make quantum computing accessible through immersive three-dimensional (3D) narrative environments. Generative world models---AI systems that synthesize interactive, navigable 3D scenes from visual data or semantic descriptions---have emerged as a transformative technology for visual storytelling and education. These models, exemplified by systems capable of generating photorealistic, physically consistent environments from image inputs~\cite{xie2024physgaussian} and from language or image prompts~\cite{worldlabs2024marble}, enable a fundamentally new approach to scientific communication: rather than manually constructing 3D assets through traditional computer graphics pipelines, we can generate explorable worlds that are both visually compelling and scientifically informative. The application of generative AI to scientific visualization has been recognized as a frontier with vast untapped potential~\cite{basole2024genai}. \textit{Quantum Cinema} harnesses this capability to create the first interactive cinematic journey through the landscape of quantum computing---weaving together Nobel Prize-winning science, generative world-model architectures, immersive exploration, and quantitative comparison into a unified four-act narrative.

The experience is structured as a \textit{four-act narrative}. Act~I presents an interactive timeline of Nobel Prize laureates in quantum science, enabling viewers to explore the foundational discoveries that shaped the field. Act~II showcases generative video worlds for each major quantum computing architecture---superconducting circuits, trapped ions, and neutral atoms---demonstrating how world models can visualize hardware-specific physics. Act~III offers an immersive 3D environment generated from \textit{World Labs} technology, allowing viewers to freely explore a navigable quantum world. Act~IV provides a quantitative comparison across architectures through interactive radar charts, supporting evidence-based understanding of trade-offs. This narrative architecture organizes fragmented
knowledge into a coherent pedagogical journey designed
to support conceptual understanding alongside visual
appreciation.The four-act structure is described in detail in Section~\ref{sec:system-design}.

This work makes two contributions tailored to the dual
audience of the paper. First, at the methodological
level, we position generative world models as a
scientific-visualization medium between manually
authored artistic 3D content and equation-based physical
simulation. We provide a reproducible pipeline that
connects scientific literature and device documentation,
structured prompt engineering, generative world
synthesis, human curation, and browser-based integration.
The resulting worlds are scientifically informed
explanatory environments, rather than validated physical
simulations.

Second, at the practical level, we provide an open-source,
zero-install browser application and a reusable
development workflow for educators, science
communicators, quantum-computing practitioners, and
world-model researchers. The platform offers a concrete
test case for future parameter-conditioned,
time-resolved, and scientifically evaluated generative
environments. This paper reports the system design,
implementation, and qualitative demonstration; it does
not yet evaluate learning outcomes through a controlled
user study.

\textbf{Data and Code Availability}: Quantum Cinema is released under the MIT License. The complete source code, documentation, and generative world templates are available on GitHub\footnote {\url{https://github.com/QuantBlockchain/quantum-cinema}}. A permanent archived version (v1.0.0) has been deposited on Zenodo~\cite{quantumcinema2026zenodo}. The repository includes installation instructions, the full AWS deployment configuration, bilingual documentation, teaching guides for all three architectures, and templates for extending the platform to additional quantum hardware modalities. No proprietary datasets are required: all device parameters are sourced from published manufacturer specifications and the AWS Braket service documentation.

The remainder of this paper is organized as follows. Section~\ref{sec:related_work} surveys related work in quantum visualization, generative world models, and AI-driven scientific communication. Section~\ref{sec:system-design} details the Quantum Cinema architecture and four-act narrative design. Section~\ref{sec:conclusion} concludes with limitations and future directions.

%% file: secs/sec2_related_work.tex
\section{Related Work}
\label{sec:related_work}
\input{tabs/tab_related_work}
We position \textit{Quantum Cinema} at the intersection of four active research areas: quantum computing visualization, immersive quantum education, generative artificial intelligence (AI) for scientific visualization, and AI for quantum science. In what follows, we review the most relevant prior work in each area and identify the key gaps our system addresses.

% ── A. Quantum Computing Visualization Tools ──────────────────────────────
\subsection{Quantum Computing Visualization Tools}
\label{subsec:vis_tools}

The earliest and most widely adopted tools for quantum computing education operate at the \textit{circuit level}, enabling users to construct and simulate quantum circuits through graphical interfaces. Quirk, developed by Gidney at Google, remains the most popular browser-based quantum circuit simulator, offering drag-and-drop construction, real-time state-vector simulation, and support for up to 16 qubits entirely within the browser \cite{gidney2016quirk}. Its accessibility and zero-installation model have made it a staple in undergraduate quantum computing courses. Similarly, the IBM Quantum Experience provides a cloud-based platform with a visual circuit composer, allowing users to execute quantum programs on real superconducting quantum processors \cite{ibmquantum2024}. While powerful, these platforms present quantum computation primarily through abstract circuit diagrams, leaving the underlying physical hardware opaque to the learner.

Recent research has introduced novel visual encodings to improve circuit interpretability. Ruan et al.\ proposed QuantumEyes, an interactive visualization system centered on a ``dandelion chart'' that maps multi-qubit states to radial visual patterns; their design was validated through 12 expert interviews \cite{ruan2023quantumeyes}. In subsequent work, the same authors introduced VENUS, a two-dimensional (2D) geometrical representation of quantum states that generalizes the conventional Bloch sphere to multi-qubit systems \cite{ruan2023venus}. Complementary efforts have focused on pedagogical notation: Norrie et al.\ developed QNotation, a visual notation translator that bridges formal Dirac notation with intuitive graphical representations for novice learners \cite{norrie2024qnotation}. For domain-specific education, Jordon et al.\ created QWalkVis, an interactive visualization tool for quantum walks designed to teach stochastic quantum processes \cite{jordon2023qwalkvis}. In the optical domain, Quantum Flytrap's Virtual Lab offers a no-code, drag-and-drop simulator for optical quantum circuits supporting up to three entangled photons \cite{migdal2022virtual}.

Despite these advances, \textit{all} existing circuit-level tools share a common limitation: they visualize quantum computation through abstract symbolic representations rather than rendering the physical hardware itself as an explorable spatial environment.

% ── B. Immersive and Interactive Quantum Education ────────────────────────
\subsection{Immersive and Interactive Quantum Education}
\label{subsec:immersive_edu}

A growing body of work has explored immersive technologies to improve quantum concept comprehension. Zable and Velloso conducted the first controlled study comparing virtual reality (VR) and desktop interfaces for quantum education, using Bloch sphere tutorials to demonstrate that VR can significantly improve spatial understanding of single-qubit states \cite{zable2020vr}. More recently, Karunathilaka et al.\ presented Intuit at ACM CHI 2025, an augmented reality (AR) system that explains quantum concepts through everyday analogies projected into the user's physical environment \cite{karunathilaka2025intuit}. Quantum Flytrap's Virtual Lab also contributes in this space by providing web-based interactive quantum simulation accessible without specialized hardware \cite{migdal2022virtual}. Song et al.\ conducted a systematic review of extended reality (XR) in quantum education, finding that while immersive modalities show promise for conceptual learning, adoption remains limited by hardware cost, setup complexity, and scalability concerns \cite{song2024immersive}.

These findings reveal a critical tension: VR and AR approaches require specialized headsets or equipment that limit accessibility, while fully web-based immersive experiences---which could reach the broadest audience---remain underexplored in quantum education.

% ── C. Generative AI for Scientific Visualization ─────────────────────────
\subsection{Generative AI for Scientific Visualization}
\label{subsec:genai_viz}

Generative world models represent a paradigm shift in how complex environments can be synthesized from natural language or structural descriptions. In scientific visualization, this capability enables the automatic generation of explorable three-dimensional (3D) scenes from high-level specifications. Xie et al.\ introduced PhysGaussian at CVPR 2024, integrating physics simulation with 3D Gaussian splatting to produce dynamic, physically grounded environments; the work has since accumulated over 470 citations \cite{xie2024physgaussian}, underscoring the community's interest in generative 3D content. Basole and Major proposed a comprehensive framework for integrating generative AI into scientific visualization pipelines, identifying data-to-scene translation as a key challenge \cite{basole2024genai}. In industry, World Labs' Marble platform---founded by Fei-Fei Li---demonstrates that generative 3D world models can produce consistent, navigable environments from single images or text prompts \cite{worldlabs2024marble}. Concurrently, a comprehensive survey by Zhu et al.\ examined whether video diffusion models such as Sora can function as world simulators, concluding that while limitations remain, these models exhibit emerging capabilities for physical reasoning and environment generation \cite{zhu2024sora}.

\textit{To our knowledge, generative world models have not yet been applied to quantum hardware visualization, leaving a significant opportunity unexplored.}

% ── D. AI for Quantum Science ─────────────────────────────────────────────
\subsection{AI for Quantum Science}
\label{subsec:ai_quantum}

The convergence of AI and quantum science has emerged as a major research direction with applications spanning simulation, optimization, and discovery. Biamonte et al.\ provided a comprehensive survey of quantum machine learning (QML), establishing the theoretical foundations and identifying near-term opportunities on noisy intermediate-scale quantum (NISQ) devices \cite{biamonte2017quantum}. On the algorithmic front, variational quantum eigensolvers (VQE) and the quantum approximate optimization algorithm (QAOA) have become flagship approaches for applying quantum computers to real-world problems in chemistry and combinatorial optimization \cite{cerezo2021variational}. Complementing these algorithmic advances, Carleo and Troyer demonstrated that neural networks can represent quantum many-body states with remarkable accuracy, introducing the paradigm of neural quantum states for simulating quantum systems that would be intractable for classical methods \cite{carleo2017solving}.

While these approaches use AI to \textit{advance} quantum science, few works use AI to \textit{make quantum science accessible} to broader audiences. Quantum Cinema occupies this unique position at the intersection of AI-driven visualization and quantum education, applying generative world models to bridge the accessibility gap identified across all three research areas above.

While much of the existing literature uses AI to advance quantum-system design, simulation, characterization, and computation, complementary systems in this broader research program explore public-facing and programming-oriented
interfaces to quantum science. QSignAI combines AI-mediated interaction with quantum-randomness-seeded identity
signatures \cite{qsignai2025}; Quantum Futures Interactive integrates quantum-risk education, participatory technology prioritization, and infrastructure
tradeoff exploration \cite{quantumfutures2026}; and Quantum Circuit Vision evaluates visual AI agents for quantum code generation under explicit cost constraints \cite{liu2026quantumcircuitvisioncostaware}. Quantum Cinema occupies a distinct position within this landscape by applying generative world models specifically to the physical hardware layer of quantum computing, transforming otherwise inaccessible architectures into browser-based, explorable environments for broad audiences.
% ── Summary Comparison ────────────────────────────────────────────────────

Table \ref{tab:related_work} summarizes the capabilities of existing tools across eight key dimensions. No prior system simultaneously supports circuit-level accuracy, hardware environment visualization, full interactivity, web-based delivery, generative AI content creation, AI-for-quantum-science framing, and no-code accessibility. To address these gaps, we present \textit{Quantum Cinema}, a unified platform that combines generative world models with quantum circuit simulation to produce interactive cinematic walkthroughs of quantum computing hardware, accessible from any modern web browser without installation or specialized equipment.

%% file: tabs/tab_related_work.tex
% Related Work Comparison Table — Quantum Cinema IEEEtran Paper
% Full-width double-column table with unified color palette
\begin{table*}[!htbp]
\centering
\small
\setlength{\tabcolsep}{0pt}
\renewcommand{\arraystretch}{1.2}
\caption{Comparison of Quantum Education and Visualization Platforms}
\label{tab:related_work}
\begin{tabular}{@{}>{\raggedright\arraybackslash}p{3.5cm}@{\hspace{4pt}}>{\raggedright\arraybackslash}p{2.2cm}@{\hspace{4pt}}>{\centering\arraybackslash}p{1.75cm}@{\hspace{4pt}}>{\centering\arraybackslash}p{1.95cm}@{\hspace{4pt}}>{\centering\arraybackslash}p{1.95cm}@{\hspace{4pt}}>{\centering\arraybackslash}p{1.55cm}@{\hspace{4pt}}>{\centering\arraybackslash}p{1.7cm}@{\hspace{4pt}}>{\centering\arraybackslash}p{1.95cm}@{}}
\toprule
\rowcolor{QCblue!12}
\textbf{Tool} & \textbf{Venue} & \cellcolor{QCteal!18} & \cellcolor{QCviolet!18} & \cellcolor{QCorange!18} & \cellcolor{QCblue!18} & \cellcolor{QCpurple!18} & \cellcolor{QCgreen!18} \\
\rowcolor{QCblue!12}
& & \cellcolor{QCteal!18}\textcolor{QCteal}{$\infty$}\;Circuit & \cellcolor{QCviolet!18}\textcolor{QCviolet}{$\blacksquare$}\;Hardware & \cellcolor{QCorange!18}\textcolor{QCorange}{$\circlearrowleft$}\;Interact. & \cellcolor{QCblue!18}\textcolor{QCblue}{$\mathbf{W}$}\;Web & \cellcolor{QCpurple!18}\textcolor{QCpurple}{$\bigstar$}\;GenAI & \cellcolor{QCgreen!18}\textcolor{QCgreen}{$\triangleright$}\;No\;Install \\
\midrule
Quirk~\cite{gidney2016quirk} & Web'16 & \textcolor{QCgreen}{\fullcirc} & \textcolor{QCrose}{\emptycirc} & \textcolor{QCgreen}{\fullcirc} & \textcolor{QCgreen}{\fullcirc} & \textcolor{QCrose}{\emptycirc} & \textcolor{QCgreen}{\fullcirc} \\
\rowcolor{AltRow}
IBM~Q~Exp.~\cite{ibmquantum2024} & IBM'24 & \textcolor{QCgreen}{\fullcirc} & \textcolor{QCrose}{\emptycirc} & \textcolor{QCgreen}{\fullcirc} & \textcolor{QCgreen}{\fullcirc} & \textcolor{QCrose}{\emptycirc} & \textcolor{QCgreen}{\fullcirc} \\
QuantumEyes~\cite{ruan2023quantumeyes} & TVCG'23 & \textcolor{QCgreen}{\fullcirc} & \textcolor{QCrose}{\emptycirc} & \textcolor{QCgreen}{\fullcirc} & \textcolor{QCrose}{\emptycirc} & \textcolor{QCrose}{\emptycirc} & \textcolor{QCrose}{\emptycirc} \\
\rowcolor{AltRow}
VENUS~\cite{ruan2023venus} & EuroVis'23 & \textcolor{QCgreen}{\fullcirc} & \textcolor{QCrose}{\emptycirc} & \textcolor{QCgreen}{\fullcirc} & \textcolor{QCrose}{\emptycirc} & \textcolor{QCrose}{\emptycirc} & \textcolor{QCrose}{\emptycirc} \\
QNotation~\cite{norrie2024qnotation} & QCE'24 & \textcolor{QCgreen}{\fullcirc} & \textcolor{QCrose}{\emptycirc} & \textcolor{QCgreen}{\fullcirc} & \textcolor{QCgreen}{\fullcirc} & \textcolor{QCrose}{\emptycirc} & \textcolor{QCgreen}{\fullcirc} \\
\rowcolor{AltRow}
QWalkVis~\cite{jordon2023qwalkvis} & QCE'23 & \textcolor{QCgreen}{\fullcirc} & \textcolor{QCrose}{\emptycirc} & \textcolor{QCgreen}{\fullcirc} & \textcolor{QCgreen}{\fullcirc} & \textcolor{QCrose}{\emptycirc} & \textcolor{QCgreen}{\fullcirc} \\
Virtual~Lab~\cite{migdal2022virtual} & SPIE'22 & \textcolor{QCgreen}{\fullcirc} & \textcolor{QCrose}{\emptycirc} & \textcolor{QCgreen}{\fullcirc} & \textcolor{QCgreen}{\fullcirc} & \textcolor{QCrose}{\emptycirc} & \textcolor{QCgreen}{\fullcirc} \\
\rowcolor{AltRow}
Intuit~\cite{karunathilaka2025intuit} & CHI'25 & \textcolor{QCamber}{\halfcirc} & \textcolor{QCrose}{\emptycirc} & \textcolor{QCgreen}{\fullcirc} & \textcolor{QCrose}{\emptycirc} & \textcolor{QCrose}{\emptycirc} & \textcolor{QCrose}{\emptycirc} \\
VR~Quantum~\cite{zable2020vr} & VRST'20 & \textcolor{QCamber}{\halfcirc} & \textcolor{QCrose}{\emptycirc} & \textcolor{QCgreen}{\fullcirc} & \textcolor{QCrose}{\emptycirc} & \textcolor{QCrose}{\emptycirc} & \textcolor{QCrose}{\emptycirc} \\
\rowcolor{AltRow}
Black~Opal~\cite{qctrl2024blackopal} & 2024 & \textcolor{QCgreen}{\fullcirc} & \textcolor{QCrose}{\emptycirc} & \textcolor{QCgreen}{\fullcirc} & \textcolor{QCgreen}{\fullcirc} & \textcolor{QCrose}{\emptycirc} & \textcolor{QCgreen}{\fullcirc} \\
\midrule
\rowcolor{QCblue!10}
\textbf{Quantum Cinema} & \textbf{2025} & \textbf{\textcolor{QCgreen}{\fullcirc}} & \textbf{\textcolor{QCgreen}{\fullcirc}} & \textbf{\textcolor{QCgreen}{\fullcirc}} & \textbf{\textcolor{QCgreen}{\fullcirc}} & \textbf{\textcolor{QCgreen}{\fullcirc}} & \textbf{\textcolor{QCgreen}{\fullcirc}} \\
\bottomrule
\end{tabular}

\vspace{6pt}
\begin{flushleft}
\footnotesize
\textbf{Legend.}~\textcolor{QCgreen}{\fullcirc}~=~full support;~\textcolor{QCamber}{\halfcirc}~=~partial support;~\textcolor{QCrose}{\emptycirc}~=~not supported.\\[2pt]
\textbf{Categories:}~{\textcolor{QCteal}{$\infty$}~\textbf{Circuit}}~--~visualization of quantum circuit diagrams and gate-level operations;~{\textcolor{QCviolet}{$\blacksquare$}~\textbf{Hardware}}~--~rendering of physical quantum processor architectures as spatial environments;~{\textcolor{QCorange}{$\circlearrowleft$}~\textbf{Interactive}}~--~user manipulation and real-time feedback;~{\textcolor{QCblue}{$\mathbf{W}$}~\textbf{Web}}~--~browser-based delivery without native application installation;~{\textcolor{QCpurple}{$\bigstar$}~\textbf{GenAI}}~--~use of generative artificial intelligence (world models, neural rendering) for content creation;~{\textcolor{QCgreen}{$\triangleright$}~\textbf{No~Install}}~--~immediate accessibility without setup, registration, or specialized hardware. Quantum Cinema is the first platform to offer all six capabilities simultaneously.
\end{flushleft}
\end{table*}

%% file: secs/sec3_system_design.tex
\section{System Design and Architecture}
\label{sec:system-design}

This section presents the end-to-end architecture of \textit{Quantum Cinema}, an interactive web application that combines generative world models with cinematic storytelling to explain quantum computing hardware. We describe the cloud deployment stack (Section~\ref{subsec:architecture}), the four-act narrative structure that guides users through the experience (Section~\ref{subsec:narrative}), and the three quantum architectures featured in the system (Section~\ref{subsec:architectures}). Detailed walkthroughs of each act, including annotated screenshots, are provided in Appendix~\ref{app:four-acts}.

\subsection{System Architecture}
\label{subsec:architecture}

\textit{Quantum Cinema} is built as a single-page application (SPA) using Next.js~16 with React~19, authored entirely in TypeScript~\cite{nextjs2024,react2024}. This stack provides server-side rendering, automatic code splitting, and a component-based architecture that supports both cinematic scroll-driven animations and interactive 3D world embedding within a unified codebase.

The application is deployed on Amazon Web Services (AWS)~\cite{aws2024} following a three-tier cloud architecture optimized for global content delivery and automatic scaling. As illustrated in Fig.~\ref{fig:architecture}, user requests first reach an AWS CloudFront Content Delivery Network (CDN) edge location, which serves cached static assets and forwards dynamic requests to an Application Load Balancer (ALB). The ALB distributes traffic across tasks running in AWS Elastic Container Service (ECS) Fargate, a serverless compute engine that eliminates the need to manage underlying virtual machine infrastructure.

\input{figs/fig1_architecture}

Each container runs the Next.js standalone build on Node.js~20 Alpine Linux with a non-root user for security hardening. The service auto-scales between one and four tasks based on 70\,\% CPU utilization, with circuit-breaker rollback to maintain availability during deployment updates. All static assets---including pre-rendered videos, Nobel laureate photographs, and architecture diagrams---are baked directly into the container image at build time. This \textit{static-first} design eliminates runtime dependencies on object storage, databases, or live quantum cloud services.

The immersive 3D environments stream from World Labs~\cite{worldlabs2024marble} via public Universal Resource Locator (URL) embeddings, allowing generative world content to render directly from the provider's infrastructure without intermediate processing. Table~\ref{tab:system-specs} summarizes the deployment parameters.

\input{tabs/tab1_system_specs}

\subsection{Four-Act Narrative Design}
\label{subsec:narrative}

The user experience follows a four-act narrative that mirrors cinematic storytelling conventions while progressively building technical understanding. Each act occupies a distinct section of the scroll-driven SPA, with smooth transitions and consistent visual theming. Fig.~\ref{fig:four-act-flow} depicts the overall flow, and Table~\ref{tab:four-act} provides the structural breakdown. Appendix~\ref{app:four-acts} presents a detailed walkthrough of each act, including annotated screenshots and pedagogical rationale.

\input{figs/fig2_four_act_flow}
\input{tabs/tab2_four_act}

\textit{Act~I---Nobel Prize} (Section~\ref{app:act1}) establishes \textit{why} quantum computing matters. Users encounter an interactive horizontal timeline centered on the 2025 Nobel Prize in Physics, with historical context connecting the laureates' contributions to the broader arc of quantum research. This creates an emotional and historical anchor for the technical content that follows.

\textit{Act~II---World Models} (Section~\ref{app:act2}) introduces the three quantum hardware architectures through curated video content. Users scroll through vertically stacked architecture cards, each containing a short looping video and a concise description of the underlying physical mechanism. At the conclusion of this act, users select one architecture to explore in depth---a choice that parameterizes the remainder of the experience.

\textit{Act~III---Explore} (Section~\ref{app:act3}) constitutes the immersive centerpiece. Upon selecting an architecture, the user enters a generative three-dimensional (3D) world representing that quantum hardware platform. These environments are AI-generated interactive scenes from World Labs~\cite{worldlabs2024marble}, not physics-based simulations. However, each world is grounded in real device parameters---cryostat geometry, vacuum chamber dimensions, laser cooling apparatus---to ensure visual fidelity and educational value. Users can orbit, zoom, and pan within the scene while annotated hotspots explain individual hardware components. Representative views of all three 3D worlds are shown in Figure~\ref{fig:teaser} of the main text.

\textit{Act~IV---Compare} (Section~\ref{app:act4}) provides an interactive
comparison across quantum architectures, grouped by computational
paradigm. Users view animated radar charts across six quantitative
metrics: coherence time, two-qubit gate fidelity, readout fidelity,
error rate, connectivity topology, and qubit count. Gate-based
devices---trapped-ion and superconducting processors---are compared
directly, whereas the neutral-atom analog Hamiltonian-simulation
device is presented separately using platform-native sequence-level
and per-atom equivalents. This grouping avoids placing gate-model
and analog hardware on a single misleading yardstick while
transforming the qualitative impressions gathered during exploration
into technically grounded comparison.

\subsection{Three Quantum Computing Architectures}
\label{subsec:architectures}

\textit{Quantum Cinema} features three leading quantum computing architectures, chosen to represent distinct physical qubit implementations with contrasting engineering trade-offs. Table~\ref{tab:architecture-comparison} presents the comparative overview, and the generative world models for each are shown in Figure~\ref{fig:teaser}. The device-level metrics reported below are representative,
dated snapshots curated from Amazon Braket device
documentation and official specifications for IonQ Aria,
Rigetti Ankaa-3, and QuEra Aquila
\cite{awsbraket2024,ionq2024,rigetti2024,quera2024}.
Architecture-level interpretation is grounded in peer-reviewed
studies of trapped-ion, neutral-atom, and superconducting
platforms
\cite{bruzewicz2019,bluvstein2024logical,krantz2019quantum}.
The frozen values used by the paper and application are
maintained in
\texttt{quantum-cinema/src/lib/data.ts}.

\input{tabs/tab3_architecture_comparison}

\textbf{Trapped-Ion} quantum computers suspend charged atoms (ions) in electromagnetic fields within an ultra-high vacuum chamber~\cite{bruzewicz2019}. Lasers tuned to specific wavelengths manipulate individual ions to perform quantum gate operations. Because all ions share a common trapping potential, trapped-ion systems offer native all-to-all connectivity and exhibit long coherence times, often exceeding one second~\cite{ionq2024}.

\textbf{Neutral Atom} systems use focused laser beams, or optical
tweezers, to arrange atoms in programmable
two-dimensional arrays \cite{bluvstein2024logical,quera2024}.
By exciting atoms to Rydberg states, engineers create
controllable many-body interactions. In QuEra Aquila,
these interactions implement analog Hamiltonian
simulation through continuous programmable evolution,
rather than a sequence of discrete two-qubit gates.

\textbf{Superconducting} quantum processors fabricate electrical circuits containing Josephson junctions and cool them to millikelvin temperatures inside dilution refrigerators~\cite{krantz2019quantum}. Microwave pulses manipulate the quantum state of each circuit element. Within the three-device snapshot used in Quantum Cinema,
QuEra Aquila has the largest addressable register, with
256 programmable atoms. Rigetti Ankaa-3 represents the
superconducting gate-based architecture, offering fast gate
operations and mature fabrication while requiring
millikelvin cryogenic infrastructure and exhibiting
substantially shorter coherence times than trapped-ion
systems \cite{rigetti2024,krantz2019quantum}.

%% file: figs/fig1_architecture.tex
% Figure 1: System Architecture of Quantum Cinema
% Full cloud deployment stack with semantic tier icons and unique-feature highlights
\begin{figure}[t]
\centering
\resizebox{\columnwidth}{!}{%
\begin{tikzpicture}[
    font=\small,
    box/.style={draw, rounded corners=8pt, minimum height=1.0cm, align=center, thick, inner sep=6pt},
    userbox/.style={draw, ellipse, minimum width=1.3cm, minimum height=0.9cm, align=center, thick, inner sep=4pt},
    dashbox/.style={draw, rounded corners=6pt, thick, dashed, inner sep=5pt},
    notebox/.style={draw, rounded corners=6pt, thick, inner sep=5pt},
    tlabel/.style={font=\tiny, text=white, circle, minimum size=0.35cm, inner sep=0pt},
    arrow/.style={->, >=Stealth, thick, shorten >=2pt, shorten <=2pt},
    label/.style={font=\scriptsize\itshape, text=QCdark},
]

% --- Tier icon row: colored circles with initials ---
\node[tlabel, fill=gray] at (0, 2.4) {\textbf{U}};
\node[tlabel, fill=QCorange] at (2.3, 2.4) {\textbf{C}};
\node[tlabel, fill=QCgreen] at (4.4, 2.4) {\textbf{L}};
\node[tlabel, fill=QCblue] at (7.0, 2.4) {\textbf{A}};
\node[tlabel, fill=QCpurple] at (10.0, 2.4) {\textbf{W}};

% --- Main horizontal pipeline ---
\node[userbox, fill=QCgray] (user) at (0, 0.8) {\textbf{User}};
\node[label, below=0.05cm of user] {\footnotesize Browser};

\node[box, fill=QCorange!12, draw=QCorange, minimum width=1.9cm] (cdn) at (2.3, 0.8) {\textbf{CloudFront}\\[-1pt]\footnotesize CDN};

\node[box, fill=QCgreen!12, draw=QCgreen, minimum width=1.4cm] (alb) at (4.4, 0.8) {\textbf{ALB}};

\node[box, fill=QCblue!8, draw=QCblue, minimum width=2.6cm] (ecs) at (7.0, 0.8) {\textbf{ECS Fargate}\\[-1pt]\footnotesize Next.js 16 SPA};

\node[box, fill=QCpurple!8, draw=QCpurple, minimum width=2.4cm] (wl) at (10.0, 0.8) {\textbf{World Labs}\\[-1pt]\footnotesize marble.wlabs.ai};

% --- Static assets (above ECS) ---
\node[dashbox, fill=gray!10, draw=gray!60, above=0.7cm of ecs] (assets) {\footnotesize\textit{Static Assets:} Videos + Images};

% --- Arrows: main pipeline ---
% Protocol labels ABOVE arrows, higher than all shape tops
\draw[arrow] (user) -- (cdn) node[pos=0.5, above=0.5cm, label] {\footnotesize HTTPS};
\draw[arrow, dashed] (cdn) -- (alb) node[pos=0.5, above=0.5cm, label] {\footnotesize secret};
\draw[arrow] (alb) -- (ecs) node[pos=0.5, above=0.5cm, label] {\footnotesize port 3000};
% iframe label BELOW arrow, lower than all shape bottoms
\draw[arrow] (ecs) -- (wl) node[pos=0.5, below=0.5cm, label] {\footnotesize iframe};

% --- Bidirectional flow: World Labs returns 3D stream ---
\draw[arrow, gray!40, ->, dashed] (wl.south) to[bend left=25] node[midway, below=2pt, label, text=gray] {\footnotesize 3D stream} (ecs.south);

% --- Arrow: assets to ECS ---
\draw[arrow, gray!50, ->] (assets.south) -- (ecs.north);

% --- Highlight: Unique features ---
\node[notebox, fill=QCamber!10, draw=QCamber, rounded corners=10pt, below=1.0cm of ecs, xshift=1.5cm] (unique) {\footnotesize\textbf{No Database} $\cdot$ \textbf{No Live QPU} $\cdot$ \textbf{Static-First}};

% --- Highlight: Security mechanism ---
\node[notebox, fill=red!5, draw=red!40, below=0.8cm of cdn, xshift=-0.5cm] (sec) {\footnotesize\textit{Missing secret $\rightarrow$ 403 Forbidden}};
\draw[->, thick, red!40] (cdn.south) -- (sec.north);

% --- Subtle braces grouping ECS and World Labs ---
\draw[thick, QCblue!30, decorate, decoration={brace, amplitude=4pt, raise=2pt}] (ecs.north west) -- (ecs.north east);
\draw[thick, QCpurple!30, decorate, decoration={brace, amplitude=4pt, raise=2pt}] (wl.north west) -- (wl.north east);

\end{tikzpicture}%
}
\caption{System Architecture of Quantum Cinema. The static-first design requires no database and places no live quantum processing unit (QPU) in the request path. All content is baked into the container at build time; 3D worlds stream from World Labs via public URL embedding. Requests without the shared CDN secret header are rejected at the edge.}
\label{fig:architecture}
\end{figure}
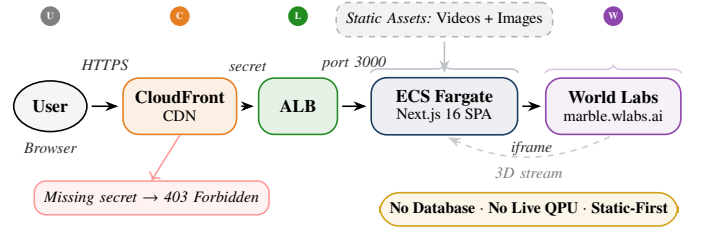

%% file: tabs/tab1_system_specs.tex
% System Specifications Table — Quantum Cinema IEEEtran Paper
% Single-column table with semantic category icons and color grouping
\begin{table}[!htbp]
\centering
\small
\setlength{\tabcolsep}{5pt}
\caption{System Specifications of Quantum Cinema}
\label{tab:system-specs}
\begin{tabular}{@{}>{\centering\arraybackslash}p{1.0cm}@{\hspace{4pt}}>{\raggedright\arraybackslash}p{3.0cm}@{\hspace{6pt}}>{\raggedright\arraybackslash}p{4.8cm}@{}}
\toprule
\rowcolor{QCblue!12}
{\textbf{Layer}} & {\textbf{Component}} & {\textbf{Technology}} \\
\midrule
\multirow{2}{*}{\shortstack{\textcolor{QCblue}{\Large$\langle\rangle$}\\[-2pt]\tiny\textbf{Frontend}}}
  & Web Framework & Next.js~16 / React~19 \\
  & Language & TypeScript \\
\midrule
\multirow{3}{*}{\shortstack{\textcolor{QCorange}{\Large$\bigstar$}\\[-2pt]\tiny\textbf{Cloud}}}
  & CDN & CloudFront \\
  & Load Balancer & ALB \\
  & Compute & ECS~Fargate (1--4 tasks, 70\%~CPU) \\
\midrule
\multirow{2}{*}{\shortstack{\textcolor{QCgreen}{\Large$\square$}\\[-2pt]\tiny\textbf{Runtime}}}
  & Container & Node.js~20 Alpine (non-root) \\
  & 3D Platform & World Labs (\url{marble.worldlabs.ai}) \\
\midrule
\multirow{2}{*}{\shortstack{\textcolor{QCamber}{\Large$\varnothing$}\\[-2pt]\tiny\textbf{Unique}}}
  & Database & \textbf{None} (static-by-design) \\
  & QPU in Path & \textbf{None} (no live quantum hardware) \\
\bottomrule
\end{tabular}

\vspace{4pt}
\begin{flushleft}
\footnotesize
\textit{Note.} The static-first architecture eliminates all runtime dependencies on databases, quantum processing units~(QPUs), and external APIs. All content is baked into the container image at build time. Icons denote architectural layers: \textcolor{QCblue}{$\langle\rangle$}~Frontend, \textcolor{QCorange}{$\bigstar$}~Cloud, \textcolor{QCgreen}{$\square$}~Runtime, \textcolor{QCamber}{$\varnothing$}~Unique~(none by design).
\end{flushleft}
\end{table}

%% file: figs/fig2_four_act_flow.tex
% Figure 2: Four-Act Narrative Flow of Quantum Cinema
% Horizontal user experience flow with semantic icons and content keywords
\begin{figure*}[t]
\centering
\resizebox{\textwidth}{!}{%
\begin{tikzpicture}[
    font=\footnotesize,
    actbox/.style={draw, rounded corners=10pt, minimum width=3.0cm, minimum height=2.2cm, align=center, thick, inner sep=4pt},
    badge/.style={font=\tiny\bfseries, text=white, circle, minimum size=0.5cm, inner sep=0pt},
    arrow/.style={->, >=Stealth, line width=2pt, QCdark},
]

% --- Act 1: Nobel Prize (amber) ---
\node[actbox, fill=QCamber!15, draw=QCamber] (act1) at (-5.5,0) {};
\node[badge, fill=QCamber] at (act1.north west) {1};
\node[font=\normalsize\bfseries, text=QCamber!80!black] at ([yshift=0.4cm]act1.center) {Nobel Prize};
\node[font=\tiny, text=QCdark, align=center] at ([yshift=-0.3cm]act1.center) {\textcolor{QCamber}{$\odot$}~2025 Laureate profiles\\The Quantum Timeline\\};
\node[font=\tiny\itshape, text=gray, below=0.05cm of act1] {Context};

% Arrow 1->2
\draw[arrow] (act1.east) -- ++(1.0,0) node[midway, above, font=\tiny\bfseries] {Why};

% --- Act 2: World Models (teal) ---
\node[actbox, fill=QCteal!15, draw=QCteal, right=1.0cm of act1] (act2) {};
\node[badge, fill=QCteal] at (act2.north west) {2};
\node[font=\normalsize\bfseries, text=QCteal!80!black] at ([yshift=0.4cm]act2.center) {World Models};
\node[font=\tiny, text=QCdark, align=center] at ([yshift=-0.3cm]act2.center) {\textcolor{QCteal}{$\blacklozenge$}~Ion trap~~~~\textcolor{QCgreen}{$\circ$}~Atom\\$\blacksquare$~Superconducting\\Video introductions};
\node[font=\tiny\itshape, text=gray, below=0.05cm of act2] {Concepts};

% Arrow 2->3
\draw[arrow] (act2.east) -- ++(1.0,0) node[midway, above, font=\tiny\bfseries] {What};

% --- Act 3: Explore (violet) ---
\node[actbox, fill=QCviolet!15, draw=QCviolet, right=1.0cm of act2] (act3) {};
\node[badge, fill=QCviolet] at (act3.north west) {3};
\node[font=\normalsize\bfseries, text=QCviolet!80!black] at ([yshift=0.4cm]act3.center) {Explore};
\node[font=\tiny, text=QCdark, align=center] at ([yshift=-0.3cm]act3.center) {\textcolor{QCviolet}{$\bigstar$}~Generative 3D worlds\\World Labs immersion\\Navigate + discover};
\node[font=\tiny\itshape, text=gray, below=0.05cm of act3] {Experience};

% Arrow 3->4
\draw[arrow] (act3.east) -- ++(1.0,0) node[midway, above, font=\tiny\bfseries] {How};

% --- Act 4: Compare (rose) ---
\node[actbox, fill=QCrose!15, draw=QCrose, right=1.0cm of act3] (act4) {};
\node[badge, fill=QCrose] at (act4.north west) {4};
\node[font=\normalsize\bfseries, text=QCrose!80!black] at ([yshift=0.4cm]act4.center) {Compare};
\node[font=\tiny, text=QCdark, align=center] at ([yshift=-0.3cm]act4.center) {\textcolor{QCrose}{$\bowtie$}~Radar charts\\6-metric paradigm-aware comparison\\Use-case matching};
\node[font=\tiny\itshape, text=gray, below=0.05cm of act4] {Decide};

% --- Bottom narrative arc ---
\node[font=\tiny\itshape, text=QCdark, below=0.4cm of act2.south east] {(a) Historical context $\rightarrow$ (b) Physical concepts $\rightarrow$ (c) Immersive exploration $\rightarrow$ (d) Informed selection};

\end{tikzpicture}%
}
\caption{The Four-Act Narrative Flow of Quantum Cinema. Each act is numbered, color-coded, and annotated with its pedagogical role and key content. Arrows are labeled with the cognitive transition they enable.}
\label{fig:four-act-flow}
\end{figure*}
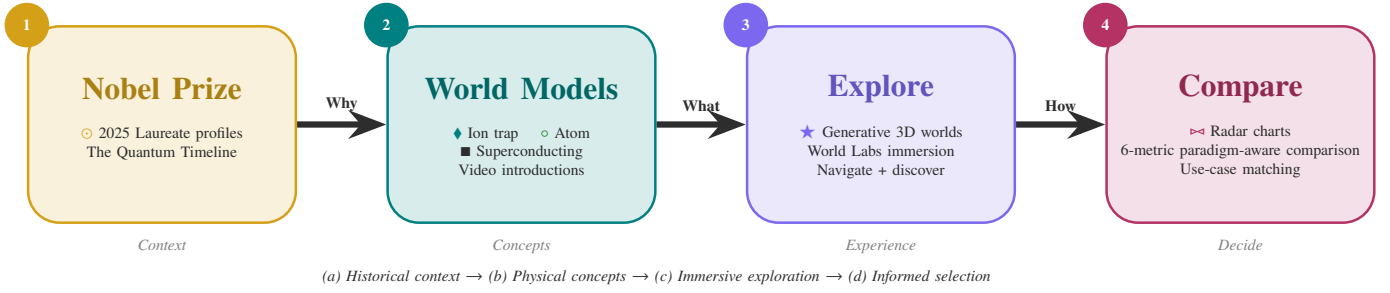

%% file: tabs/tab2_four_act.tex
% Four-Act Narrative Structure Table — Quantum Cinema IEEEtran Paper
% Wide double-column table with color-coded acts and semantic icons
\begin{table*}[t]
\centering
\small
\setlength{\tabcolsep}{8pt}
\renewcommand{\arraystretch}{1.3}
\caption{The Four-Act Narrative Structure of Quantum Cinema}
\label{tab:four-act}
\begin{tabular}{@{}>{\centering\arraybackslash}p{0.8cm}@{\hspace{6pt}}>{\raggedright\arraybackslash}p{2.6cm}@{\hspace{8pt}}>{\raggedright\arraybackslash}p{3.6cm}@{\hspace{8pt}}>{\raggedright\arraybackslash}p{3.0cm}@{\hspace{8pt}}>{\raggedright\arraybackslash}p{5.8cm}@{}}
\toprule
\rowcolor{QCblue!10}
{\textbf{Act}} & {\textbf{Name}} & {\textbf{Purpose}} & {\textbf{Component}} & {\textbf{Key Content}} \\
\midrule
{\cellcolor{QCamber!20}\textcolor{QCamber!80!black}{\textbf{1}}} & \textcolor{QCamber!80!black}{$\odot$}~\textbf{Nobel Prize} & Establish why quantum matters & NobelPrizeStep & 2025 Nobel Prize in Physics with interactive history timeline of quantum research \\
{\cellcolor{QCteal!20}\textcolor{QCteal}{\textbf{2}}} & \textcolor{QCteal}{$\blacklozenge$}~\textbf{World Models} & Introduce architectures & VideoShowcaseStep & Curated videos per architecture: ion trap, neutral atom, superconducting \\
{\cellcolor{QCviolet!20}\textcolor{QCviolet}{\textbf{3}}} & \textcolor{QCviolet}{$\bigstar$}~\textbf{Explore} & Immersive 3D experience & WorldModelStep & Generative world (World Labs): navigable 3D environment with scientific annotations \\
{\cellcolor{QCrose!20}\textcolor{QCrose}{\textbf{4}}} & \textcolor{QCrose}{$\bowtie$}~\textbf{Compare} & Hardware comparison & ComparisonStep & Radar charts grouped by paradigm
(6 metrics: coherence, 2-qubit fidelity, readout fidelity,
error rate, connectivity, and qubit count)
+ use-case matching \\
\bottomrule
\end{tabular}

\vspace{6pt}
\begin{flushleft}
\footnotesize
\textit{Note.} Each act is color-coded and icon-tagged to match Figure~\ref{fig:four-act-flow}. The narrative follows a ``why $\rightarrow$ what $\rightarrow$ how $\rightarrow$ which'' cognitive progression: Act~I motivates through the 2025 Nobel Prize and historical context, Act~II introduces physical concepts through video, Act~III enables embodied learning through immersive 3D exploration, and Act~IV supports decision-making through quantitative comparison. The Component column names the React component implementing each act in the source code.
\end{flushleft}
\end{table*}

%% file: tabs/tab3_architecture_comparison.tex
% Three Quantum Architectures Comparison Table — Quantum Cinema IEEEtran Paper
% Updated to distinguish:
% (1) computational model: gate-based vs. analog Hamiltonian simulation
% (2) physical architecture: trapped-ion, superconducting, neutral-atom
%
% Required packages:
% \usepackage{booktabs}
% \usepackage{array}
% \usepackage[table]{xcolor}
% \usepackage{amssymb}

\begin{table*}[t]
\centering
\footnotesize
\setlength{\tabcolsep}{6pt}
\renewcommand{\arraystretch}{1.22}

\caption{Comparison of Quantum Computing Architectures and Computational Models in Quantum Cinema}
\label{tab:architecture-comparison}

\begin{tabular}{
@{}
>{\raggedright\arraybackslash}p{2.8cm}
@{\hspace{6pt}}
>{\centering\arraybackslash}p{4.0cm}
@{\hspace{6pt}}
>{\centering\arraybackslash}p{4.0cm}
@{\hspace{6pt}}
>{\centering\arraybackslash}p{4.0cm}
@{}
}
\toprule

% ------------------------------------------------------------------
% Computational-model grouping
% ------------------------------------------------------------------
\rowcolor{QCblue!10}
\textbf{Computational model}
&
\multicolumn{2}{c}{
  \cellcolor{QCblue!14}
  \textbf{Gate-Based Quantum Processors}
}
&
\cellcolor{QCorange!18}
\textbf{Analog Hamiltonian Simulation (AHS)}
\\

\cmidrule(lr){2-3}
\cmidrule(lr){4-4}

% ------------------------------------------------------------------
% Physical architectures
% ------------------------------------------------------------------
\rowcolor{QCblue!10}
\textbf{Physical architecture}
&
\cellcolor{QCteal!18}
\textcolor{QCteal}{$\blacklozenge$}~
\textbf{Trapped-Ion}
&
\cellcolor{QCviolet!18}
\textcolor{QCviolet}{$\blacksquare$}~
\textbf{Superconducting}
&
\cellcolor{QCorange!18}
\textcolor{QCorange}{$\circ$}~
\textbf{Neutral Atoms}
\\

% ------------------------------------------------------------------
% Representative devices
% ------------------------------------------------------------------
\rowcolor{QCblue!10}
\textbf{Representative device}
&
\cellcolor{QCteal!18}
\textbf{IonQ Aria}
&
\cellcolor{QCviolet!18}
\textbf{Rigetti Ankaa-3}
&
\cellcolor{QCorange!18}
\textbf{QuEra Aquila}
\\

\midrule

Execution mechanism
&
Discrete quantum gates and circuits
&
Discrete quantum gates and circuits
&
Continuous programmable Hamiltonian evolution
\\

\rowcolor{AltRow}
Coherence / evolution time
&
1--10~s
&
20--100~$\mu$s
&
1--10~$\mu$s\textsuperscript{a}
\\

2-Qubit / sequence fidelity
&
99.5\%
&
99.0\%
&
97--99\%\textsuperscript{b}
\\

\rowcolor{AltRow}
Readout fidelity
&
$\sim$99.7\%
&
$\sim$97--99\%
&
$\sim$99\% per atom
\\

Error rate
&
$\sim$0.5\%
&
$\sim$1\%
&
$\sim$1--3\%
\\

\rowcolor{AltRow}
Connectivity
&
All-to-all
&
Nearest-neighbor
&
Programmable geometry
\\

Physical qubits / atoms
&
25
&
84
&
256
\\

\rowcolor{AltRow}
Operating temperature
&
Room temperature (vacuum)
&
10--15~mK
&
Room temperature (vacuum)
\\

Key visualized phenomenon
&
Linear ion chain with Raman-laser addressing
&
Josephson-junction chip in a dilution refrigerator
&
Programmable atom array controlled by optical tweezers
\\

\bottomrule
\end{tabular}

\vspace{5pt}

\begin{minipage}{0.99\textwidth}
\footnotesize
\textit{Note.}
The table distinguishes the \emph{physical architecture} of
each platform from its \emph{computational model}. IonQ Aria
and Rigetti Ankaa-3 are gate-based quantum processors that
execute discrete quantum circuits. QuEra Aquila is a
neutral-atom quantum processor implementing analog
Hamiltonian simulation through continuous programmable
quantum evolution. The computational-model labels apply to
the representative devices shown here; neutral-atom technology
as a broader hardware family is not necessarily restricted to
analog Hamiltonian simulation.

\textit{Note.}
Values are representative, dated device snapshots rather than
live calibration measurements. IonQ Aria values are curated
from Amazon Braket and IonQ specifications
\cite{awsbraket2024,ionq2024}; Rigetti Ankaa-3 values are
curated from Amazon Braket and Rigetti specifications
\cite{awsbraket2024,rigetti2024}; and QuEra Aquila values are
curated from Amazon Braket and QuEra documentation
\cite{awsbraket2024,quera2024}. Architecture-level context is
drawn from peer-reviewed studies of trapped-ion,
neutral-atom, and superconducting platforms
\cite{bruzewicz2019,bluvstein2024logical,krantz2019quantum}.

For QuEra Aquila, the reported microsecond value denotes a
platform-native analog-Hamiltonian-simulation evolution or
sequence window rather than a gate-model $T_2$ measurement.
Because Aquila does not expose discrete two-qubit gates, its
reported fidelity is a platform-native sequence-level
equivalent and is not directly interchangeable with
two-qubit gate fidelity on gate-based devices.

The frozen device snapshot used throughout the paper and
application is maintained in
\texttt{quantum-cinema/src/lib/data.ts}. The frozen device snapshot used by the paper and
application is maintained in
\texttt{quantum-cinema/src/lib/data.ts}. It supplies the
raw values reported in Table~IV and used by the
architecture-comparison interface. No cell is designated as universally
``best,'' because suitability depends on the computational
model, target workload, and dashboard scoring convention.

The visual encodings are used consistently throughout the
paper:
\textcolor{QCteal}{$\blacklozenge$}~trapped-ion,
\textcolor{QCviolet}{$\blacksquare$}~superconducting, and
\textcolor{QCorange}{$\circ$}~neutral-atom.
\end{minipage}

\end{table*}

%% file: secs/sec4_world_model_pipeline.tex
\section{Generative World Model Pipeline}
\label{sec:world_model_pipeline}

This section describes the pipeline for creating the 3D immersive environments that form the experiential core of Quantum Cinema. We detail the five-step world creation methodology, discuss the scientific accuracy of generative visualizations, and explain how developers can extend the platform with new quantum architectures.

\subsection{World Creation Methodology}
\label{subsec:world_creation}

Each 3D world in Quantum Cinema is created through a five-step pipeline that transforms scientific specifications into navigable, photorealistic environments. The pipeline bridges quantum hardware documentation and generative 3D scene synthesis, enabling rapid prototyping of educational environments without manual 3D modeling.

\input{figs/fig3_world_pipeline}  % Pipeline diagram

\textbf{Step 1 -- Scientific Concept Extraction.}
For each quantum architecture, we identify key physical phenomena and structural details from peer-reviewed literature and Amazon Web Services (AWS) Braket device specifications \cite{awsbraket2024}. For example, the trapped-ion world is grounded in the physical description of a \textit{linear chain of ytterbium ions} confined in a \textit{Paul trap} (an oscillating electromagnetic field configuration that confines charged particles) and addressed by \textit{Raman laser beams} (lasers tuned to induce stimulated Raman transitions between atomic energy levels, enabling qubit operations). Similarly, the superconducting world captures the visual character of dilution refrigerators housing quantum processors built from \textit{Josephson junctions} (superconducting devices consisting of two superconducting electrodes separated by a thin insulating barrier, serving as the fundamental qubit element).

\textbf{Step 2 -- Prompt Engineering.}
We craft detailed text prompts that balance scientific accuracy with visual storytelling. Each prompt incorporates three elements: (1) the physical layout of the hardware (e.g., chandelier structure of a superconducting processor, hexagonal lattice of neutral atoms), (2) salient visual features that distinguish the architecture (e.g., gold-plated coaxial lines, vacuum chamber windows), and (3) reference device photographs from published hardware teardowns and manufacturer documentation to ensure structural fidelity. The prompt for the trapped-ion world, for instance, specifies ``a linear chain of ytterbium ions suspended in a vacuum chamber, illuminated by intersecting Raman laser beams, with gold-plated electrodes of a Paul trap visible along the axis.''

\textbf{Step 3 -- Generative World Synthesis.}
The engineered prompts are submitted to the World Labs Marble platform \cite{worldlabs2024marble}, a generative 3D world creation system that produces persistent, navigable environments from text and image inputs. The resulting environments are fully explorable via keyboard and mouse, with spatial audio and dynamic lighting, creating an embodied sense of presence within quantum hardware facilities.

\textbf{Step 4 -- Curated Refinement.}
Generated worlds are iteratively refined using the World Labs Chisel editor, an interactive curation tool that allows authors to adjust camera angles, lighting conditions, material properties, and spatial composition. This step ensures that the environments accurately reflect hardware topology -- for example, verifying that the superconducting world conveys the vertical ``chandelier'' hierarchy of control electronics above the cryostat, or that the neutral-atom world correctly depicts the two-dimensional array of traps created by \textit{optical tweezers} (focused laser beams that trap and manipulate individual atoms) and the spatial patterns induced by the \textit{Rydberg blockade} (a phenomenon where excitation of one atom to a Rydberg state shifts the energy levels of nearby atoms, preventing simultaneous excitation within a critical radius).

\textbf{Step 5 -- Integration.}
The refined world is exported as a public URL via the World Labs viewer and embedded directly into the React frontend component. The viewer handles all rendering, navigation, and event propagation, requiring only a single iframe or WebView integration point. This architecture decouples world creation from application development, allowing pedagogical content to be authored independently of the 3D pipeline.

\subsection{Scientific Accuracy of Generative Worlds}
\label{subsec:scientific_accuracy}

It is essential to emphasize that the 3D worlds in Quantum Cinema are \textit{generative visualizations} -- AI-generated scenes informed by quantum-hardware
documentation and curated device parameters, not exact physical simulations. They make otherwise invisible quantum phenomena (decoherence, laser cooling, energy loss during gate operations) observable as visual narrative, but should not be interpreted as precise physical models. Their intended pedagogical value lies in supporting more
concrete explanatory models of hardware structure and
operating principles, rather than in providing
computational fidelity to quantum-mechanical equations.

To maintain a meaningful connection to real hardware, we
curate six device metrics from Amazon Braket documentation
and official manufacturer specifications
\cite{awsbraket2024,ionq2024,rigetti2024,quera2024}:
coherence or platform-native evolution time, two-qubit or
sequence-level fidelity, readout fidelity, error rate,
connectivity topology, and physical qubit or atom count.

For gate-based devices, these quantities correspond to
standard calibration metrics. For QuEra Aquila, an analog
Hamiltonian-simulation device, the temporal and fidelity
axes use explicitly labelled platform-native equivalents
rather than a gate-model $T_2$ measurement or a discrete
two-qubit-gate fidelity. These metrics anchor the
architecture comparison in quantitative device
characteristics while avoiding a misleading direct
equivalence between gate-based and analog platforms.

The frozen raw values used by the paper and application
are maintained in
\texttt{quantum-cinema/src/lib/data.ts}.

\input{tabs/tab4_generative_prompts}  % Table showing prompts and outputs

Table~\ref{tab:generative_prompts} presents representative prompts and the corresponding visual outputs for each quantum architecture, illustrating how physical specifications are translated into generative scene descriptions.

\subsection{Extensibility for New Architectures}
\label{subsec:extensibility}

Adding a new architecture follows a modular workflow designed to separate pedagogical content from application logic. First, the developer reviews the scientific literature for the target hardware platform and extracts key physical phenomena, structural parameters, and visual features that distinguish the architecture. Second, these specifications are translated into structured text prompts for the World Labs Marble generative world model platform, following the prompt engineering methodology described in Section~\ref{subsec:world_creation}. Third, the generated world is iteratively refined through manual curation to ensure scientific accuracy and pedagogical clarity. Fourth, the developer authors a teaching guide specifying learning objectives, key concepts, discussion questions, and cross-references to existing architectures. Finally, the new world and its teaching content are registered in the application configuration, and the platform is redeployed through its continuous delivery pipeline. This separation of concerns---content, world assets, and application logic---ensures that domain experts can contribute new quantum hardware visualizations without modifying core application code, a design decision that supports community-driven expansion to emerging architectures such as photonic and topological qubit systems.

This modular structure ensures that domain experts can contribute new worlds without modifying core application code. The separation of pedagogical content (Markdown files), 3D world assets (World Labs URLs), and application logic (React components) follows established software engineering principles and lowers the barrier to community contributions. As new modalities such as photonic quantum computing and topological qubits mature \cite{awsbraket2024}, they can be incorporated into the platform through this same standardized workflow.

%% file: figs/fig3_world_pipeline.tex
% Figure 3: Generative World Model Creation Pipeline
% 5-step horizontal pipeline with semantic icons and bidirectional feedback
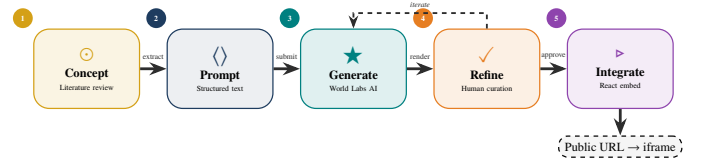
\begin{figure}[t]
\centering
\resizebox{\columnwidth}{!}{%
\begin{tikzpicture}[
    font=\footnotesize,
    stepbox/.style={draw, rounded corners=10pt, minimum width=2.4cm, minimum height=1.8cm, align=center, thick, inner sep=4pt},
    numcircle/.style={font=\tiny\bfseries, text=white, circle, minimum size=0.45cm, inner sep=0pt},
    arrow/.style={->, >=Stealth, line width=1.8pt, QCdark},
    feedback/.style={->, >=Stealth, line width=1.2pt, QCdark, dashed},
]

% --- Step 1: Scientific Concept (amber) ---
\node[stepbox, fill=QCamber!12, draw=QCamber] (s1) at (0,0) {\textcolor{QCamber}{\Large$\odot$}\\[2pt]\textbf{Concept}\\[-1pt]\tiny Literature review};
\node[numcircle, fill=QCamber, above left=2pt and 2pt of s1.north west] {1};

% --- Step 2: Prompt Engineering (blue) ---
\node[stepbox, fill=QCblue!8, draw=QCblue, right=0.6cm of s1] (s2) {\textcolor{QCblue}{\Large$\langle\rangle$}\\[2pt]\textbf{Prompt}\\[-1pt]\tiny Structured text};
\node[numcircle, fill=QCblue, above left=2pt and 2pt of s2.north west] {2};

% --- Step 3: World Labs Gen (teal) ---
\node[stepbox, fill=QCteal!12, draw=QCteal, right=0.6cm of s2] (s3) {\textcolor{QCteal}{\Large$\bigstar$}\\[2pt]\textbf{Generate}\\[-1pt]\tiny World Labs AI};
\node[numcircle, fill=QCteal, above left=2pt and 2pt of s3.north west] {3};

% --- Step 4: Curated Refinement (orange) ---
\node[stepbox, fill=QCorange!12, draw=QCorange, right=0.6cm of s3] (s4) {\textcolor{QCorange}{\Large$\checkmark$}\\[2pt]\textbf{Refine}\\[-1pt]\tiny Human curation};
\node[numcircle, fill=QCorange, above left=2pt and 2pt of s4.north west] {4};

% --- Step 5: Integration (purple) ---
\node[stepbox, fill=QCpurple!10, draw=QCpurple, right=0.6cm of s4] (s5) {\textcolor{QCpurple}{\Large$\triangleright$}\\[2pt]\textbf{Integrate}\\[-1pt]\tiny React embed};
\node[numcircle, fill=QCpurple, above left=2pt and 2pt of s5.north west] {5};

% --- Arrows with labels ---
\draw[arrow] (s1) -- (s2) node[midway, above=2pt, font=\tiny] {extract};
\draw[arrow] (s2) -- (s3) node[midway, above=2pt, font=\tiny] {submit};
\draw[arrow] (s3) -- (s4) node[midway, above=2pt, font=\tiny] {render};
\draw[arrow] (s4) -- (s5) node[midway, above=2pt, font=\tiny] {approve};

% --- Feedback loop: step 4 back to step 3 ---
\draw[feedback] (s4.north) -- ++(0,0.35) -| node[pos=0.25, above, font=\tiny\itshape] {iterate} (s3.north);

% --- Output: below step 5 ---
\node[draw, rounded corners=6pt, thick, dashed, fill=gray!8, below=0.6cm of s5, minimum width=2.4cm, align=center, inner sep=4pt] (out) {\footnotesize Public URL $\rightarrow$ iframe};

\draw[arrow] (s5.south) -- (out.north);

\end{tikzpicture}%
}
\caption{The five-stage generative world model pipeline in Quantum Cinema. Each architecture's immersive 3D environment progresses from scientific literature review (Step~1) through structured prompt engineering (Step~2), AI synthesis via World Labs~\cite{worldlabs2024marble} (Step~3), human curation with iterative refinement (Step~4), and frontend integration (Step~5). The feedback loop between Steps 3 and 4 ensures scientific accuracy before publication.}
\label{fig:world_pipeline}
\end{figure}

%% file: tabs/tab4_generative_prompts.tex
% Generative World Concepts Table — Quantum Cinema IEEEtran Paper
% Wide double-column table with color-coded architectures
\begin{table*}[t]
\centering
\footnotesize
\setlength{\tabcolsep}{8pt}
\renewcommand{\arraystretch}{1.25}
\caption{Generative World Model Concepts by Architecture}
\label{tab:generative_prompts}
\begin{tabular}{@{}>{\raggedright\arraybackslash}p{4.5cm}@{\hspace{8pt}}>{\raggedright\arraybackslash}p{5.5cm}@{\hspace{8pt}}>{\raggedright\arraybackslash}p{6.5cm}@{}}
\toprule
\rowcolor{QCblue!10}
\textbf{Architecture} & \textbf{Scientific Concept} & \textbf{Key Visual Elements in Generated World} \\
\midrule
\cellcolor{QCteal!18}\textcolor{QCteal}{$\blacklozenge$}~\textbf{Trapped-Ion} (IonQ) & Linear chain of Yb$^+$ ions confined in a Paul trap, addressed by intersecting Raman laser beams for quantum gate operations & Glowing blue-white ytterbium ions in perfect equilibrium; gold-violet Raman beams entering from multiple angles; faint golden standing-wave field representing shared vibrational mode; dark cylindrical vacuum chamber \\
\cellcolor{QCorange!18}\textcolor{QCorange}{$\circ$}~\textbf{Neutral Atoms} (QuEra) & Programmable two-dimensional array of Rb atoms held by optical tweezers, with Rydberg interactions enabling programmable many-body
dynamics and analog Hamiltonian simulation & Red optical tweezer beams crisscrossing to form atom grid; soft blue glow of individual rubidium atoms; Rydberg excitation halo around targeted atoms; reconfigurable geometric patterns (triangular, square) \\
\cellcolor{QCviolet!18}\textcolor{QCviolet}{$\blacksquare$}~\textbf{Superconducting} (Rigetti) & Josephson junction circuits cooled to $\sim$15~mK in a dilution refrigerator, controlled by microwave pulses & Golden microwave waveguides routing control signals; superconducting processor chip with circuit traces; frost and ice crystals on copper cooling stages; tall cylindrical dilution refrigerator vessel \\
\bottomrule
\end{tabular}

\vspace{6pt}
\begin{flushleft}
\footnotesize
\textit{Note.} Each row describes the scientific concept grounding the generative prompt and the resulting visual elements in the AI-generated 3D world. Colors are consistent with Table~\ref{tab:architecture-comparison} and Figure~\ref{fig:teaser}. All three environments are synthesized via World Labs' generative world-model pipeline from combinations of scientific illustrations and reference device photographs (Appendix~\ref{app:world-models}).
\end{flushleft}
\end{table*}

%% file: secs/sec5_use_cases.tex
\section{Use Cases and Demonstration}
\label{sec:use_cases}

This section presents four use cases illustrating how \textit{Quantum Cinema} serves its dual target communities: educators, researchers, and science communicators seeking an intuitive tool for explaining quantum hardware, and developers seeking to replicate or extend the platform.

\subsection{Use Case 1: Teaching Quantum Entanglement}
\label{subsec:uc1_educators}

Consider an undergraduate physics instructor preparing a lesson on quantum entanglement for a classroom of students with no prior exposure to quantum computing hardware. The instructor directs students to \textit{Quantum Cinema}, where each student progresses through the four-act narrative at their own pace.

In \textit{Act~1---Nobel Prize}, students first encounter the
2025 Nobel Prize in Physics as the contemporary hardware
anchor for the experience
\cite{nobel2025physics}. The interactive timeline then
connects this milestone to the 2022 Nobel Prize in
Physics awarded to Alain Aspect, John Clauser, and Anton
Zeilinger for foundational experiments on quantum
entanglement and Bell inequalities
\cite{nobel2022physics}. This sequence links the physical
hardware of contemporary quantum processors to the
entanglement principles introduced in the subsequent
acts.

In \textit{Act~2---World Models}, the student selects the \textit{trapped-ion} architecture card. A short looping video introduces the core physical concept: individual charged atoms suspended in an electromagnetic trap and manipulated by laser beams. The student learns that trapped-ion systems are one of the leading platforms for realizing entangled quantum states in a controlled, repeatable manner~\cite{ionq2024}.

\textit{Act~3---Explore} delivers the immersive centerpiece. The student enters a generative three-dimensional world depicting a linear chain of trapped ytterbium ions suspended in an ultra-high vacuum chamber. Gold-violet \textit{Raman laser beams} enter from multiple directions, and a faint golden standing-wave structure represents the \textit{collective phonon mode}---the shared vibrational motion of the entire chain that serves as the quantum bus coupling distant qubits. Two highlighted ions at opposite ends of the chain are phase-locked to this shared field. An annotation delivers the key teaching moment: these ions are entangled not through any physical wire, but through their shared coupling to the collective motion of the ion chain. This makes abstract textbook descriptions of entanglement concrete and observable. Teaching guides with discussion questions and conceptual checkpoints accompany the scene~\cite{norrie2024qnotation}.

In \textit{Act~4---Compare}, the student observes that trapped-ion systems offer \textit{all-to-all connectivity}---any qubit interacts directly with any other---in contrast to the limited nearest-neighbor connectivity of superconducting architectures or the geometry-constrained connectivity of neutral-atom systems. This observation reinforces why trapped-ion platforms have been central to entanglement research: their native connectivity mirrors the non-local correlations that entanglement produces.

\subsection{Use Case 2: Architecture Comparison for Quantum Researchers}
\label{subsec:uc2_researchers}

A quantum-computing researcher evaluating hardware
platforms can access Act IV directly, bypassing the
narrative Acts I--III. The comparison dashboard shown in
Fig.~\ref{fig:radar_chart} groups devices by
computational paradigm. The gate-based view directly
compares IonQ Aria and Rigetti Ankaa-3, whereas QuEra
Aquila is presented in a separate analog
Hamiltonian-simulation view using platform-native
sequence-level and per-atom quantities.

The dashboard reports six metrics: coherence or evolution
time, two-qubit or sequence-level fidelity, readout
fidelity, error rate, connectivity, and physical qubit or
atom count. Users may switch between paradigm views and
toggle the displayed devices. The interface deliberately
avoids producing a single universal ranking because
hardware suitability depends on computational model and
target workload.

The current deployment instead provides illustrative
workload matches. IonQ Aria is associated with
small-molecule simulation, where long coherence and high
fidelity are important; Rigetti Ankaa-3 is associated with
iterative optimization applications such as power-grid
planning, where fast gate operations are valuable; and
QuEra Aquila is associated with materials and
carbon-capture modelling that can exploit native analog
Hamiltonian simulation. These examples are explanatory
workload mappings rather than claims of benchmarked
application superiority.

This capability transforms architecture comparison from a
fragmented documentation exercise into an interactive,
visually grounded exploration of hardware trade-offs.

\input{figs/fig5_radar_chart}  % Radar comparison chart

\subsection{Use Case 3: Science Communication and Public Engagement}
\label{subsec:uc3_communication}

A science journalist preparing an article on the competitive landscape of quantum computing needs to understand the differences between hardware architectures but lacks a physics background. Existing technical documentation assumes familiarity with concepts such as cryogenic cooling, electromagnetic confinement, and laser addressing---barriers that prevent accurate reporting. \textit{Quantum Cinema} addresses this gap through its four-act narrative structure, which requires no quantum computing background and progressively builds understanding through visual metaphors.

The \textit{freezing temperature} required for superconducting qubits---approximately 15 millikelvin, colder than outer space---is rendered as shimmering ice crystals descending through the cryogenic stages of the dilution refrigerator. The \textit{laser beams} that control trapped-ion qubits appear as golden threads of light, making visible the invisible electromagnetic fields that perform quantum gate operations. The \textit{optical tweezers} that arrange neutral atoms are depicted as delicate pink beams sculpting a programmable lattice, conveying the programmable reconfigurability of this architecture.

These visual metaphors are intended to provide
accessible explanatory representations that journalists
can translate into prose for general readerships. The
experience is shareable through a single URL and can be
embedded into web articles as an iframe, avoiding the
installation and headset requirements associated with
many virtual-reality approaches. Prior immersive and
augmented-reality research motivates the use of spatial
interaction for quantum communication
\cite{zable2020vr,karunathilaka2025intuit}; however, the
present paper does not measure engagement, retention, or
learning gains. Controlled evaluation of these outcomes
is therefore identified as future work.

\subsection{D. Use Case 4: Extending the Platform}
\label{subsec:uc4_developers}

For developers and systems researchers who wish to replicate Quantum Cinema or adapt its pipeline to other domains of scientific infrastructure, the platform provides a complete, documented path from source code to deployed application. The replication workflow is designed to require minimal configuration: the application runs locally with a single command after dependency installation, and all static assets are bundled at build time, eliminating external service dependencies during development.

The extension workflow for adding new quantum architectures follows the modular pipeline described in Section~\ref{subsec:world_creation}. Developers begin by conducting a scientific literature review for the target hardware, extract key physical phenomena and structural features, engineer structured text prompts for the generative world model platform, curate the resulting environment for accuracy, author pedagogical content, and register the new world in the application configuration. This separation of pedagogical content, three-dimensional world assets, and application logic ensures that domain experts can contribute without modifying core code.

The deployment pipeline provisions the CloudFront content delivery network, Application Load Balancer, and Elastic Container Service Fargate cluster through infrastructure-as-code configuration. The static-first architecture ensures predictable scaling behavior and low operational overhead, making the platform suitable for classroom deployment, public outreach events, and integration into institutional learning management systems.

%% file: figs/fig5_radar_chart.tex
% Figure 5: Paradigm-Aware Quantum Architecture Comparison
%
% Six axes, normalized to 0--100:
% 1. Coherence / evolution time
% 2. Entangling / sequence fidelity
% 3. Readout fidelity
% 4. Error rate
% 5. Connectivity
% 6. Scale
%
% Gate-based devices are compared directly.
% QuEra Aquila is displayed separately because it implements
% Analog Hamiltonian Simulation and uses platform-native
% evolution- and sequence-level equivalents.
%
% Existing project colors assumed:
% QCteal   = trapped-ion / IonQ
% QCviolet = superconducting / Rigetti
% QCorange = neutral-atom / QuEra

\begin{figure*}[t]
\centering

\resizebox{\textwidth}{!}{%
\begin{tikzpicture}[
    font=\footnotesize,
    axis label/.style={
        font=\scriptsize\bfseries,
        align=center
    },
    panel title/.style={
        font=\small\bfseries,
        align=center
    },
    panel subtitle/.style={
        font=\scriptsize,
        text=gray!70,
        align=center
    },
    grid label/.style={
        font=\tiny,
        text=gray!65
    },
    legend entry/.style={
        font=\scriptsize
    },
]

% ============================================================
% Global configuration
% ============================================================

% Radar radius corresponding to a score of 100
\def\radarR{3.20}

% Distance between the centers of the two paradigm panels
\def\panelSep{10.20}

% ============================================================
% Radar-series drawing macro
%
% Arguments:
% #1 = color
% #2 = line style
% #3 = coherence / evolution score
% #4 = entangling / sequence fidelity score
% #5 = readout fidelity score
% #6 = error-rate score
% #7 = connectivity score
% #8 = scale score
% ============================================================

\def\PlotRadar#1#2#3#4#5#6#7#8{%

    % Convert 0--100 scores into radar radii
    \pgfmathsetmacro{\rA}{(#3/100)*\radarR}
    \pgfmathsetmacro{\rB}{(#4/100)*\radarR}
    \pgfmathsetmacro{\rC}{(#5/100)*\radarR}
    \pgfmathsetmacro{\rD}{(#6/100)*\radarR}
    \pgfmathsetmacro{\rE}{(#7/100)*\radarR}
    \pgfmathsetmacro{\rF}{(#8/100)*\radarR}

    % Soft outer glow
    \draw[
        #1!35,
        line width=3.6pt,
        opacity=0.18,
        #2
    ]
        ({\rA*cos(90)},   {\rA*sin(90)})   --
        ({\rB*cos(30)},   {\rB*sin(30)})   --
        ({\rC*cos(-30)},  {\rC*sin(-30)})  --
        ({\rD*cos(-90)},  {\rD*sin(-90)})  --
        ({\rE*cos(-150)}, {\rE*sin(-150)}) --
        ({\rF*cos(150)},  {\rF*sin(150)})  -- cycle;

    % Transparent polygon fill
    \fill[
        #1!18,
        opacity=0.36
    ]
        ({\rA*cos(90)},   {\rA*sin(90)})   --
        ({\rB*cos(30)},   {\rB*sin(30)})   --
        ({\rC*cos(-30)},  {\rC*sin(-30)})  --
        ({\rD*cos(-90)},  {\rD*sin(-90)})  --
        ({\rE*cos(-150)}, {\rE*sin(-150)}) --
        ({\rF*cos(150)},  {\rF*sin(150)})  -- cycle;

    % Main polygon border
    \draw[
        #1,
        line width=1.5pt,
        #2
    ]
        ({\rA*cos(90)},   {\rA*sin(90)})   --
        ({\rB*cos(30)},   {\rB*sin(30)})   --
        ({\rC*cos(-30)},  {\rC*sin(-30)})  --
        ({\rD*cos(-90)},  {\rD*sin(-90)})  --
        ({\rE*cos(-150)}, {\rE*sin(-150)}) --
        ({\rF*cos(150)},  {\rF*sin(150)})  -- cycle;

    % White-ring data markers
    \foreach \ang/\rad in {
        90/\rA,
        30/\rB,
        -30/\rC,
        -90/\rD,
        -150/\rE,
        150/\rF
    }{
        \fill[white]
            ({\rad*cos(\ang)},{\rad*sin(\ang)})
            circle (2.5pt);

        \draw[
            #1,
            line width=0.8pt
        ]
            ({\rad*cos(\ang)},{\rad*sin(\ang)})
            circle (2.5pt);

        \fill[#1]
            ({\rad*cos(\ang)},{\rad*sin(\ang)})
            circle (1.45pt);
    }
}

% ============================================================
% Shared six-axis grid
% ============================================================

\def\RadarGrid{%

    % Concentric hexagons at 20-point intervals
    \foreach \level in {1,2,3,4,5}{

        \pgfmathsetmacro{\rr}{\level*\radarR/5}
        \pgfmathtruncatemacro{\pct}{20*\level}

        \draw[
            gray!22,
            very thin
        ]
            ({\rr*cos(90)},   {\rr*sin(90)})   --
            ({\rr*cos(30)},   {\rr*sin(30)})   --
            ({\rr*cos(-30)},  {\rr*sin(-30)})  --
            ({\rr*cos(-90)},  {\rr*sin(-90)})  --
            ({\rr*cos(-150)}, {\rr*sin(-150)}) --
            ({\rr*cos(150)},  {\rr*sin(150)})  -- cycle;

        \node[
            grid label,
            anchor=east
        ] at (-0.08,\rr) {\pct};
    }

    % Six radial axes
    \foreach \ang in {90,30,-30,-90,-150,150}{
        \draw[
            gray!32,
            thin
        ]
            (0,0) --
            ({\radarR*cos(\ang)},{\radarR*sin(\ang)});
    }

    % Axis labels
    \node[
        axis label,
        above
    ] at (0,3.82)
        {\shortstack{
            Coherence /\\[-1pt]
            Evolution
        }};

    \node[
        axis label,
        anchor=west
    ] at (3.25,1.90)
        {\shortstack{
            Entangling /\\[-1pt]
            Sequence Fid.
        }};

    \node[
        axis label,
        anchor=west
    ] at (3.25,-1.90)
        {\shortstack{
            Readout\\[-1pt]
            Fidelity
        }};

    \node[
        axis label,
        below
    ] at (0,-3.82)
        {Error Rate$^{\dagger}$};

    \node[
        axis label,
        anchor=east
    ] at (-3.25,-1.90)
        {Connectivity};

    \node[
        axis label,
        anchor=east
    ] at (-3.25,1.90)
        {\shortstack{
            Scale\\[-1pt]
            (Qubits / Atoms)
        }};
}

% ============================================================
% Panel (a): Gate-based quantum processors
% ============================================================

\begin{scope}[xshift=0cm]

    \node[
        panel title
    ] at (0,5.55)
        {(a) Gate-Based Quantum Processors};

    \node[
        panel subtitle
    ] at (0,5.20)
        {Direct comparison: IonQ Aria vs. Rigetti Ankaa-3};

    % Legend placed above chart to avoid axis-label collision
    \draw[
        QCteal,
        line width=1.6pt
    ]
        (-3.05,4.72) -- (-2.45,4.72);

    \fill[QCteal]
        (-2.75,4.72)
        circle (1.5pt);

    \node[
        legend entry,
        anchor=west
    ] at (-2.34,4.72)
        {\textcolor{QCteal}{$\blacklozenge$}~IonQ Aria};

    \draw[
        QCviolet,
        line width=1.6pt,
        densely dashed
    ]
        (0.35,4.72) -- (0.95,4.72);

    \fill[QCviolet]
        (0.65,4.72)
        circle (1.5pt);

    \node[
        legend entry,
        anchor=west
    ] at (1.06,4.72)
        {\textcolor{QCviolet}{$\blacksquare$}~Rigetti Ankaa-3};

    \RadarGrid

    % --------------------------------------------------------
    % Current deployed dashboard scores, normalized to 0--100
    %
    % IonQ Aria:
    % Coherence       = 95
    % Fidelity        = 90
    % Readout         = 95
    % Error-rate score= 85
    % Connectivity    = 95
    % Scale           = 12
    % --------------------------------------------------------

    \PlotRadar
        {QCteal}
        {solid}
        {95}
        {90}
        {95}
        {85}
        {95}
        {12}

    % --------------------------------------------------------
    % Rigetti Ankaa-3:
    % Coherence       = 25
    % Fidelity        = 80
    % Readout         = 70
    % Error-rate score= 70
    % Connectivity    = 30
    % Scale           = 45
    % --------------------------------------------------------

    \PlotRadar
        {QCviolet}
        {densely dashed}
        {25}
        {80}
        {70}
        {70}
        {30}
        {45}

\end{scope}

% ============================================================
% Separator between computational paradigms
% ============================================================

\draw[
    gray!25,
    very thin
]
    (5.10,-4.40) -- (5.10,5.60);

% ============================================================
% Panel (b): Analog Hamiltonian Simulation
% ============================================================

\begin{scope}[xshift=\panelSep cm]

    \node[
        panel title
    ] at (0,5.55)
        {(b) Analog Hamiltonian Simulation};

    \node[
        panel subtitle
    ] at (0,5.20)
        {QuEra Aquila; platform-native AHS equivalents};

    % Legend
    \draw[
        QCorange,
        line width=1.6pt
    ]
        (-1.55,4.72) -- (-0.95,4.72);

    \fill[QCorange]
        (-1.25,4.72)
        circle (1.5pt);

    \node[
        legend entry,
        anchor=west
    ] at (-0.84,4.72)
        {\textcolor{QCorange}{$\circ$}~QuEra Aquila};

    \RadarGrid

    % --------------------------------------------------------
    % QuEra Aquila:
    % Coherence/evolution = 20
    % Sequence fidelity   = 65
    % Readout fidelity    = 80
    % Error-rate score    = 55
    % Connectivity        = 70
    % Scale               = 100
    % --------------------------------------------------------

    \PlotRadar
        {QCorange}
        {solid}
        {20}
        {65}
        {80}
        {55}
        {70}
        {100}

\end{scope}

% ============================================================
% Shared methodological note
% ============================================================

\node[
    font=\scriptsize\itshape,
    text=gray!72,
    align=center,
    anchor=north
] at (5.10,-4.55) {%
    Scores are displayed on a 0--100 scale and oriented so that
    higher is better.\\[-1pt]
    $^{\dagger}$Only the error-rate axis is inverted; QuEra uses
    platform-native evolution- and sequence-level equivalents.%
};

\end{tikzpicture}%
}

\caption{Paradigm-aware comparison of the three
representative quantum devices across six 0--100 dashboard display scores. The gate-based panel
directly compares IonQ Aria and Rigetti Ankaa-3, while QuEra
Aquila is shown separately because analog Hamiltonian simulation
does not expose all gate-model calibration quantities. Scores
reproduce the current deployed comparison dashboard on a
0--100 scale; only the error-rate axis is inverted so that higher
values are preferable. QuEra's temporal and fidelity axes use
explicitly labelled platform-native evolution-window and
sequence-level equivalents. Raw device values are curated from
Amazon Braket and official manufacturer specifications
\cite{awsbraket2024,ionq2024,rigetti2024,quera2024}.
}
\label{fig:radar_chart}

\end{figure*}
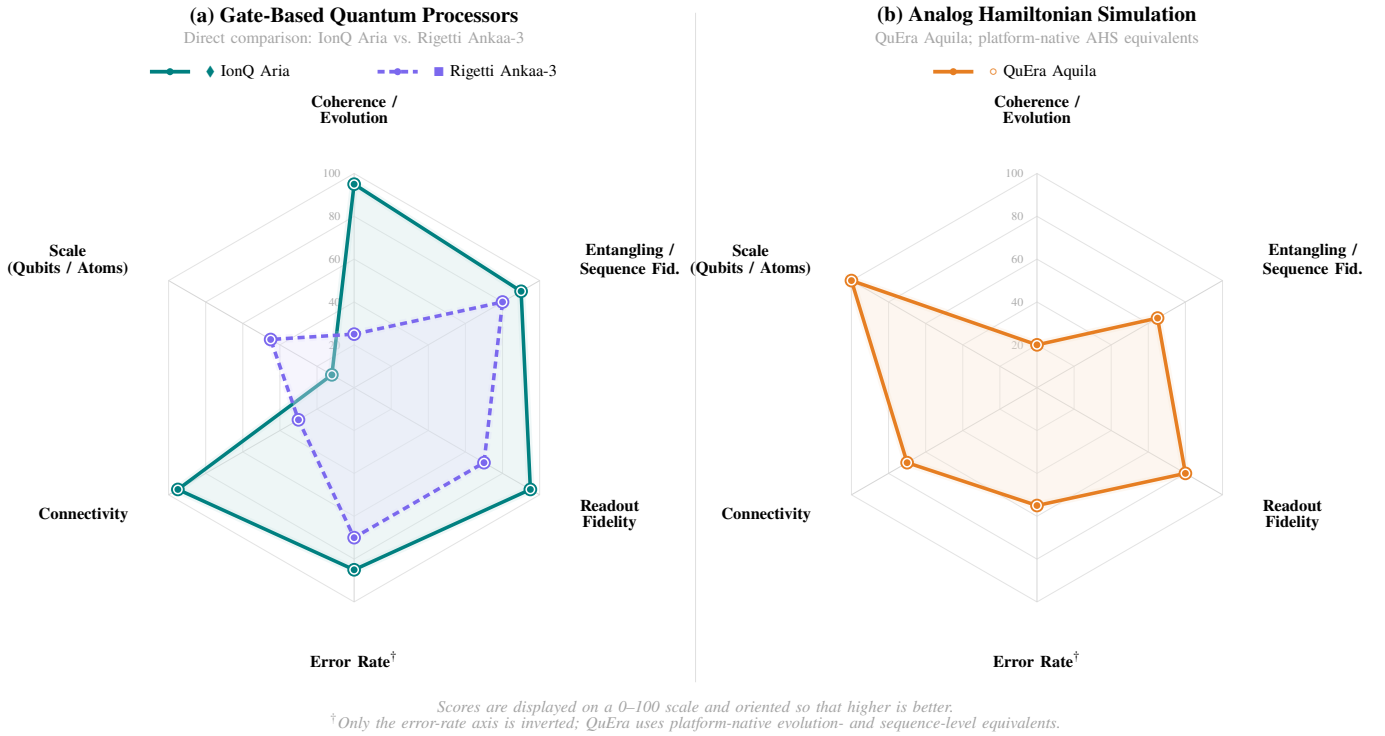

%% file: secs/sec6_conclusion.tex
\section{Conclusion, Limitations, and Future Work}
\label{sec:conclusion}

Quantum Cinema represents the first interactive system to leverage generative world models---neural networks that learn to simulate virtual environments---for the visualization of quantum computing hardware, directly addressing the ``imagination gap'' between quantum computing's transformative potential and public understanding. By rendering the invisible subatomic machinery of quantum processors as explorable cinematic worlds, we bridge a critical communication barrier that has long impeded the broader adoption and comprehension of quantum technologies.

The platform's four-act cinematic structure---spanning from a Nobel Prize historical narrative through curated video introductions for conceptual grounding, into freeform 3D exploration, and culminating in side-by-side hardware comparison---makes quantum hardware accessible to non-expert audiences while preserving the scientific depth required by researchers. Each featured architecture is accompanied by a dated, curated device snapshot drawn from Amazon Braket and official manufacturer documentation. These values support the quantitative architecture comparison but do not constitute live calibration measurements or physical validation of the generated scenes.The complete system is open-source, runs entirely in the browser without installation, and is freely accessible to a global audience regardless of technical background or computational resources.

We acknowledge several limitations of the current system and outline corresponding future directions across three areas.

\paragraph{Fidelity and Coverage} 

The generative worlds are explanatory visualizations
informed by device documentation and curated parameters,
not physical simulations. They are designed to support
more concrete understanding of hardware structure and
operating principles; users seeking computational
fidelity should consult dedicated quantum-simulation
frameworks. Device parameters further represent static snapshots rather than real-time data, and the platform is currently limited to three architectures (superconducting, trapped-ion, and neutral atom), with photonic and topological qubit systems under active development. The reliance on a commercial generative platform (World Labs) also introduces a dependency, which we mitigate by documenting our complete prompt engineering methodology so that worlds can be regenerated using alternative platforms as the ecosystem evolves. Future work will pursue live integration with AWS Braket to dynamically update hardware parameters, incorporate additional architectures including photonic and topological qubits, and maintain platform-agnostic documentation for reproducibility.

\paragraph{Interactivity and Pedagogy} The immediate research
priority is a controlled user study with students,
educators, and science communicators to evaluate changes
in quantum-hardware understanding, conceptual retention,
usability, and engagement. Such an evaluation is
necessary before making causal claims about educational
effectiveness. The platform also does not currently support interactive
quantum-circuit execution within the 3D environments,
limiting users to observational rather than experimental
exploration. Subsequent engineering work will embed
circuit simulators within the immersive worlds so that
users can observe how program-level operations relate to
hardware-level representations. Integration with
established frameworks and curricula, including Qiskit,
Cirq, and PennyLane, will further support adoption in
existing educational settings.

\paragraph{Accessibility and Community.} Future releases will
extend global accessibility through multi-language
support and collaborative multiplayer exploration modes
that enable group learning and shared scientific
discovery.

\medskip
\noindent\textbf{Broader Impact.}
Quantum Cinema is intended to contribute to United
Nations Sustainable Development Goal 4 through
browser-based access to quantum-science education, and
to the inclusive-infrastructure and access objectives of
SDGs 9 and 10 through its open-source, zero-install, and
no-headset design
\cite{unitednations2015agenda}. The platform can be
reused by educators and science communicators without
requiring local quantum hardware. These connections
describe intended accessibility contributions; the
present study does not report measured learning gains,
low-bandwidth performance, or the comparative energy
efficiency of quantum and classical computing.

In closing, Quantum Cinema establishes a new paradigm at
the intersection of generative artificial intelligence
(AI) and quantum education.

%% file: secs/sec8_appendix.tex
\section{Glossary}
\label{app:glossary}
Quantum Cinema brings together concepts from four distinct knowledge domains: quantum computing hardware, generative artificial intelligence, web application infrastructure, and foundational quantum science. To serve the interdisciplinary readership of this paper---spanning computer scientists, quantum physicists, educators, and science communicators---we provide below a comprehensive glossary organized by domain. Each table is visually distinguished by a unique color and icon to facilitate quick navigation.

Table~\ref{tab:glossary-hardware} (\textcolor{QCteal}{$\blacklozenge$}~teal) defines the \textit{physical vocabulary} of quantum computing: the three hardware architectures featured in Quantum Cinema (trapped-ion, neutral-atom, and superconducting), their constituent components (Josephson junctions, optical tweezers, Paul traps), and the fundamental quantum mechanical phenomena (entanglement, superposition, decoherence) that make quantum computation possible. These definitions directly inform the generative world models of Act~III, ensuring that every 3D environment is grounded in empirically validated hardware descriptions.

Table~\ref{tab:glossary-ai} (\textcolor{QCpurple}{$\bigstar$}~purple) covers the \textit{generative AI technologies} that enable Quantum Cinema's immersive visualizations: the world model pipeline that transforms scientific specifications into explorable 3D environments, the Gaussian splatting technique used for real-time neural rendering, and the World Labs platform that powers the scene synthesis. These terms bridge the hardware definitions of Table~\ref{tab:glossary-hardware} with the navigable 3D experiences presented to the user.

Table~\ref{tab:glossary-infra} (\textcolor{QCblue}{$\blacksquare$}~navy) documents the \textit{web engineering stack}: the serverless cloud architecture (AWS ECS Fargate, CloudFront CDN), the front-end framework (Next.js, React, TypeScript), and the single-page application model that together enable Quantum Cinema's zero-installation, globally accessible delivery. Understanding this infrastructure is essential for researchers and developers seeking to replicate or extend the platform.

Table~\ref{tab:glossary-science} (\textcolor{QCamber!80!black}{$\odot$}~amber) situates the system within its broader scientific
context through the Nobel Prizes, foundational quantum
concepts, and representative quantum algorithms. The
algorithm entries provide general background and are not
used to rank hardware in the current Act~IV dashboard.

\input{tabs/tab_glossary_hardware}
\input{tabs/tab_glossary_ai}
\input{tabs/tab_glossary_infra}
\input{tabs/tab_glossary_science}

%% file: tabs/tab_glossary_hardware.tex
% Glossary Table I: Quantum Computing Hardware
% Color: teal; Icon: diamond
\begin{table*}[t]
\centering
\small
\setlength{\tabcolsep}{5pt}
\caption{Glossary of Technical Terms: Quantum Computing Hardware}
\label{tab:glossary-hardware}
\begin{tabular}{@{}>{\raggedright\arraybackslash}p{4.0cm}@{\hspace{0.5cm}}>{\raggedright\arraybackslash}p{12.6cm}@{}}
\toprule
\rowcolor{QCteal!20}
\multicolumn{2}{@{}l}{\textcolor{QCteal}{$\blacklozenge$}~\textbf{Quantum Computing Hardware}} \\
\midrule
\rowcolor{AltRow}
\textit{Term} & \textit{Definition} \\
\midrule
Ion Trap & A quantum computing platform that confines charged atomic ions in electromagnetic fields within an ultra-high vacuum chamber, using precisely tuned laser pulses to perform quantum gate operations on individual ions with high fidelity [\cite{bruzewicz2019}]. \\
\rowcolor{AltRow}
Neutral Atom & An atom with no net electrical charge that is confined and manipulated using focused laser beams called optical tweezers, forming the basis of a quantum computing platform that offers programmable two-dimensional geometries and flexible connectivity patterns [\cite{bluvstein2024logical}]. \\
Superconducting Qubit & A quantum bit implemented using superconducting electrical circuits, typically containing one or more Josephson junctions, that are cooled to millikelvin temperatures to preserve quantum coherence and enable gate operations [\cite{krantz2019quantum}]. \\
\rowcolor{AltRow}
Josephson Junction & A superconducting device consisting of two superconducting electrodes separated by a thin insulating barrier, serving as the fundamental nonlinear circuit element that enables superconducting qubit operation through the Josephson effect [\cite{krantz2019quantum}]. \\
Paul Trap & An ion trap design that uses oscillating radio-frequency electromagnetic fields to confine charged particles in three-dimensional space without the need for physical walls or containers, named after Wolfgang Paul who shared the 1989 Nobel Prize in Physics for its invention [\cite{bruzewicz2019}]. \\
\rowcolor{AltRow}
Optical Tweezer & A tightly focused laser beam that creates a trapping potential capable of holding and manipulating microscopic particles, used extensively in neutral-atom quantum computing to arrange individual atoms in programmable two-dimensional arrays [\cite{bluvstein2024logical}]. \\
Rydberg Blockade & A quantum phenomenon in which the excitation of one atom to a highly excited Rydberg state shifts the energy levels of nearby atoms within a critical radius, preventing their simultaneous excitation and thereby enabling controlled entangling operations [\cite{bluvstein2024logical}]. \\
\rowcolor{AltRow}
Raman Laser & A laser tuned to stimulate Raman transitions between atomic energy levels, a technique widely used in trapped-ion quantum computing to implement both single-qubit rotations and multi-qubit entangling gate operations [\cite{bruzewicz2019}]. \\
Quantum Bit (Qubit) & The fundamental unit of quantum information that can exist in a superposition of basis states, enabling quantum parallelism and computational advantages over classical binary computation for certain problem classes [\cite{nielsen2010quantum}]. \\
\rowcolor{AltRow}
Coherence Time & The characteristic duration during which a quantum system maintains its superposition state before environmental interactions cause decoherence, representing a fundamental limit on the length of quantum computations that can be performed reliably [\cite{biamonte2017quantum}]. \\
Decoherence & The irreversible loss of quantum mechanical properties, including superposition and entanglement, that occurs when a quantum system interacts with its surrounding environment; decoherence represents the primary obstacle to scaling quantum computers [\cite{nielsen2010quantum}]. \\
\rowcolor{AltRow}
Entanglement & A quantum mechanical phenomenon in which two or more particles become correlated such that the quantum state of each particle cannot be described independently of the others, regardless of the spatial separation between them; the 2022 Nobel Prize in Physics was awarded for experimental demonstrations of this phenomenon [\cite{nobel2022physics}]. \\
Fidelity & A quantitative measure of the accuracy with which a quantum gate operation or quantum state preparation is performed, defined mathematically as the overlap between the intended and actual quantum states, expressed as a value between zero and unity [\cite{cerezo2021variational}]. \\
\rowcolor{AltRow}
Error Rate & The probability that a quantum gate operation produces an incorrect output, typically quantified through randomized benchmarking protocols and reported as an aggregate figure of merit for comparing quantum device performance [\cite{cerezo2021variational}]. \\
Gate & A quantum logic operation that manipulates one or more qubits through unitary transformations, analogous to classical logic gates but operating on quantum amplitudes rather than binary values [\cite{nielsen2010quantum}]. \\
\rowcolor{AltRow}
Bloch Sphere & A geometric representation of a single qubit's quantum state as a point on the surface of a unit sphere, providing an intuitive visualization of superposition, quantum phase, and the effects of single-qubit rotations [\cite{nielsen2010quantum}]. \\
Superposition & A fundamental quantum mechanical principle stating that a quantum system can exist in multiple states simultaneously until a measurement is performed, which collapses the system into a single definite state [\cite{nielsen2010quantum}]. \\
\rowcolor{AltRow}
Millikelvin (mK) & A unit of temperature equal to one thousandth of a kelvin, representing the operating temperature regime for superconducting qubits where thermal noise is suppressed below the energy scale of the quantum computational states [\cite{krantz2019quantum}]. \\
Dilution Refrigerator & A specialized cryogenic cooling system that uses a mixture of helium-3 and helium-4 isotopes to reach temperatures in the millikelvin range---approximately fifteen thousandths of a degree above absolute zero---which is required for operating superconducting quantum processors [\cite{krantz2019quantum}]. \\
\bottomrule
\end{tabular}

\vspace{4pt}
\begin{flushleft}
\footnotesize
\textit{Note.} These eighteen terms constitute the physical vocabulary of quantum computing as presented in Quantum Cinema. Each definition grounds the corresponding generative world model---the immersive 3D environments of Act~III---in empirically validated hardware descriptions drawn from peer-reviewed reviews of trapped-ion [\cite{bruzewicz2019}], neutral-atom [\cite{bluvstein2024logical}], and superconducting [\cite{krantz2019quantum}] architectures. Readers seeking a comprehensive introduction to quantum computation may consult Nielsen and Chuang [\cite{nielsen2010quantum}].
\end{flushleft}
\end{table*}

%% file: tabs/tab_glossary_ai.tex
% Glossary Table II: Generative AI and Scientific Visualization
% Color: purple; Icon: star
\begin{table*}[t]
\centering
\small
\setlength{\tabcolsep}{5pt}
\caption{Glossary of Technical Terms: Generative AI and Scientific Visualization}
\label{tab:glossary-ai}
\begin{tabular}{@{}>{\raggedright\arraybackslash}p{4.0cm}@{\hspace{0.5cm}}>{\raggedright\arraybackslash}p{12.6cm}@{}}
\toprule
\rowcolor{QCpurple!20}
\multicolumn{2}{@{}l}{\textcolor{QCpurple}{$\bigstar$}~\textbf{Generative AI and Scientific Visualization}} \\
\midrule
\rowcolor{AltRow}
\textit{Term} & \textit{Definition} \\
\midrule
Generative World Model & An artificial intelligence system that learns to synthesize realistic, interactive virtual environments by predicting the spatial structure, physical dynamics, and visual appearance of scenes from high-level descriptions or partial observations [\cite{ding2025worldmodels}]. \\
\rowcolor{AltRow}
Gaussian Splatting & A neural rendering technique that represents three-dimensional scenes as collections of three-dimensional Gaussian primitives, enabling real-time photorealistic novel-view synthesis from sparse input photographs or text descriptions [\cite{xie2024physgaussian}]. \\
World Labs & A company founded by Fei-Fei Li that develops generative artificial intelligence systems for creating persistent, explorable three-dimensional virtual environments from text descriptions and images, providing the platform that powers Quantum Cinema's immersive scenes [\cite{worldlabs2024marble}]. \\
\bottomrule
\end{tabular}

\vspace{4pt}
\begin{flushleft}
\footnotesize
\textit{Note.} These three terms describe the AI substrate of Quantum Cinema. The generative world model pipeline (Section~\ref{sec:world_model_pipeline}) translates the hardware concepts of Table~\ref{tab:glossary-hardware} into navigable 3D environments, bridging the ``imagination gap'' between abstract quantum physics and public understanding. For a comprehensive survey of world models, see Ding et al. [\cite{ding2025worldmodels}].
\end{flushleft}
\end{table*}

%% file: tabs/tab_glossary_infra.tex
% Glossary Table III: Web Application Infrastructure
% Color: navy; Icon: square
\begin{table*}[t]
\centering
\small
\setlength{\tabcolsep}{5pt}
\caption{Glossary of Technical Terms: Web Application Infrastructure}
\label{tab:glossary-infra}
\begin{tabular}{@{}>{\raggedright\arraybackslash}p{4.0cm}@{\hspace{0.5cm}}>{\raggedright\arraybackslash}p{12.6cm}@{}}
\toprule
\rowcolor{QCblue!20}
\multicolumn{2}{@{}l}{\textcolor{QCblue}{$\blacksquare$}~\textbf{Web Application Infrastructure}} \\
\midrule
\rowcolor{AltRow}
\textit{Term} & \textit{Definition} \\
\midrule
Amazon Web Services (AWS) Braket & A fully managed quantum computing service provided by Amazon Web Services that offers access to quantum hardware from multiple vendors, including trapped-ion, neutral-atom, and superconducting quantum processors, along with classical simulation tools and quantum algorithm development environments [\cite{awsbraket2024}]. \\
\rowcolor{AltRow}
Content Delivery Network (CDN) & A geographically distributed system of proxy servers that caches and delivers web content to end users from the nearest edge location, thereby reducing latency, improving load times, and enhancing global availability [\cite{aws2024}]. \\
Elastic Container Service (ECS) Fargate & A serverless compute engine provided by Amazon Web Services for running containerized applications without requiring the user to provision or manage underlying server infrastructure, enabling automatic scaling and fault-tolerant deployment [\cite{aws2024}]. \\
\rowcolor{AltRow}
Next.js & An open-source React framework that provides server-side rendering, automatic code splitting, and hybrid static site generation, designed for building production-grade web applications with optimized performance [\cite{nextjs2024}]. \\
React & An open-source JavaScript library for building user interfaces through a component-based architecture that enables declarative, efficient, and flexible front-end development, maintained by Meta Platforms [\cite{react2024}]. \\
\rowcolor{AltRow}
Single-Page Application (SPA) & A web application architecture that dynamically updates content through JavaScript without loading entire new pages from the server, providing a fluid, responsive user experience similar to native desktop applications [\cite{nextjs2024}]. \\
TypeScript & A typed superset of JavaScript developed by Microsoft that adds static type checking and advanced language features, improving developer productivity and code reliability for large-scale web applications [\cite{nextjs2024}]. \\
\bottomrule
\end{tabular}

\vspace{4pt}
\begin{flushleft}
\footnotesize
\textit{Note.} These seven terms describe the software engineering stack that enables Quantum Cinema's zero-installation, globally accessible delivery model (Section~\ref{sec:system-design}). The static-first, serverless architecture was chosen specifically to eliminate barriers to adoption---no quantum hardware access, no software installation, and no user account are required.
\end{flushleft}
\end{table*}

%% file: tabs/tab_glossary_science.tex
% Glossary Table IV: Foundational Science and Algorithms
% Color: amber; Icon: sun
\begin{table*}[t]
\centering
\small
\setlength{\tabcolsep}{5pt}
\caption{Glossary of Technical Terms: Foundational Science and Algorithms}
\label{tab:glossary-science}
\begin{tabular}{@{}>{\raggedright\arraybackslash}p{4.0cm}@{\hspace{0.5cm}}>{\raggedright\arraybackslash}p{12.6cm}@{}}
\toprule
\rowcolor{QCamber!20}
\multicolumn{2}{@{}l}{\textcolor{QCamber!80!black}{$\odot$}~\textbf{Foundational Science and Algorithms}} \\
\midrule
\rowcolor{AltRow}
\textit{Term} & \textit{Definition} \\
\midrule
Nobel Prize in Physics 2022 & Awarded to Alain Aspect, John Clauser, and Anton Zeilinger for experiments with entangled photons that established the violation of Bell inequalities and pioneered the field of quantum information science [\cite{nobel2022physics}]. \\
\rowcolor{AltRow}
Nobel Prize in Physics 2024 & Awarded to John Hopfield and Geoffrey Hinton for foundational discoveries and inventions that enable machine learning with artificial neural networks, underscoring the transformative role of artificial intelligence in scientific discovery [\cite{nobel2024physics}]. \\
Nobel Prize in Physics 2025 & Recognized advances at the intersection of quantum science and quantum computing, cementing the field's position at the forefront of modern physics and highlighting the growing societal importance of quantum technologies [\cite{nobel2025physics}]. \\
\rowcolor{AltRow}
Quantum Computing & A paradigm of computation that exploits quantum mechanical phenomena---superposition, entanglement, and quantum interference---to process information in ways that can provide exponential speedups over classical computers for specific tasks [\cite{nielsen2010quantum}]. \\
Shor's Algorithm & A quantum algorithm for integer factorization that runs in polynomial time, offering an exponential speedup over the best known classical algorithms and demonstrating the transformative potential of quantum computing for cryptography [\cite{nielsen2010quantum}]. \\
\rowcolor{AltRow}
Quantum Approximate Optimization Algorithm (QAOA) & A variational quantum algorithm designed for combinatorial optimization problems, which prepares approximate ground states of problem Hamiltonians by alternating between application of a phase separator and a mixer operator [\cite{cerezo2021variational}]. \\
Variational Quantum Eigensolver (VQE) & A hybrid quantum-classical algorithm that uses a quantum computer to prepare trial states and a classical optimizer to adjust parameters, finding approximate ground state energies of molecular Hamiltonians [\cite{cerezo2021variational}]. \\
\bottomrule
\end{tabular}

\vspace{4pt}
\begin{flushleft}
\footnotesize
\textit{Note.} 
These entries provide foundational scientific and
algorithmic context for Quantum Cinema. The current
Act~IV dashboard uses paradigm-aware,
application-oriented workload examples rather than
algorithm-specific claims of hardware superiority.

\end{flushleft}
\end{table*}

%% file: secs/sec9_appendix_acts.tex
\section{The Four Acts of Quantum Cinema}
\label{app:four-acts}

This appendix provides a detailed walkthrough of each act in Quantum Cinema's narrative, with annotated screenshots for Acts I, II, and IV. Act III (the immersive 3D world exploration) is illustrated in Figure~\ref{fig:teaser} of the main text.

% ============================================================
% Act 1: Nobel Prize
% ============================================================
\subsection{Act I: Nobel Prize -- Establishing Historical Context}
\label{app:act1}

\begin{figure}[t]
\centering
\includegraphics[width=0.85\columnwidth]{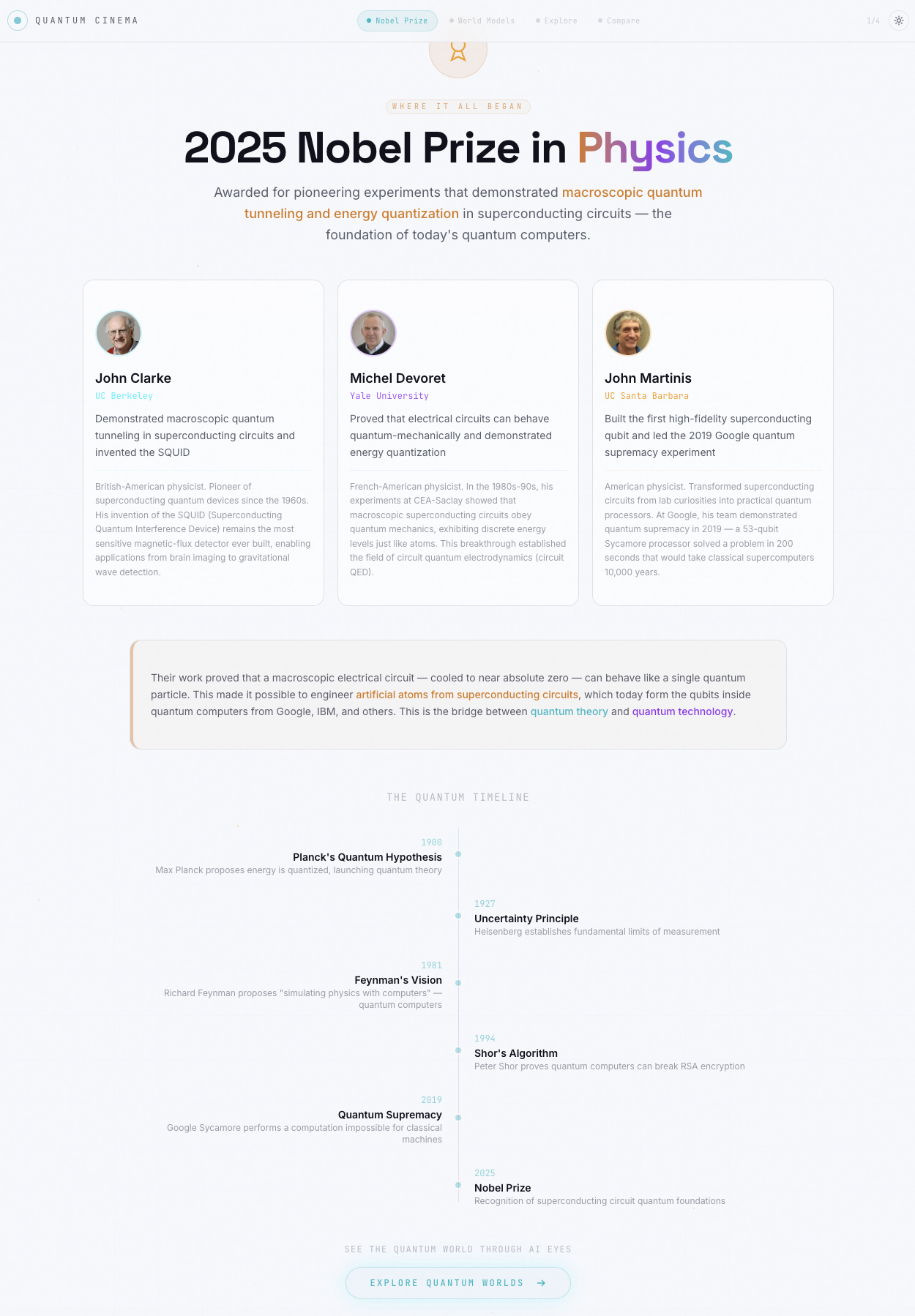}
\caption{Act I: Nobel Prize timeline. Users interact with laureate profiles to understand the historical significance of quantum entanglement, neural networks, and quantum computing advances.}
\label{fig:act1}
\end{figure}

Act I grounds the user in the historical and scientific significance of quantum mechanics through an interactive timeline of Nobel Prize laureates (Figure~\ref{fig:act1}). The screen presents three Nobel Prizes in Physics: the 2022 award to Aspect, Clauser, and Zeilinger for experimental entanglement; the 2024 award to Hopfield and Hinton for foundational machine learning; and the 2025 award recognizing quantum computing advances. Each laureate entry includes a portrait, citation text, and a one-sentence explanation of their contribution's relevance to quantum technology. Users scroll through the timeline at their own pace, building the ``why''---the motivational foundation that answers why quantum computing matters.

\noindent\textbf{Pedagogical rationale.} Research in science communication emphasizes that historical narrative increases engagement and retention when introducing complex technical topics~\cite{dede2009immersive}. By beginning with Nobel Prize laureates rather than abstract physics, Quantum Cinema leverages the authority and familiarity of these awards to build trust and curiosity in non-expert users.

% ============================================================
% Act 2: World Models
% ============================================================
\subsection{Act II: World Models -- Introducing Architectures}
\label{app:act2}

\begin{figure}[t]
\centering
\includegraphics[width=0.85\columnwidth]{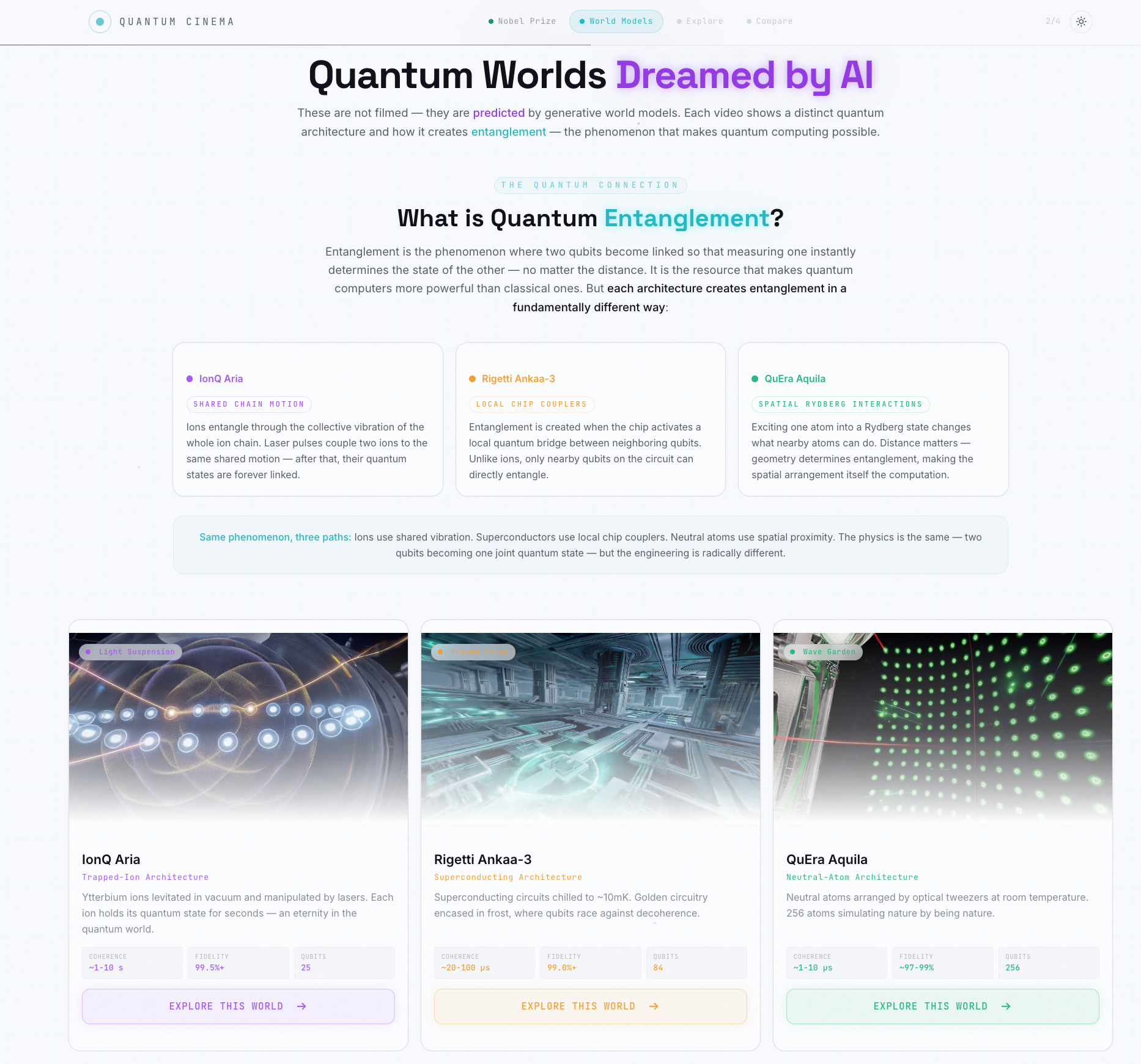}
\caption{Act II: World Models video showcase. Users select an architecture to watch its introductory video, building conceptual understanding before entering the immersive 3D environment.}
\label{fig:act2}
\end{figure}

Act II transitions from historical context to technical content through curated video introductions for each of the three quantum architectures (Figure~\ref{fig:act2}). The screen presents a horizontal selector: trapped-ion (teal), neutral-atom (orange), and superconducting (violet). Selecting an architecture plays a short video that visually introduces its key physical features---linear ion chains, optical tweezer arrays, or Josephson junction circuits---without requiring prior quantum physics knowledge. Users may watch all three videos in any order before proceeding.

\noindent\textbf{Pedagogical rationale.} The video-first sequence introduces key terminology and
visual motifs before immersive exploration, providing
conceptual orientation for the subsequent
three-dimensional experience.

% ============================================================
% Act 3: Explore (referenced from teaser)
% ============================================================
\subsection{Act III: Explore -- Immersive 3D World Exploration}
\label{app:act3}

Act III is the centerpiece of Quantum Cinema. After selecting an architecture in Act II, the user enters a full-screen, navigable 3D world generated by World Labs' Gaussian splatting pipeline. Figure~\ref{fig:teaser} of the main text shows fifteen representative views across all three architectures.

\noindent\textbf{Trapped-Ion World (teal).} Users explore a linear chain of ytterbium ions confined in a Paul trap. Gold-violet Raman laser beams enter from multiple directions. A faint golden standing-wave field represents the shared vibrational mode. Two highlighted ions demonstrate entanglement through the shared medium---they are phase-locked not through a physical wire but through collective motion of the ion chain.

\noindent\textbf{Neutral-Atom World (orange).} Users navigate a two-dimensional array of rubidium atoms held by red optical tweezers. A Rydberg excitation glow surrounds targeted atoms. The programmable geometry---atoms arranged in triangular, square, or arbitrary patterns---is visible and manipulable.

\noindent\textbf{Superconducting World (violet).} Users explore a superconducting processor chip mounted at the base of a dilution refrigerator. Golden microwave waveguides route control signals. Frost and ice crystals on copper stages visualize the cryogenic environment. Circuit traces show the Josephson junction patterns.

\noindent\textbf{Pedagogical rationale.} Immersive three-dimensional environments can support
spatial understanding of quantum concepts in ways that
static diagrams cannot\cite{dede2009immersive}. The generative world model approach makes invisible quantum phenomena---decoherence, laser cooling, energy loss---observable as visual narrative, directly addressing the ``imagination gap'' described in Section~\ref{sec:introduction}.

% ============================================================
% Act 4: Compare
% ============================================================
\subsection{Act IV: Compare---Paradigm-Aware Architecture Comparison}
\label{app:act4}

\begin{figure}[!t]
    \centering
    \includegraphics[width=0.92\columnwidth]{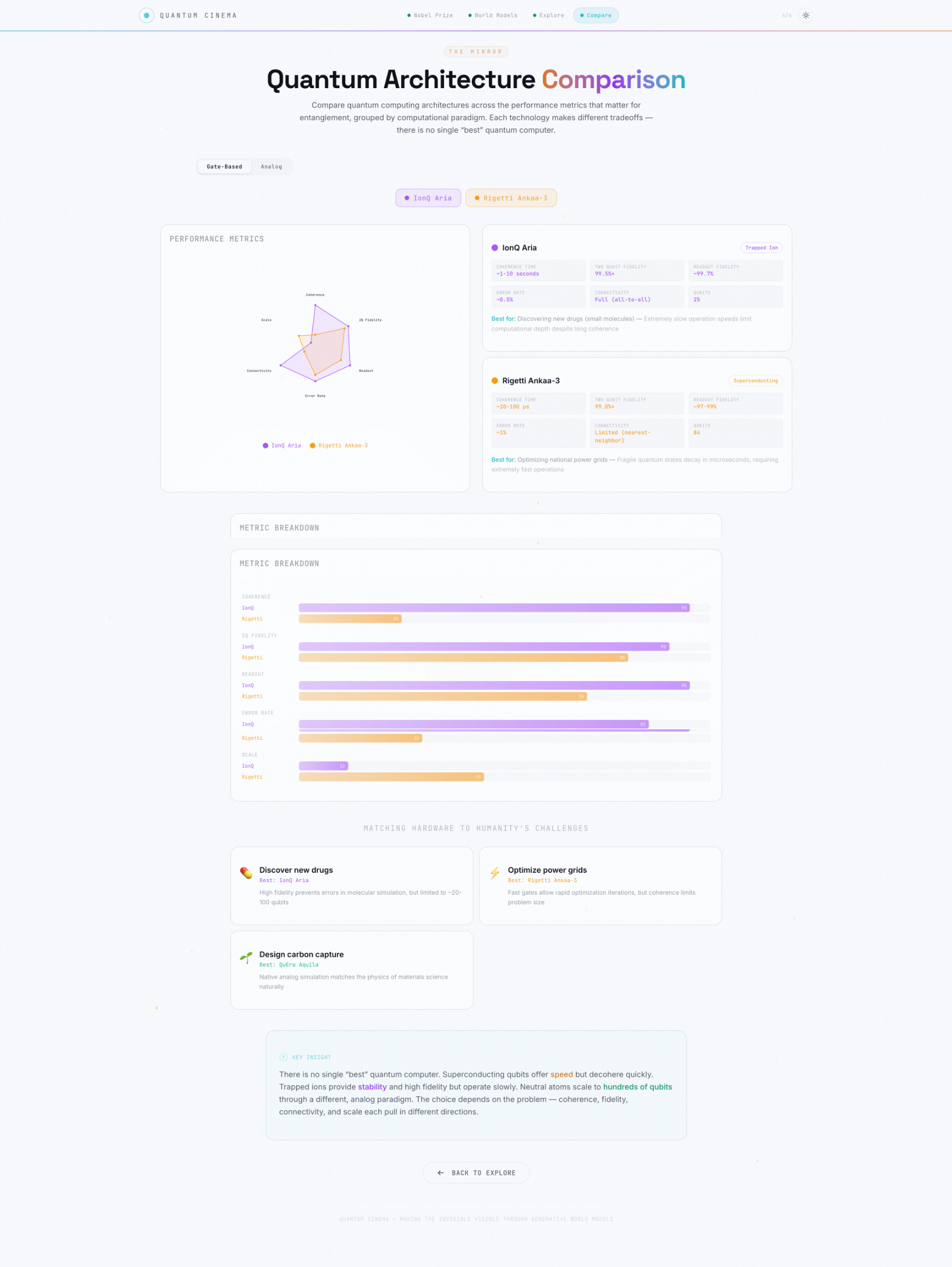}
    \caption{Act IV: paradigm-aware architecture-comparison
    dashboard. The gate-based tab directly compares IonQ Aria
    and Rigetti Ankaa-3, while the analog tab presents QuEra
    Aquila separately as an Analog Hamiltonian Simulation
    device. The dashboard reports six 0--100 display
    scores---coherence or evolution time, two-qubit or
    sequence-level fidelity, readout fidelity, error rate,
    connectivity, and physical qubit or atom count---together
    with illustrative workload-matching recommendations. Only
    the error-rate axis is inverted; QuEra's temporal and
    fidelity axes use platform-native AHS equivalents.}
    \label{fig:act4}
\end{figure}

Act IV provides the final integrative stage of the Quantum
Cinema experience: an interactive architecture-comparison
dashboard organized by computational paradigm
(Fig.~\ref{fig:act4}). The gate-based view directly compares
IonQ Aria and Rigetti Ankaa-3, whereas the analog view
presents QuEra Aquila separately as an Analog Hamiltonian
Simulation (AHS) device. This separation avoids treating
gate-model calibration quantities and analog platform-native
quantities as directly interchangeable.

The dashboard reports six metrics: coherence or evolution
time, two-qubit or sequence-level fidelity, readout fidelity,
error rate, connectivity, and physical qubit or atom count.
The corresponding radar and bar-chart values are displayed
on a 0--100 scale, with only the error-rate axis inverted so
that a higher value is preferable. For QuEra Aquila, the
temporal and fidelity axes use explicitly labelled
platform-native evolution-window, sequence-level, and
per-atom equivalents rather than a gate-model $T_2$
measurement or discrete two-qubit-gate fidelity. The same
six-axis comparison framework is summarized in
Fig.~\ref{fig:radar_chart}.

Rather than producing a single universal ranking, the
dashboard presents illustrative workload matches. In the
current deployment, IonQ Aria is associated with
small-molecule and drug-discovery applications, where long
coherence and high fidelity are valuable; Rigetti Ankaa-3 is
associated with iterative optimization applications such as
power-grid planning, where fast gate operations are useful;
and QuEra Aquila is associated with materials and
carbon-capture modelling that can exploit native analog
Hamiltonian simulation. These mappings are explanatory
examples and should not be interpreted as claims of
benchmarked application superiority.

\noindent\textbf{Pedagogical rationale.}
The comparison stage requires users to integrate the
historical, physical, and architectural concepts introduced
in the preceding acts and apply them to a structured
technology-selection problem. Grouping devices by
computational paradigm helps non-expert users recognize
hardware trade-offs without implying that all metrics have
identical physical meanings across gate-based and analog
systems. The underlying raw values use the same dated device
snapshot reported in Table~\ref{tab:architecture-comparison}
and maintained in the repository's canonical
\texttt{data.ts} file, with values curated from Amazon Braket
and official manufacturer documentation
\cite{awsbraket2024,ionq2024,rigetti2024,quera2024}.
The dashboard scores support visual explanation and
comparative reasoning; they do not constitute a universal
hardware ranking or an independently benchmarked measure
of application performance.

%% file: secs/sec10_appendix_worldmodels.tex
\section{Generative World Model Details}
\label{app:world-models}

This appendix details the generative world model creation process for each of the three quantum computing architectures in Quantum Cinema. For each architecture, we present: (i)~the scientific concepts and reference device photographs that inform the prompt, (ii)~the AI-generated immersive world output, and (iii)~five representative navigable views (Figure~\ref{fig:teaser} of the main text). The generation pipeline follows the five-step process described in Section~\ref{sec:world_model_pipeline} and illustrated in Figure~\ref{fig:world_pipeline}.

% ============================================================
% Ion Trap
% ============================================================
\subsection{Trapped-Ion World Model}
\label{app:ion-trap}

\textbf{Scientific basis.} Trapped-ion quantum computers confine charged atoms (ions) in electromagnetic fields within an ultra-high vacuum chamber~\cite{bruzewicz2019}. Individual ions are addressed by precisely tuned laser beams to perform quantum gate operations. The key visualized phenomena include: the linear ion chain suspended in the trap, intersecting Raman laser beams, and the shared vibrational mode that mediates entanglement between ions.

\begin{figure}[ht]
\centering
\subfloat[Concept: ionization process]{\includegraphics[width=0.32\columnwidth]{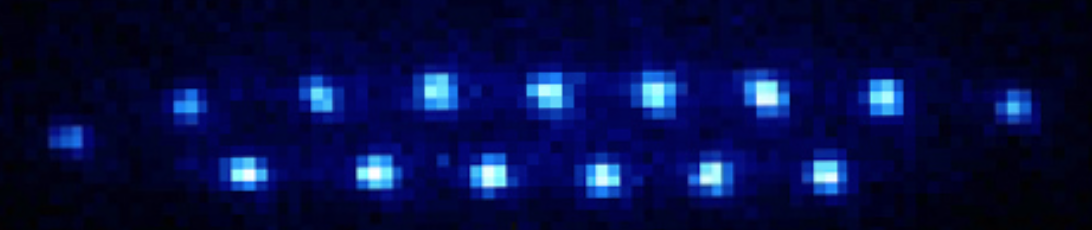}}\hfill
\subfloat[Device: IonQ trapped-ion system]{\includegraphics[width=0.32\columnwidth]{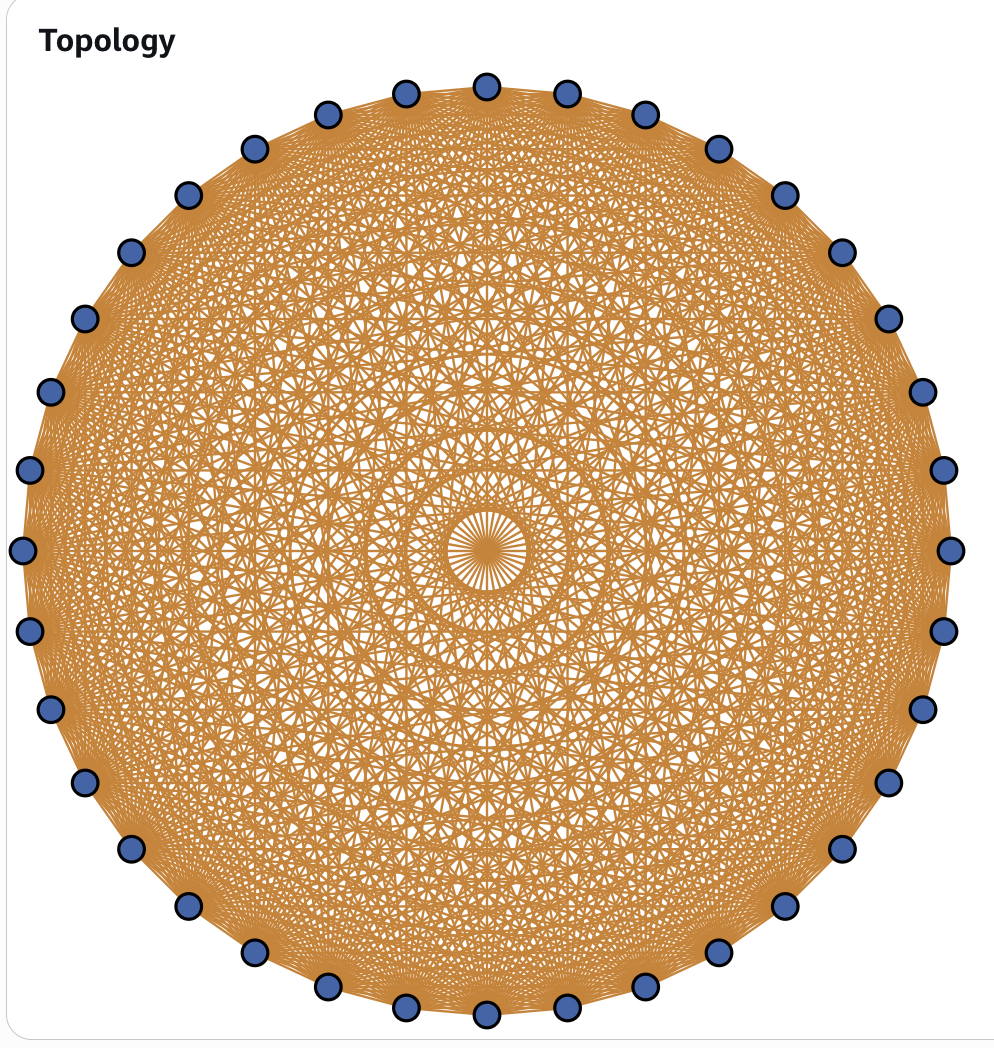}}\hfill
\subfloat[Original: ion trap apparatus]{\includegraphics[width=0.32\columnwidth]{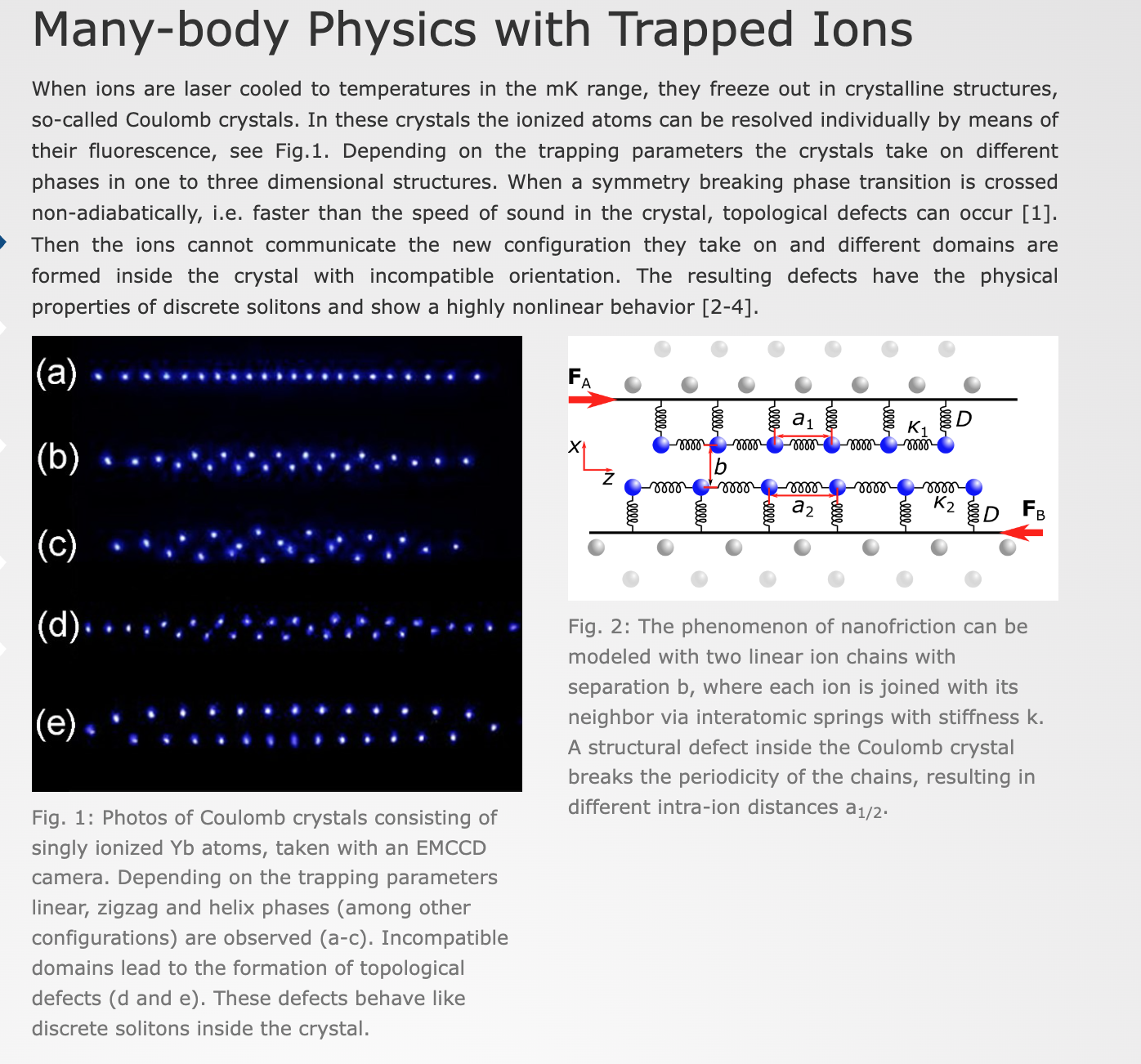}}
\caption{Input materials for the trapped-ion generative world model. (a)~Scientific concept illustration of ionization. (b)~Reference photograph of the IonQ trapped-ion device. (c)~Original reference image of the ion trap apparatus.}
\label{fig:ion-input}
\end{figure}

\textbf{Input.} The generative prompt combines a scientific concept illustration of ionization (the process of creating charged ions from neutral atoms), a reference photograph of an IonQ trapped-ion device, and an original reference image of the ion trap apparatus (Figure~\ref{fig:ion-input}). These inputs establish the structural fidelity and physical accuracy of the generated scene.

\begin{figure}[ht]
\centering
\includegraphics[width=0.85\columnwidth]{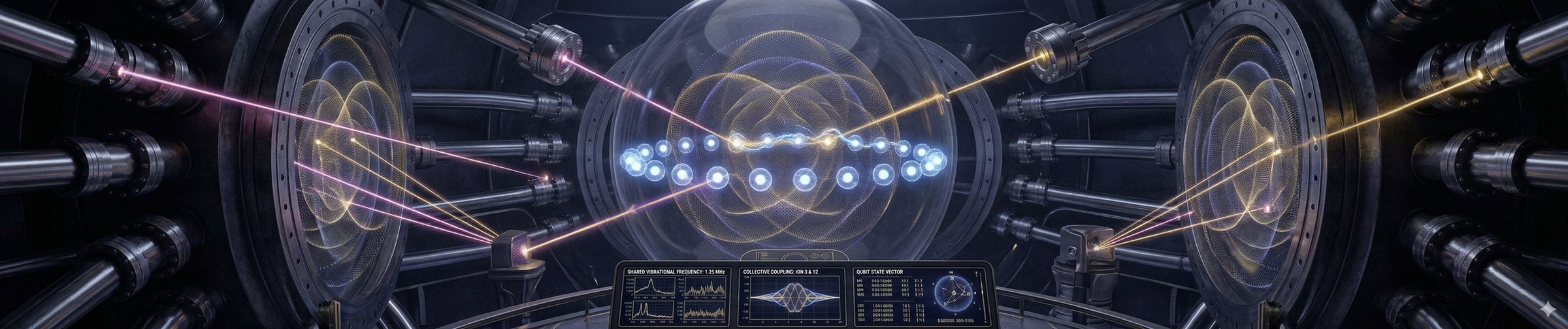}
\caption{AI-generated trapped-ion world model output. The scene shows a linear chain of ions in a Paul trap with intersecting Raman laser beams, synthesized from the inputs in Figure~\ref{fig:ion-input}.}
\label{fig:ion-output}
\end{figure}

\textbf{Output.} World Labs' generative world-model pipeline synthesizes a persistent, navigable 3D environment from these inputs (Figure~\ref{fig:ion-output}). The resulting world features a linear chain of ytterbium ions suspended in a Paul trap, with gold-violet Raman laser beams entering from multiple directions. A faint golden standing-wave field represents the shared vibrational mode. Two highlighted ions demonstrate entanglement through the shared medium.

\textbf{Navigable views.} Five representative viewpoints from the immersive 3D environment are shown in the top row of Figure~\ref{fig:teaser}.

% ============================================================
% Neutral Atoms
% ============================================================
\subsection{Neutral-Atom World Model}
\label{app:neutral-atoms}

\textbf{Scientific basis.} Neutral-atom quantum computers use focused laser beams (optical tweezers) to arrange individual neutral atoms in programmable two-dimensional arrays~\cite{bruzewicz2019}. By exciting atoms to highly excited Rydberg states, engineers exploit the Rydberg blockade effect---in which nearby atoms cannot simultaneously be excited---Rydberg interactions enabling programmable many-body
dynamics and analog Hamiltonian simulation. Key visualized phenomena include: the optical tweezer array, the Rydberg excitation glow, and the programmable atom geometry.

\begin{figure}[h]
\centering
\subfloat[Concept: atomic structure]{\includegraphics[width=0.32\columnwidth]{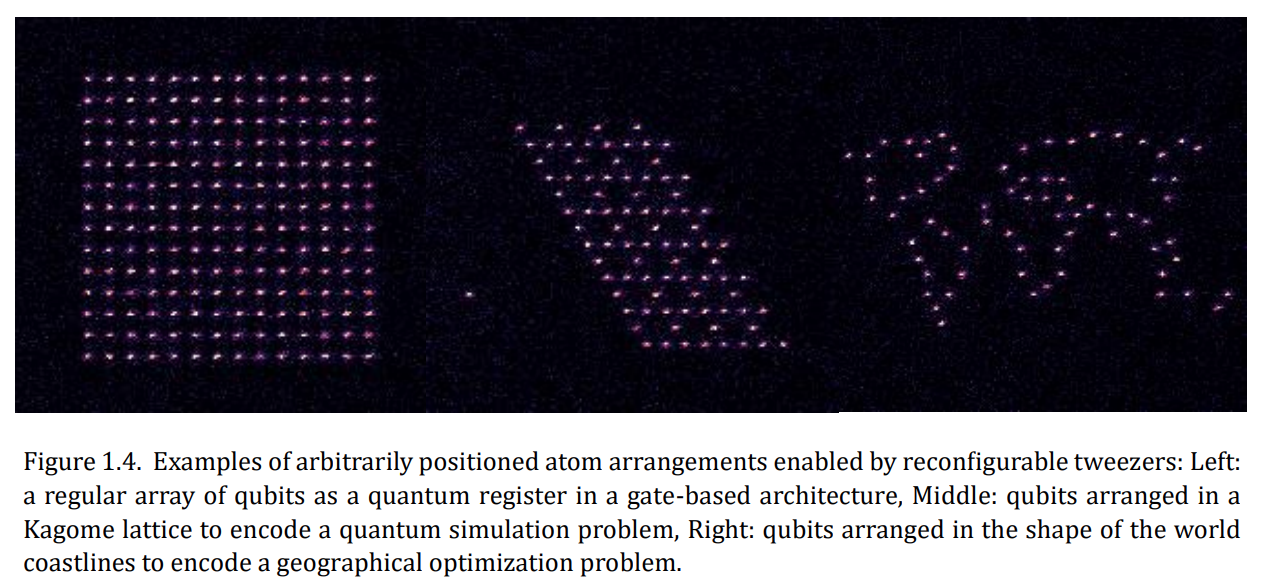}}\hfill
\subfloat[Device: QuEra on AWS Braket]{\includegraphics[width=0.32\columnwidth]{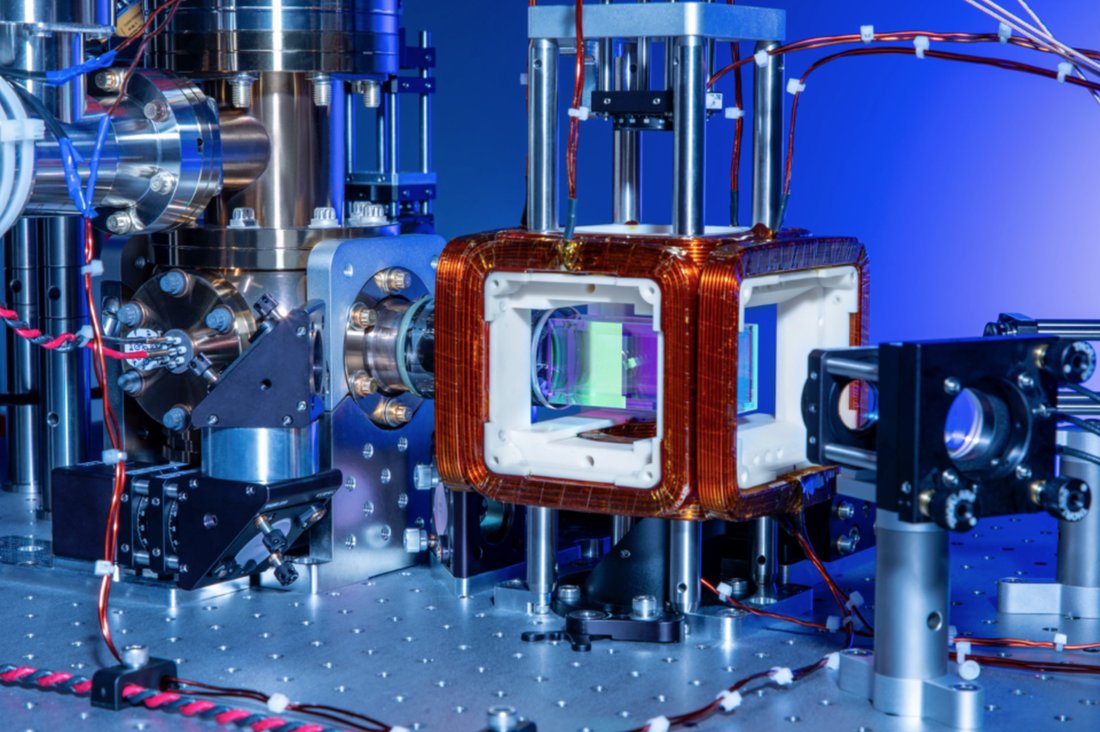}}\hfill
\subfloat[Device: QuEra HPCWire feature]{\includegraphics[width=0.32\columnwidth]{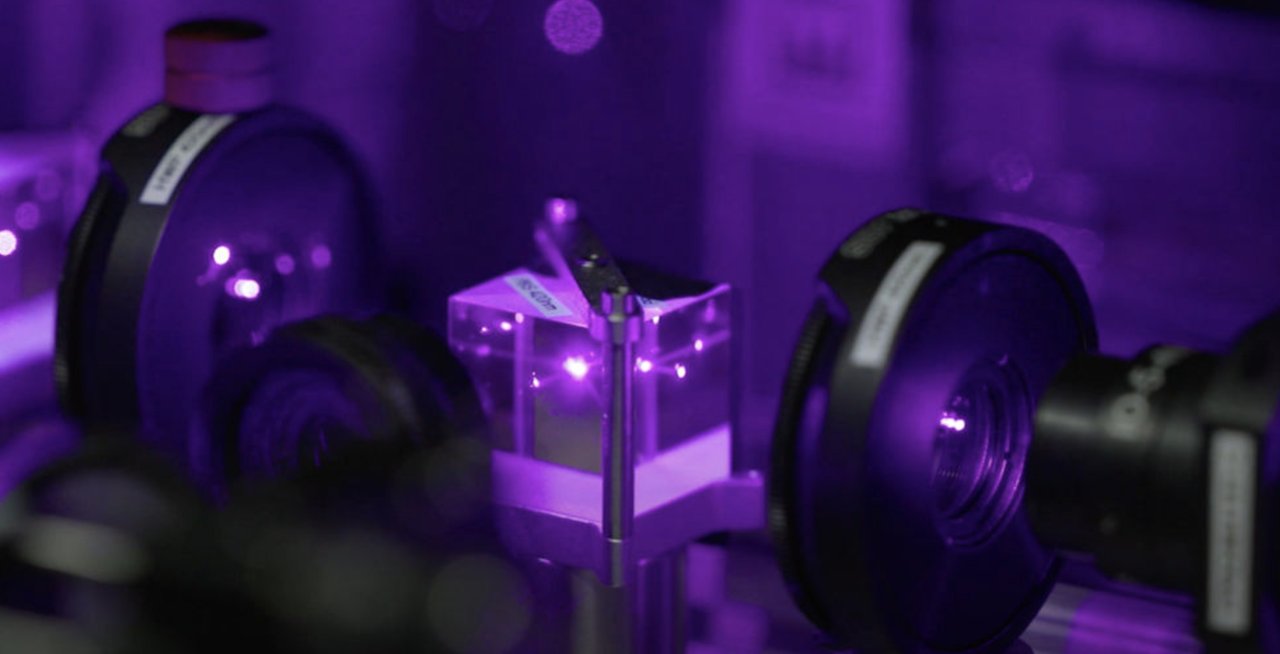}}
\caption{Input materials for the neutral-atom generative world model. (a)~Scientific concept illustration of atomic arrangements. (b)~Reference photograph of the QuEra neutral-atom device on AWS Braket. (c)~QuEra device as featured in HPCWire.}
\label{fig:atom-input}
\end{figure}

\textbf{Input.} The prompt combines a scientific concept illustration of atomic structure with multiple reference photographs of QuEra's neutral-atom device, including the AWS Braket deployment and the HPCWire-featured system (Figure~\ref{fig:atom-input}).

\begin{figure}[h]
\centering
\includegraphics[width=0.85\columnwidth]{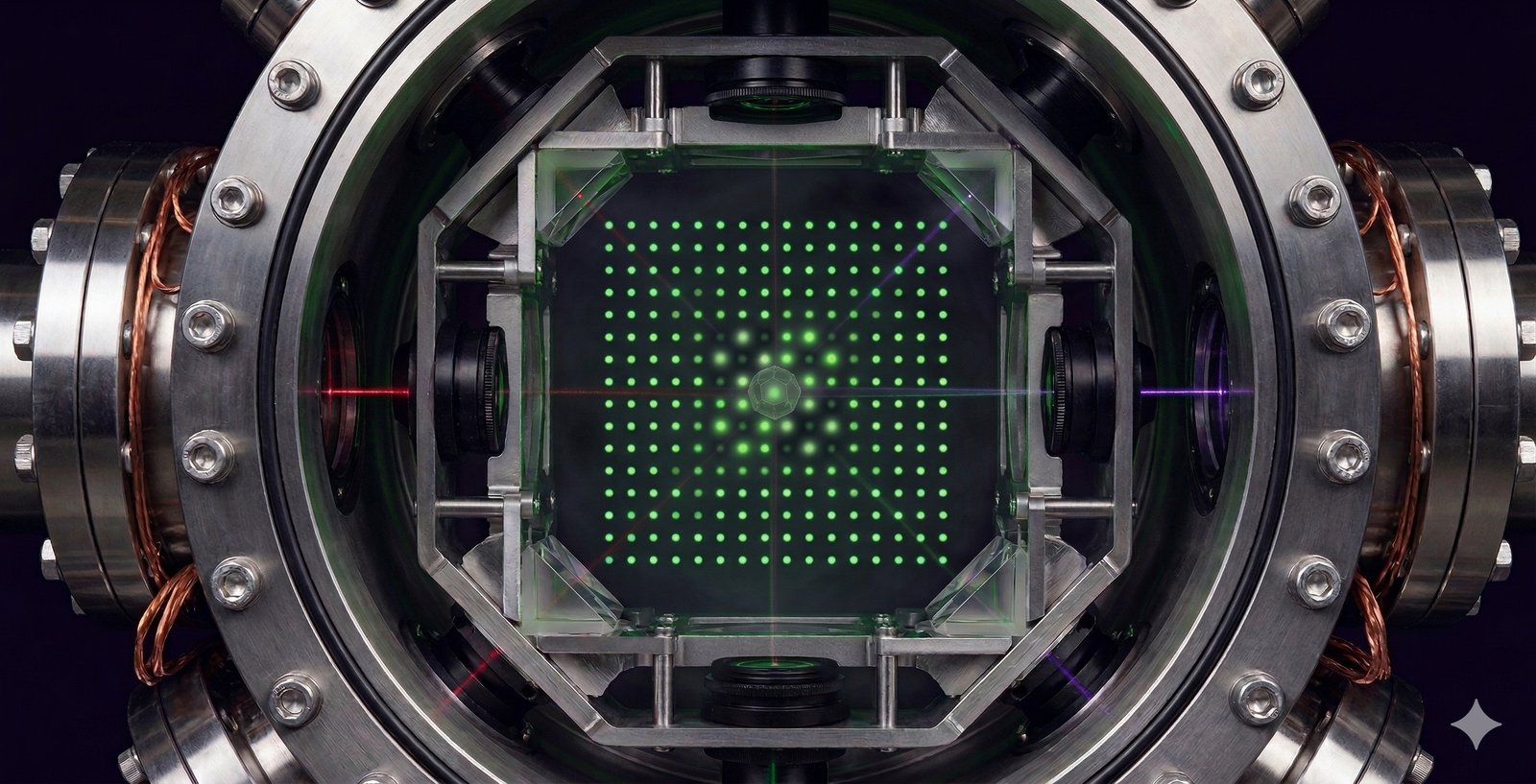}
\caption{AI-generated neutral-atom world model output. The scene shows a programmable rubidium atom array with optical tweezers, synthesized from the inputs in Figure~\ref{fig:atom-input}.}
\label{fig:atom-output}
\end{figure}

\textbf{Output.} The generated world (Figure~\ref{fig:atom-output}) presents a two-dimensional array of rubidium atoms held by red optical tweezers. A Rydberg excitation glow surrounds targeted atoms, and the programmable geometry---atoms arranged in various patterns---is visible and explorable.

\textbf{Navigable views.} Five representative viewpoints are shown in the middle row of Figure~\ref{fig:teaser}.

% ============================================================
% Superconducting
% ============================================================
\subsection{Superconducting World Model}
\label{app:superconducting}

\textbf{Scientific basis.} Superconducting quantum processors fabricate electrical circuits containing Josephson junctions---nanoscale superconducting weak links---and cool them to millikelvin temperatures inside dilution refrigerators~\cite{krantz2019quantum}. Microwave pulses transmitted through on-chip control lines manipulate the quantum state of each circuit element. Key visualized phenomena include: the Josephson junction circuits, the dilution refrigerator cryostat, the golden microwave waveguides, and the frost/ice crystals that form at cryogenic temperatures.

\begin{figure}[h]
\centering
\subfloat[Concept: Josephson effect]{\includegraphics[width=0.48\columnwidth]{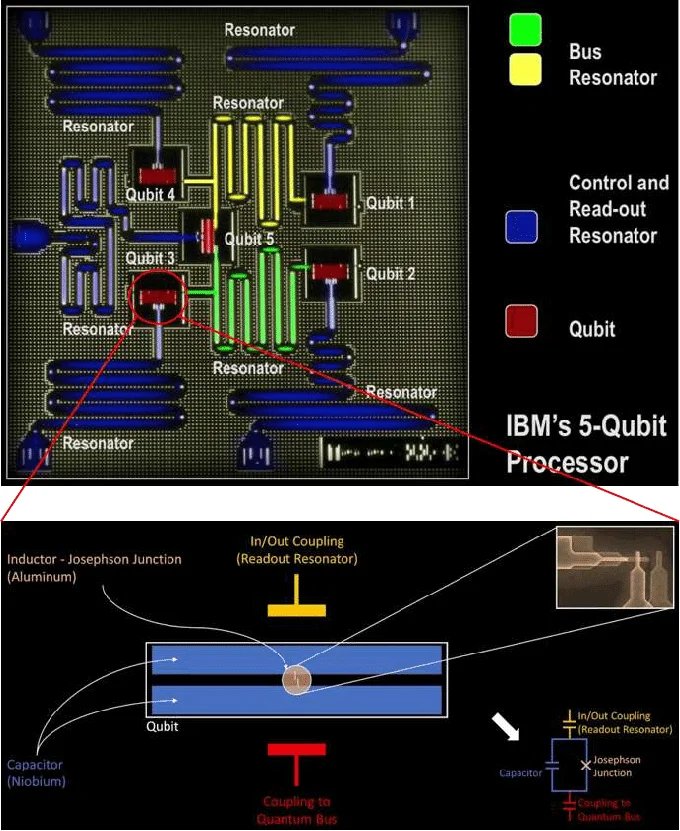}}\hfill
\subfloat[Device: Rigetti superconducting processor]{\includegraphics[width=0.48\columnwidth]{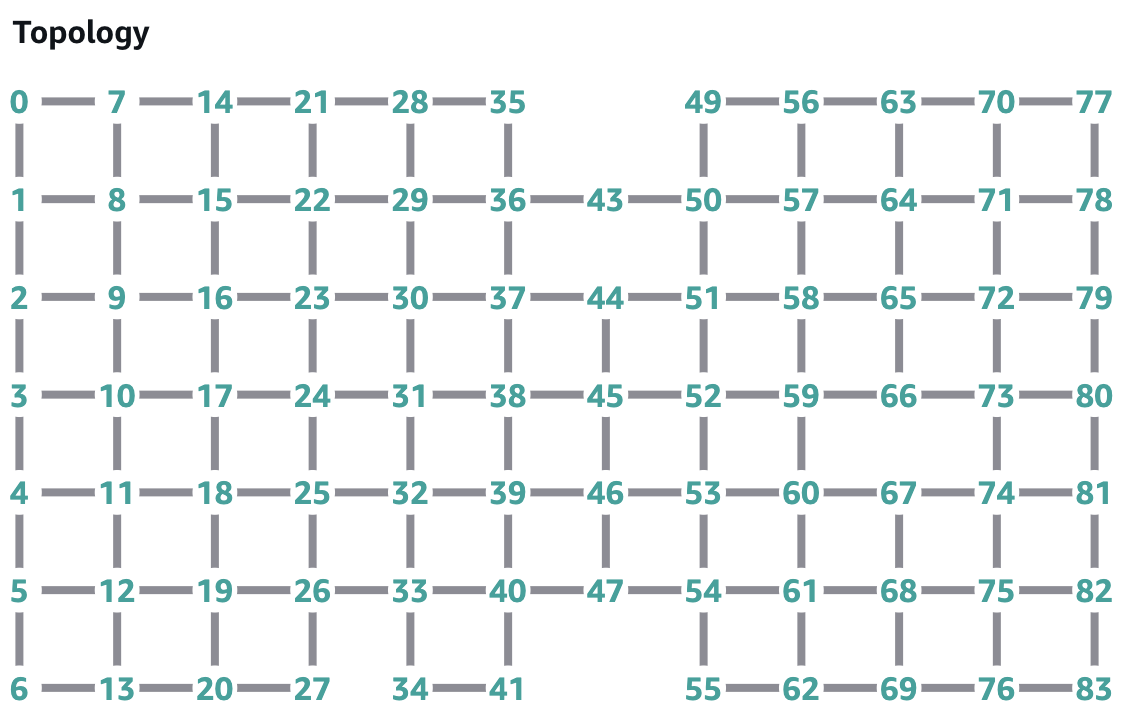}}
\caption{Input materials for the superconducting generative world model. (a)~Scientific concept illustration of the Josephson effect. (b)~Reference photograph of the Rigetti superconducting processor.}
\label{fig:super-input}
\end{figure}

\textbf{Input.} The prompt combines a scientific concept illustration of the Josephson effect with a reference photograph of Rigetti's superconducting processor (Figure~\ref{fig:super-input}).

\begin{figure}[h]
\centering
\includegraphics[width=0.85\columnwidth]{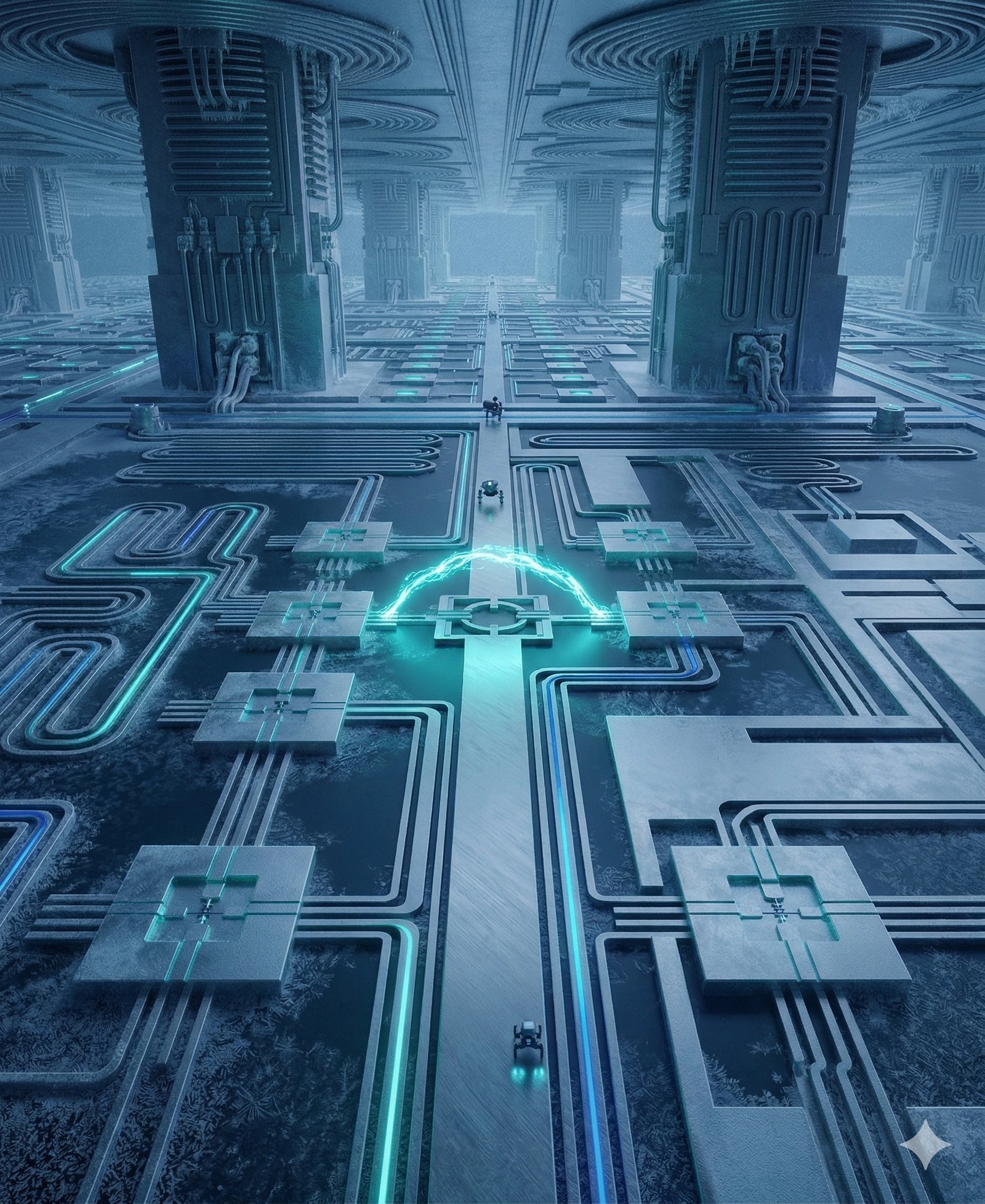}
\caption{AI-generated superconducting world model output. The scene shows a Josephson-junction chip in a dilution refrigerator with microwave waveguides and cryogenic infrastructure, synthesized from the inputs in Figure~\ref{fig:super-input}.}
\label{fig:super-output}
\end{figure}

\textbf{Output.} The generated world (Figure~\ref{fig:super-output}) shows a superconducting quantum processor chip mounted at the base of a dilution refrigerator. Golden microwave waveguides route control signals to individual qubits, and frost crystals on copper cooling stages visualize the cryogenic environment.

\textbf{Navigable views.} Five representative viewpoints are shown in the bottom row of Figure~\ref{fig:teaser}.